\documentclass[12pt, draftclsnofoot, onecolumn]{IEEEtran}

\usepackage{tabularx}
\usepackage[T1]{fontenc}
\usepackage{siunitx}
\usepackage{cite}
\usepackage{amsmath}
\usepackage{amssymb}
\usepackage{mathtools}
\usepackage{makecell}
\usepackage{boldline}
\usepackage{adjustbox}
\usepackage{caption}
\usepackage{algorithm}
\usepackage{algorithmic}
\usepackage{multirow}
\usepackage{url}

\usepackage{tikz}
\usepackage{pgfplots}
\usepackage{graphicx}
\pgfplotsset{compat=newest}
\usetikzlibrary{shapes,arrows,fit,positioning,calc}
\usetikzlibrary{plotmarks}
\usetikzlibrary{decorations.pathreplacing}
\usetikzlibrary{patterns}
\usetikzlibrary{dsp,chains,arrows,shapes,spy}

\usetikzlibrary{automata}
\usepackage{xcolor}

\usepackage{ragged2e}
\usepackage{hhline}

\usepackage[mathscr]{eucal}

\definecolor{b}{rgb}{0, 0, 0}
\definecolor{c}{rgb}{0, 0, 0}
\definecolor{r}{rgb}{0, 0, 0}
\definecolor{r2}{rgb}{0,0,0}

\definecolor{r3}{rgb}{0,0,0}

\definecolor{r4}{rgb}{0,0,0}

\DeclareMathAlphabet\mathbfcal{OMS}{cmsy}{b}{n}

\usepackage{bm}

\usepackage{comment}

\ifCLASSINFOpdf
\else
\fi
\usepackage{amsmath}
\begin{document}
%

%

\title{Multiuser-MIMO Systems Using Comparator Network-Aided 
Receivers With 1-Bit Quantization}

%
%

\author{Ana~Beatriz~L.~B.~Fernandes,
		{Zhichao~Shao},~\IEEEmembership{Member,~IEEE},	{Lukas~T.~N.~Landau,~\IEEEmembership{Senior Member,~IEEE}
		and
		{Rodrigo~C.~de~Lamare,~\IEEEmembership{Senior Member,~IEEE}}
		}
\thanks{A.~B.~L.~B.~Fernandes, L.~T.~N.~Landau and R.~C.~de Lamare are with the Centro de Estudos em Telecomunica\c{c}\~{o}es, Pontif\'{i}cia Universidade Cat\'{o}lica do Rio de Janeiro, Rio de Janeiro CEP 22453-900, Brazil, (e-mail:  anafernandes@cetuc.puc-rio.br;
lukas.landau@cetuc.puc-rio.br;
delamare@cetuc.puc-rio.br).
Z.~Shao is
with the National Key Laboratory of Science and Technology on Communications, University of Electronic Science and Technology of
China, Chengdu 611731, China (e-mail: zhichao.shao@uestc.edu.cn).
Parts of this work, have been presented on the 2020 54th Asilomar Conference on Signals, Systems, and Computers \cite{9443453} and on the IEEE Statistical Signal Processing Workshop 2021 (SSP) \cite{SSP2021}. 
\textcolor{r4}{This work is based on the thesis
\cite{mastersthesis_Fernandes}
of
A.~B.~L.~B.~Fernandes.}
This work has been supported by the {ELIOT ANR18-CE40-0030 and FAPESP 2018/12579-7} project and FAPERJ.
}
}
\maketitle


\begin{abstract}
Low-resolution analog-to-digital converters (ADCs) are promising for reducing energy consumption and costs of multiuser multiple-input multiple-output (MIMO) systems with many antennas. 
We propose low-resolution multiuser MIMO receivers where the signals are simultaneously processed by 1-bit ADCs and a comparator network, which can be interpreted as additional virtual channels with binary outputs. 
We distinguish the proposed comparator networks in fully and partially connected. For such receivers, we develop the low-resolution aware linear minimum mean-squared error (LRA-LMMSE) channel estimator and detector according to the Bussgang theorem. We also develop a robust detector which takes into account the channel state information (CSI) mismatch statistics. By exploiting knowledge of the channel coefficients we devise a mean-square error (MSE) greedy search and a sequential signal-to-interference-plus-noise ratio (SINR) search for optimization of partially connected networks. Numerical results show that a system with extra virtual channels can outperform a system with additional receive antennas, in terms of bit error rate (BER). Furthermore, by employing the proposed channel estimation with its error statistics, we construct a lower bound on the ergodic sum rate for a linear receiver. Simulation results confirm that the proposed approach outperforms the conventional 1-bit MIMO system in terms of BER, MSE and sum rate.
\end{abstract}

\begin{IEEEkeywords}
\textcolor{r3}{
Multiuser MIMO, 1-bit ADCs, Comparator Network, Channel Estimation, Signal Detection}
\end{IEEEkeywords}


%
\IEEEpeerreviewmaketitle

\section{Introduction}

Large-scale Multiple-Input Multiple-Output (MIMO) technology is a promising approach for future wireless communication networks that scale up in speed and bandwidth.
There are many advantages compared to current systems including high spectral efficiency and less propagation loss caused by channel fading, among others \cite{Larsson}. However, there are still some practical challenges when it comes to the deployment of a large number of antennas at the base station (BS), such as the hardware cost and power consumption. For instance, the power consumption of ADCs scales exponentially with the number of quantization bits \cite{Walden}. Therefore, the use of current high-speed and high-resolution ADCs (8-12 bits) for each antenna array would become a great burden to the BS. Consequently, the use of low-cost and low-resolution ADCs (1-3 bits) are promoted as a potential solution to this problem \cite{Jacobsson,8337813,8732599}.

\subsection{Motivation and Prior Works}

Many works have studied large-scale multiuser MIMO systems with low resolution ADCs at the front-end. Specifically, 1-bit ADCs are of interest in such systems due to their low power consumption. A commonly used technique to mitigate the performance loss caused by the coarse quantization is oversampling, where the received signal is sampled at a rate faster than the Nyquist rate \cite{7837644}. In this regard, the studies in \cite{8907816,8450809,9085989, 9354165} have considered temporal oversampling at the receiver in order to achieve better channel estimation and detection performance. However, for temporal oversampling, the computational complexity is relatively high for receive processing.  

Channel estimation is another problem that currently limits the performance of coarsely quantized systems. Several papers have investigated channel estimation for quantized massive MIMO systems such as least squares (LS) \cite{risi2014massive},  recursive least squares (RLS) \cite{8385500}, approximate message passing (AMP) \cite{8171203} and generalized approximate message passing (GAMP) \cite{7925515}. Another sophisticated channel estimator is given by the near maximum likelihood (nML) estimator devised in \cite{7439790}. The authors in \cite{7931630} have developed a Bussgang linear minimum mean squared error (BLMMSE) channel estimator, where lower bounds on the theoretical achievable rate for maximum ratio combiner (MRC) and zero-forcing (ZF) receivers are derived.
The linear Bussgang based channel estimation was generalized for 1-bit ADCs with nonzero thresholds in \cite{Wan2020}, where the authors also devised the optimization of the thresholds.
Moreover, maximum a-posteriori probability (MAP) channel estimators and the corresponding performance analysis have been studied in \cite{8314750, Kim2019}.
The work in \cite{Kim2019_2} develops a channel estimator by taking spatial and temporal correlations into consideration. Furthermore, the low complexity channel estimator and its corresponding achievable rate from \cite{7600443} rely on a model with infinite number of channel taps, and independent and identically distributed quantization noise.

Several works related to signal detection in systems with low-resolution ADCs are: iterative detection and decoding (IDD) \cite{8240730}, sphere decoding \cite{8345169} and nML \cite{7439790}. Moreover, coarse quantization in large-scale MIMO systems has been considered in a number of different aspects and approaches. Recently, the authors in \cite{9110866,8682842} have considered spatial oversampling by employing a 1-bit Sigma-Delta ($\Sigma\Delta$) sampling scheme which has shown large performance gains on channel estimation and signal detection.
For the case of receivers with multi-bit ADCs, the design of a proper automatic gain control (AGC) can be relevant in order to minimize the signal distortion due to the coarse quantization \cite{8491040}.
Furthermore, the studies in \cite{7472304,8010806,9097110,9084109} have devised different precoding techniques which rely on coarse quantization at the transmitter.

\subsection{Main Contributions}
The scope of this study lies in the design of a novel multiuser MIMO receiver architecture that uses 1-bit ADCs. The proposed multiuser MIMO receiver includes a comparator network with binary outputs which can compare signals from different antennas in terms of real and imaginary parts. The additional comparator network can be interpreted as additional virtual channels that aid further receive processing such as
detection and channel estimation.
As the virtual channels can be seen as an alternative to additional receive antennas the proposed approach is especially promising for receivers with limited number of antennas due to space constraints.

Two types of comparator networks are proposed, i.e., fully and partially connected networks. 
The design of partially connected networks
is a combinatorial problem, which can be computationally very costly.
Two suboptimal design methods are proposed in terms of a MSE based Greedy Search-based algorithm and a sequential SINR based search algorithm, which require much less comparators to approach a comparable detection performance as the performance with fully connected networks. Numerical results confirm that detection and channel estimation performance can be significantly improved when adding a comparator network to the conventional 1-bit multiuser MIMO receiver. Based on the proposed detector, a sum rate analysis is presented. Simulation results show that the proposed
comparator network based-systems yield significantly higher sum rate performance in comparison to the conventional 1-bit multiuser MIMO receiver.

The contributions of this work can be summarized as follows:
\begin{itemize}
    \item Proposal of a novel multiuser MIMO receiver architecture with 1-bit ADCs and partially or fully connected comparator networks.
    \item Development of the statistically equivalent linear model for the novel receiver with comparator network.
    \item Development of a linear MSE based channel estimation for the novel receiver with comparator network and calculation of the channel estimation error statistics.
    \item Development of a linear MSE based detector for the novel receiver and a robust version that takes into account channel estimation error statistics.
    \item Sum rate analysis in terms of construction of a lower bound.
    \item A study of cost and power consumption of the proposed architecture.
\end{itemize}
Note also that contributions of the proposed study go significantly beyond the contributions of the associated conference papers 
\cite{9443453}, \cite{SSP2021} and the thesis \cite{mastersthesis_Fernandes}, in terms of the generalized robust detection method in Section~\ref{sec:Robust_general}, a more practical approach for optimization of comparator networks in Section~\ref{sec:SINR_seq_search}, the study about the power consumption in Section~\ref{sec:power} and simulation results with more realistic scenarios.


\subsection{Organization of the Paper and Notation}

The remainder of this paper is organized as follows: Section~\ref{sec:system_model} describes the general system model, whereas Section~\ref{sec:channel_estimation} derives  the LRA-LMMSE channel estimator for the proposed system. Section \ref{sec:linear_detection} details the derivation of the LRA-LMMSE detector and a robust version of it, while Section \ref{sec:design} describes the insight and the design of the comparator network and Section \ref{sec:complexity} shows the complexity analysis of the proposed system. Section \ref{sec:achievable_rates} presents the construction of the lower bound on the sum rate. Then, Section~\ref{sec:numerical_results} presents and discusses the numerical results. Finally, Section \ref{sec:conclusion} gives the conclusions.

Regarding the notation, bold upper and lower case letters such as $\mathbf{A}$ and $\mathbf{a}$ denote matrices and vectors, respectively. $\mathbf{I}_n$ is a $n\times n$ identity matrix. The vector or matrix transpose and conjugate transpose are represented by $(\cdot)^T$ and $(\cdot)^H$. $\mathfrak{R}\{\cdot\}$ and $\mathfrak{I}\{\cdot\}$ get the real and imaginary part from the corresponding vector or matrix, respectively. Additionally, $\text{diag}(\mathbf{A})$ is a diagonal matrix only containing the diagonal elements of $\mathbf{A}$. The inverse of the sine function is denoted by $\text{sin}^{-1}(\cdot)$. Moreover, $\text{vec}(\mathbf{A})$ is the vectorized form of $\mathbf{A}$ obtained by stacking its columns, while the inverse of this operation is $\text{unvec}(\mathbf{a})$, depending on the context. The expectation operator is denoted as $E[\cdot]$. The Kronecker product is denoted by $\otimes$. 

\section{System Model}
\label{sec:system_model}
\begin{figure*}[t]
	\centering
	\input{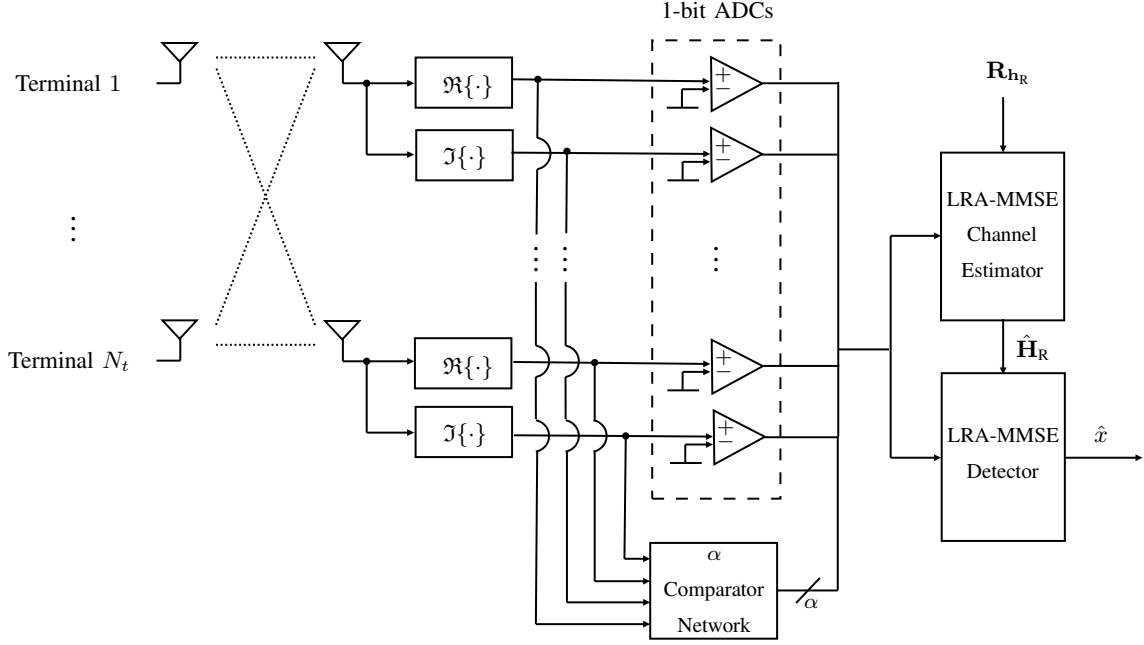}
	\caption{System model of comparator network based 1-bit MIMO systems.}
	\label{fig:transmitter}
\end{figure*}

The general system model for the proposed comparator network-aided 1-bit multiuser MIMO system is illustrated with a block diagram in Fig.~\ref{fig:transmitter},
where the received signal $\mathbf{y}$, of dimensions ${(N_r\times 1)}$, for the uplink of a single-cell multiuser MIMO system with $N_t$ single-antenna users and $N_r$ receive antennas is written as
\begin{equation}
\mathbf{y}=\mathbf{H}\mathbf{x}+\mathbf{n}.
\label{equ_model}
\end{equation}
The vector $\mathbf{x}$ contains complex transmit symbols of the $N_t$ users with transmit power $\mathrm{E}\{  \mathfrak{R}\{x_i \}^2 \}=  \mathrm{E}\{  \mathfrak{I}\{x_i \}^2 \}=   \frac{1}{2}  \sigma_x^2$, $\mathbf{H}\in\mathbb{C}^{N_r\times N_t}$ is the channel matrix and $\mathbf{n}\in\mathbb{C}^{N_r\times 1}$ is the noise vector where each element is complex Gaussian distributed with $\mathcal{CN}(0,\sigma_n^2)$.

\begin{figure}[!htbp]
	\centering
	\scalebox{0.82}{\tikzset{%
	harddecision/.style={draw, 
		path picture={
			\pgfpointdiff{\pgfpointanchor{path picture bounding box}{north east}}%
			{\pgfpointanchor{path picture bounding box}{south west}}
			\pgfgetlastxy\x\y
			\tikzset{x=\x*.4, y=\y*.4}
			%
			\draw (-0.5,-0.5)--(0,-0.5)--(0,0.5)--(0.5,0.5);  
			\draw (-0.25,0)--(0.25,0);
	}}
}

	\begin{tikzpicture}
	\node (c0) {};
	\node[coordinate,below=1.4cm of c0] (c1) {};
	\node[coordinate,right=1cm of c1] (c2) {};
	\node[coordinate,right=0.5cm of c2] (c3) {};
	\node[coordinate,right=0.5cm of c3] (c6) {};
	
	\node[coordinate,below=0.4cm of c1] (c02) {};
	\node[coordinate,right=1cm of c02] (c22) {};
	\node[coordinate,right=0.5cm of c22] (c62) {};
	\node[coordinate,right=1cm of c22, yshift=-0.2cm] (c32) {};
	\node[coordinate,right=0.5cm of c32] (c4) {};
	\node[coordinate,above=0.3cm of c4] (c43) {};
	\node[coordinate,below=0.25cm of c4] (c45) {};
	\node[coordinate,right=1.1cm of c45] (c46) {};
	\node[coordinate,right=1cm of c46] (c47) {};
	\node[coordinate,right=0.1cm of c4] (c48) {};
	\node[coordinate,right=0.1cm of c48] (c49) {};
	\node[coordinate,right=0.2cm of c48] (c5) {};
	\node[coordinate,above=0.1cm of c49] (c492) {};
	\node[coordinate,below=0.1cm of c49] (c493) {};
	
	\node[coordinate,below=0.4cm of c02] (c03) {};
	\node[coordinate,right=1cm of c03] (c23) {};
	\node[coordinate,right=0.5cm of c23] (c64) {};
	\node[coordinate,right=1cm of c23, yshift=-0.25cm] (c33) {};
	\node[coordinate,right=0.5cm of c33] (c42) {};
	\node[coordinate,below=0.3cm of c42] (c44) {};
	\node[coordinate,right=0.1cm of c42] (c494) {};
	\node[coordinate,right=0.2cm of c494] (c495) {};

	\node[coordinate,below=0.8cm of c03] (c04) {};
	\node[coordinate,right=1cm of c04] (c24) {};
	\node[coordinate,right=0.5cm of c24] (c34) {};
	\node[coordinate,right=0.5cm of c34] (c63) {};
	
	\node[coordinate,right=0.4cm of c0,yshift=-1cm] (c8) {};
	\node[coordinate,right=3.6cm of c8] (c82) {};
	\node[coordinate,below=2.6cm of c8] (c83) {};
	\node[coordinate,right=3.6cm of c83] (c84) {};
	\node[coordinate,below=2.6cm of c82] (c85) {};
	
	\node[coordinate,right=0.9cm of c0,yshift=-0.5cm] (c86) {};
	\node[coordinate,right=3.5cm of c86] (c87) {};
	\node[coordinate,below=0.5cm of c86] (c88) {};
	\node[coordinate,below=2.6cm of c87] (c89) {};
	\node[coordinate,left=0.4cm of c89] (c9) {};
	
	\node[coordinate,below=0.8cm of c82] (c97) {};
	\node[coordinate,right=0.9cm of c97] (c98) {};
	
	\node[coordinate,right=1.3cm of c0] (c92) {};
	\node[coordinate,right=3.5cm of c92] (c93) {};
	\node[coordinate,below=0.5cm of c92] (c94) {};
	\node[coordinate,below=2.6cm of c93] (c95) {};
	\node[coordinate,left=0.4cm of c95] (c96) {};
	
	\node[coordinate,below=0.8cm of c87] (c99) {};
	\node[coordinate,right=0.8cm of c99] (c10) {};

	\draw[thick] (c1) -- node[] {} (c2);
	\draw node at (1,-1.58) {\scriptsize \textbullet};
	\draw[thick] (c02) -- node[] {} (c22);
	\draw node at (1,-1.98) {\scriptsize \textbullet};
	\draw[thick] (c03) -- node[] {} (c23);
	\draw node at (1,-2.37) {\scriptsize \textbullet};
	\draw[thick] (c04) -- node[] {} (c24);
	\draw node at (1,-3.17) {\scriptsize \textbullet};
	
	\draw node at (1,-2.6) {\tiny \textbullet};
	\draw node at (1,-2.76) {\tiny \textbullet};
	\draw node at (1,-2.92) {\tiny \textbullet};
	
	\draw node at (2,-2.15) {\scriptsize \textbullet};
	\draw[thick] (c3) -- node[] {} (c32);
	\draw node at (2,-2.62) {\scriptsize \textbullet};
	\draw[thick] (c33) -- node[] {} (c34);
	\draw[thick] (c32) -- node[] {} (c4);
	\draw[thick] (c33) -- node[] {} (c42);
	\draw[thick] (c43) -- node[] {} (c44);
	\draw[thick] (c43) -- node[] {} (c46);
	\draw[thick] (c44) -- node[] {} (c46);
	\draw[thick] (c46) -- node[] {} (c47);
	\draw node at (4.6,-2.42) {\scriptsize \textbullet};
	
	\draw[thick] (c48) -- node[] {} (c5);
	\draw[thick] (c492) -- node[] {} (c493);
	\draw[thick] (c494) -- node[] {} (c495);
	
	\draw[-stealth,shorten >= 2pt] (c6.north) to[bend right] node[midway,above] {} (c62.north);
	\draw[-stealth,shorten >= 2pt] (c63.north) to[bend left] node[midway,above] {} (c64.north);
	
	\draw[thick] (c8) -- node[] {} (c82);
	\draw[thick] (c8) -- node[] {} (c83);
	\draw[thick] (c83) -- node[] {} (c84);
	\draw[thick] (c82) -- node[] {} (c85);

	\draw[thick] (c86) -- node[] {} (c87);
	\draw[thick] (c86) -- node[] {} (c88);
	\draw[thick] (c87) -- node[] {} (c89);
	\draw[thick] (c89) -- node[] {} (c9);
	\draw[thick] (c97) -- node[] {} (c98);
	\draw node at (5,-1.82) {\scriptsize \textbullet};
	
	\draw[thick] (c92) -- node[] {} (c93);
	\draw[thick] (c92) -- node[] {} (c94);
	\draw[thick] (c93) -- node[] {} (c95);
	\draw[thick] (c95) -- node[] {} (c96);
	\draw[thick] (c99) -- node[] {} (c10);
	\draw node at (5.3,-1.32) {\scriptsize \textbullet};
	
	\draw node at (1.3,-0.15) {\tiny \textbullet};
	\draw node at (1.2,-0.25) {\tiny \textbullet};
	\draw node at (1.1,-0.35) {\tiny \textbullet};

	\end{tikzpicture}}
	\caption{Schematic of the comparator network.}
	\label{fig:comparator}
\end{figure}
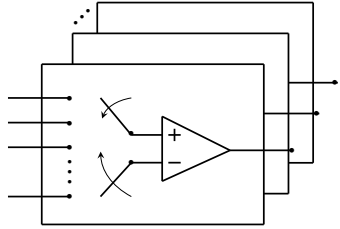

The real-valued system model can be written as 
\begin{equation}
\begin{bmatrix}
\mathfrak{R}\{\mathbf{y}\} \\
\mathfrak{I}\{\mathbf{y}\}
\end{bmatrix}=\begin{bmatrix} 
\mathfrak{R}\{\mathbf{H}\} &-\mathfrak{I}\{\mathbf{H}\}\\
\mathfrak{I}\{\mathbf{H}\} & \mathfrak{R}\{\mathbf{H}\}
\end{bmatrix}\begin{bmatrix} 
\mathfrak{R}\{\mathbf{x}\} \\
\mathfrak{I}\{\mathbf{x}\}
\end{bmatrix}+\begin{bmatrix} 
\mathfrak{R}\{\mathbf{n}\} \\
\mathfrak{I}\{\mathbf{n}\}
\end{bmatrix}\text{.}
\label{equ_real_sys}
\end{equation}
A more compact notation for (\ref{equ_real_sys}) reads as
\begin{equation}
    \mathbf{y}_{\text{R}} = \mathbf{H}_{\text{R}}\mathbf{x}_{\text{R}}+\mathbf{n}_{\text{R}},
\label{real_sys_compact}
\end{equation}
where $\mathbf{y}_{\text{R}}\in\mathbb{R}^{2N_r\times 1}$, $\mathbf{H}_{\text{R}}\in\mathbb{R}^{2N_r\times 2N_t}$ and $\mathbf{x}_{\text{R}}\in\mathbb{R}^{2N_t\times 1}$.
The received signal is then forwarded to the 1-bit ADCs and the comparator network (shown in Fig. \ref{fig:comparator}). Each comparator compares two received signals and quantizes the difference as $\left\{\pm 1 \right\}$.
Accordingly, each comparator process can be divided into a subtraction and a quantization step. The subtractions are described by the multiplication of $\mathbf{y}_{\text{R}}$ with the matrix $\mathbf{B}'$ and the quantizations are equivalent to the quantization operations that describe the 1-bit ADCs.

Letting $\mathcal{Q}(\cdot)$ represent the 1-bit quantization, the output of the ADCs and the comparator network is described by
\begin{equation}
\mathbf{z}_{\mathcal{Q}}= \mathcal{Q}\left(\begin{bmatrix}
\mathbf{y}_{\text{R}} \\
\mathbf{B}'\mathbf{y}_{\text{R}}\end{bmatrix}\right)=\mathcal{Q}\left(\begin{bmatrix}
\mathbf{I}_{2N_r} \\
\mathbf{B}'\end{bmatrix}\mathbf{y}_{\text{R}}\right),
\label{system_model}
\end{equation}
where $\mathbf{B}'\in\mathbb{R}^{\alpha\times 2N_r}$ refers to the comparator network with the form
\begin{equation}
    \mathbf{B}'=
    \begin{bmatrix}
    \mathbf{B}'_{\text{R}}  &  \mathbf{B}'_{\text{I}}    \end{bmatrix}
 =   {\frac{1}{\sqrt{2}}}
    \begin{bmatrix}
 1&-1&0&0&\cdots&0\\  0&0&0&-1&\cdots&1\\ \vdots&\vdots&\vdots&\vdots&\vdots&\vdots\\
 0&1&0&0&\cdots&-1\\
 \end{bmatrix}.
 \label{B_prime}
\end{equation}
In each row of $\mathbf{B}'$, there is only one pair of 1 and -1 and the remaining entries are zeros.
Each row is associated with a comparator which compares the received signal addressed with the position of the 1 with another received signal addressed by the position of the -1.
With $\mathbf{B}=\begin{bmatrix}
\mathbf{I}_{2N_r}; \mathbf{B}'\end{bmatrix}$, (\ref{system_model}) reads as
\begin{equation}
\mathbf{z}_\mathcal{Q}= \mathcal{Q}(\mathbf{z}_{\text{R}}) =\mathcal{Q}\left(\mathbf{By}_{\text{R}}\right).
\label{new_system_model}
\end{equation}


\section{Channel Estimation}
\label{sec:channel_estimation}
In this section, we describe pilot-based channel estimation and 
propose the statistically equivalent linear model for the novel receiver with 1-bit ADCs and the comparator network, which is then utilized for the proposed LRA-MMSE channel estimation.
We then characterize the MSE performance associated with the channel estimates.

\subsection{Pilot-Based Channel Estimation and the Proposed Linear Model}

A common technique for channel estimation is to let the users transmit orthogonal pilot sequences with length $\tau$ and evaluate the effect of the channel on these symbols at the BS. During the training phase, 
the received signal at the BS can be described by
\begin{equation}
\mathbf{Y}_p= \mathbf{H}\mathbf{\Phi}^T+\mathbf{N}_p,
\label{eq:Yp}
\end{equation} 
where $\mathbf{Y}_p \in \mathbb{C}^{N_r\times \tau}$ is the matrix containing the unquantized received signal, $\mathbf{\Phi} \in \mathbb{C}^{\tau\times N_t}$ is the matrix of pilot symbols and $\mathbf{N}_p\in \mathbb{C}^{N_r\times \tau}$ is the noise matrix. Vectorizing the signal in \eqref{eq:Yp} yields
\begin{equation}
\text{vec}(\mathbf{Y}_p) = \mathbf{y}_p = \tilde{\mathbf{\Phi}}\mathbf{h} +\mathbf{n}_p,
\end{equation} 
where $\tilde{\mathbf{\Phi}}= (\mathbf{\Phi}\otimes\mathbf{I}_{N_r})\in\mathbb{C}^{\tau N_r\times N_rN_t}$, $\mathbf{h} = \text{vec}(\mathbf{H})\in\mathbb{C}^{N_rN_t \times 1}$ and  $\mathbf{n}_p = \text{vec}(\mathbf{N}_p)\in\mathbb{C}^{\tau N_r \times 1}$. A real-valued representation of the system is defined by
\begin{equation}
\begin{split}
\mathbf{y}_{\text{R}_p} & =
\begin{bmatrix} 
\mathfrak{R}\{\tilde{\mathbf{\Phi}}\} &-\mathfrak{I}\{\tilde{\mathbf{\Phi}}\}\\
\mathfrak{I}\{\tilde{\mathbf{\Phi}}\} & \mathfrak{R}\{\tilde{\mathbf{\Phi}}\}
\end{bmatrix}
\begin{bmatrix}
\mathfrak{R}\{\mathbf{h}\} \\
\mathfrak{I}\{\mathbf{h}\}
\end{bmatrix} +
\begin{bmatrix}
\mathfrak{R}\{\mathbf{n}_p\} \\
\mathfrak{I}\{\mathbf{n}_p\}
\end{bmatrix} = \tilde{\mathbf{\Phi}}_{\text{R}}\mathbf{h}_{\text{R}} +\mathbf{n}_{{\text{R}}_p},
\end{split}
\label{eq_real}
\end{equation} 
where $\mathbf{y}_{\text{R}_p} \in \mathbb{R}^{\tau(2 N_r) \times 1}$ is the real-valued received signal vector. Then, when we multiply (\ref{eq_real}) with an effective comparator network matrix $\mathbf{B}_{\text{eff}}\in \mathbb{R}^{\tau(2 N_r + \alpha)\times \tau(2 N_r)}$, we obtain the vector
\begin{equation}
\mathbf{z}_{{\text{R}}_p} =\mathbf{B}_{\text{eff}}\mathbf{y}_{{\text{R}}_p}=\mathbf{B}_{\text{eff}}\tilde{\mathbf{\Phi}}_{\text{R}}\mathbf{h}_{\text{R}} +\mathbf{B}_{\text{eff}}\mathbf{n}_{{\text{R}}_p},
\label{eq:zrp}
\end{equation} 
\textcolor{r3}{
of dimensions ${(\tau(2 N_r+\alpha)\times 1)}$},
where $\mathbf{B_{eff}}$ is described by 
\begin{equation}
\mathbf{B}_{\text{eff}} = \begin{bmatrix}
\mathbf{I}_{\tau(2 N_r)} \\
\mathbf{B}'_{\text{eff}}\end{bmatrix} =
\begin{bmatrix}
\mathbf{I}_{\tau(2 N_r)} \\
\begin{bmatrix}
 \mathbf{B}'_{\text{eff, R}}  &  \mathbf{B}'_{\text{eff, I}}    \end{bmatrix}\end{bmatrix},
\end{equation} 
with $\mathbf{B}'_{\text{eff, R}}= (\mathbf{B}'_{\text{R}}\otimes\mathbf{I_{\tau}})\in \mathbb{R}^{\alpha\tau\times N_r\tau}$ and $\mathbf{B}'_{\text{eff, I}}= (\mathbf{B}'_{\text{I}}\otimes\mathbf{I_{\tau}})\in \mathbb{R}^{\alpha\tau\times N_r\tau}$.
Moreover, $\mathbf{B}'_{\text{R}}$ and $\mathbf{B}'_{\text{I}}$ denote the parts of the matrix $\mathbf{B'}$ which are associated to the real and imaginary parts of the received signal, as shown in (\ref{B_prime}). 
According to \eqref{new_system_model}, the quantized signal, \textcolor{r3}{
of dimensions ${(\tau(2 N_r+\alpha)\times 1)}$},
is given by
\begin{equation}
\mathbf{z}_{\mathcal{Q}_p}= \mathcal{Q} (\mathbf{z}_{{\text{R}}_p}) = \hat{\mathbf{\Phi}}_{\text{R}} \mathbf{h}_{\text{R}} + \tilde{\mathbf{n}}_{{\text{R}}_p},
\label{eq:zqp}
\end{equation} 
where the right hand side corresponds to a linear model that relies on the Bussgang decomposition approach,
and the linear operator is
 $\hat{\mathbf{\Phi}}_{\text{R}} = \mathbf{A}_{{\text{R}}_p} \mathbf{B}_{\text{eff}} \tilde{\mathbf{\Phi}}_{\text{R}}\in \mathbb{R}^{\tau(2N_r+\alpha)\times 2N_rN_t}$. The effective noise vector $\tilde{\mathbf{n}}_{{\text{R}}_p} = \mathbf{A}_{{\text{R}}_p} \mathbf{B}_{\text{eff}}\mathbf{n}_{{\text{R}}_p} + \mathbf{n}_{q, p}$ relies on a well-chosen square matrix $\mathbf{A}_{{\text{R}}_p}\in\mathbb{R}^{\tau(2 N_r+\alpha)\times \tau(2 N_r+\alpha)}$ and the quantization noise $\mathbf{n}_{q, p}$. The matrix $\mathbf{A}_{{\text{R}}_p}$ makes the quantization noise uncorrelated with $\mathbf{z}_{{\text{R}}_p}$ \cite{Mezghani2012CapacityLB}, and is given by
\begin{equation}
\mathbf{A}_{{\text{R}}_p} = \mathbf{C}_{\mathbf{z}_{{\text{R}}_p}\mathbf{z}_{\mathcal{Q}_p}}^H \mathbf{C}_{\mathbf{z}_{{\text{R}}_p}}^{-1} = \sqrt{\frac{2}{\pi}}\mathbf{K}_{{\text{R}}_p} ,
\label{eq:A_rp}
\end{equation} 
with 
$\mathbf{K}_{{\text{R}}_p}=
\text{diag}(\mathbf{C}_{\mathbf{z}_{{\text{R}}_p}})^{-\frac{1}{2}}
$.
The cross-correlation matrix between the received signal $\mathbf{z}_{{\text{R}}_p}$ and its quantized signal $\mathbf{z}_{\mathcal{Q}_p}$, \textcolor{r3}{
of dimensions ${(\tau(2 N_r+\alpha)\times (\tau(2 N_r+\alpha))}$},
reads as
\begin{equation}
\mathbf{C}_{\mathbf{z}_{{\text{R}}_p}\mathbf{z}_{\mathcal{Q}_p}} = \sqrt{\frac{2}{\pi}}\mathbf{K}_{{\text{R}}_p}\mathbf{C}_{\mathbf{z}_{{\text{R}}_p}} .
\end{equation} 
The matrix $\mathbf{C}_{\mathbf{z}_{{\text{R}}_p}}$, \textcolor{r3}{
of dimensions ${(\tau(2 N_r+\alpha)\times (\tau(2 N_r+\alpha))}$},
denotes the auto-correlation matrix of $\mathbf{z}_{{\text{R}}_p}$
\begin{equation}
\begin{split}
\mathbf{C}_{\mathbf{z}_{{\text{R}}_p}} =  \mathbf{B}_{\text{eff}}
(
\tilde{\mathbf{\Phi}}_{\text{R}}
\mathbf{R_{h_{\text{R}}}}
\tilde{\mathbf{\Phi}}_{\text{R}}^T
+ \frac{1}{2} \sigma_n^2   \mathbf{I}_{\tau(2 N_r)}  )
\mathbf{B}_{\text{eff}} ^T,
\label{eq:Czrp}
\end{split}
\end{equation} 
where $\mathbf{R_{h_{\text{R}}}} = E \left[ {\mathbf{h}_{\text{R}} \mathbf{h}_{\text{R}}^T} \right]$.  In \eqref{eq:Czrp} it is considered $E \left[ {\mathbf{n}_{{\text{R}}_p} \mathbf{n}_{{\text{R}}_p}^T} \right] = \frac{1}{2} \sigma_n^2 \mathbf{I}_{\tau(2 N_r)}$,  due to noise samples in real-valued notation with variance $\frac{1}{2} \sigma_n^2$.
Note that knowledge about $\mathbf{R_{h_{\text{R}}}}$ might not be available at the receiver. 
For this case,
a sample channel correlation
matrix can be estimated based on 1-bit measurements using the method
devised in \cite{Kim2019_2}. Assuming that the channel is a Rayleigh fading channel without spatial correlation implies $\mathbf{R_{h_{\text{R}}}}=\frac{1}{2} \mathbf{I}$. For this case \eqref{eq:Czrp} simplifies to
\begin{equation}
\mathbf{C}_{\mathbf{z}_{{\text{R}}_p}} =  \mathbf{B}_{\text{eff}}
\left( \frac{1}{2}
\tilde{\mathbf{\Phi}}_{\text{R}}
\tilde{\mathbf{\Phi}}_{\text{R}}^T
+ \frac{1}{2}\sigma_n^2   \mathbf{I}_{2 \tau N_r}  \right)
\mathbf{B}_{\text{eff}} ^T .
\end{equation} 
Considering that $\tau=N_t$ we have $\tilde{\mathbf{\Phi}}_{\text{R}}
\tilde{\mathbf{\Phi}}_{\text{R}}^T = N_t \sigma_x^2  \mathbf{I}_{2 \tau N_r} $ which yields 
\begin{equation}
\mathbf{C}_{\mathbf{z}_{{\text{R}}_p}} =  \mathbf{B}_{\text{eff}}
\left( \frac{1}{2}  N_t \sigma_x^2
  \mathbf{I}_{2 \tau N_r}
+ \frac{1}{2} \sigma_n^2  \mathbf{I}_{2 \tau N_r}  \right)
\mathbf{B}_{\text{eff}} ^T .
\end{equation} 
This can be written more compactly as
\begin{equation}
\mathbf{C}_{\mathbf{z}_{{\text{R}}_p}} =  
\frac{1}{2}  \left(N_t \sigma_x^2+\sigma_n^2 \right)
\mathbf{B}_{\text{eff}}
\mathbf{B}_{\text{eff}} ^T .
\end{equation} 
The matrix $\mathbf{B}_{\text{eff}}
\mathbf{B}_{\text{eff}} ^T$ has only ones on its diagonal. With this, one can conclude that
\begin{equation}
\mathbf{A}_{{\text{R}}_p} =\sqrt{\frac{2}{\pi}  
\frac{2}{N_t \sigma_x^2+\sigma_n^2}} \mathbf{I}_{\tau(2 N_r+\alpha)}.
\label{eq:bussgang_factor}
\end{equation}


\subsection{Proposed LRA-LMMSE Channel Estimator}
Employing the proposed statistically equivalent linear model in \eqref{eq:zqp} allows the adoption of the established technique for linear channel estimation with quantization error \cite{7931630,Wan2020,Kim2019_2}.
Based on \eqref{eq:zqp} the LRA-LMMSE estimator can be obtained through the optimization problem formulated as
\begin{equation}
\begin{split}
\mathbf{W}_{\text{R, LRA-LMMSE}}  = \arg\min_{\mathbf{W}}\enspace E\left[\left\vert\left\vert \mathbf{h}_{\text{R}}-\mathbf{W}\mathbf{z}_{\mathcal{Q}_p} \right\vert\right\vert^2_2\right] =  \mathbf{R_{h_{\text{R}}}} \hat{\mathbf{\Phi}}_{\text{R}}^T \mathbf{C}_{\mathbf{z}_{\mathcal{Q}p}}^{-1} ,
\label{eq:WLMMSE}
\end{split}
\end{equation} 
where the auto-correlation of the quantized signal is calculated as \cite{Jacovitti}
\begin{equation}
\mathbf{C}_{\mathbf{z}_{\mathcal{Q}_p}} =\frac{2}{\pi}\text{sin}^{-1}\left(\mathbf{K}_{{\text{R}}_p}\mathbf{C}_{\mathbf{z}_{{\text{R}}_p}}\mathbf{K}_{{\text{R}}_p}\right) .
\label{Czqp}
\end{equation}
For the Rayleigh fading channel with $\mathbf{R_{h_{\text{R}}}}=\frac{1}{2} \mathbf{I}_{2 N_t N_r} $ and $\tau=N_t$
we have
\begin{align}
\mathbf{C}_{\mathbf{z}_{\mathcal{Q}_p}} =\frac{2}{\pi} \text{sin}^{-1}
\left(\mathbf{B}_{\text{eff}}
\mathbf{B}_{\text{eff}} ^T\right) \approx \frac{2}{\pi} \left( 
\mathbf{B}_{\text{eff}}
\mathbf{B}_{\text{eff}} ^T+ (\frac{\pi}{2}-1) \mathbf{I}_{2 \tau N_r}
\right)=\frac{  \pi-2 }{\pi }
\left( \frac{2}{(\pi-2)}
\mathbf{B}_{\text{eff}}
\mathbf{B}_{\text{eff}} ^T+  \mathbf{I}_{2 \tau N_r}
\right) \text{.}
\label{eq:Czqp_approx}
\end{align}
The resulting LRA-LMMSE channel estimator is given by
\begin{equation}
\hat{\mathbf{h}}_{\text{R, LRA-LMMSE}} = \begin{bmatrix}
\hat{\mathbf{h}}_{\text{R, LRA-LMMSE, a}} \\
\hat{\mathbf{h}}_{\text{R, LRA-LMMSE, b}}
\end{bmatrix} = \mathbf{W}_{\text{R, LRA-LMMSE}}~\mathbf{z}_{\mathcal{Q}_p},
\label{eq:h_estimated}
\end{equation} 
where $\hat{\mathbf{h}}_{\text{R, LRA-LMMSE, a}}\in \mathbb{R}^{N_rN_t\times 1}$ is the first half of the channel estimate which corresponds to the real part and $\hat{\mathbf{h}}_{\text{R, LRA-LMMSE, b}}\in \mathbb{R}^{N_rN_t\times 1}$ is the second half which corresponds to the imaginary part.
A detailed derivation is shown in Appendix \ref{subsec:derivation_of_W_LMMSE}.

\subsection{Error Correlation Matrix and MSE of the Channel Estimate}
Inserting the estimated channel vector in the cost function of the MSE problem (\ref{eq:WLMMSE}) yields 
\begin{align}
\mathcal{M}_{\text{R, LRA-MMSE}} &= E\left[\left\vert\left\vert   \mathbf{h}_{\text{R}} -  \hat{\mathbf{h}}_{\text{R, LRA-MMSE}} \right\vert\right\vert^2_2 \right] \\
&=    \mathrm{tr} \left(    E\left[    \mathbf{\epsilon}_{\text{R}}
\mathbf{\epsilon}_{\text{R}}^T  \right]  \right) \text{,} 
\end{align}
with $\mathbf{\epsilon}_{\text{R}}=  
\mathbf{h}_{\text{R}}
- \hat{\mathbf{h}}_{\text{R, LRA-MMSE}}
$. The channel estimation error correlation matrix is then given by 
\begin{equation}
E\left[    \mathbf{\epsilon}_{\text{R}}
\mathbf{\epsilon}_{\text{R}}^T  \right]
= E\left[  \hat{\mathbf{h}}_{\text{R, LRA-MMSE}}
\hat{\mathbf{h}}_{\text{R, LRA-MMSE}}^T -2 \hat{\mathbf{h}}_{\text{R, LRA-MMSE}}\mathbf{h}_{\text{R}}^T  +\mathbf{h}_{\text{R}}\mathbf{h}_{\text{R}}^T  \right].
\label{eq:mse_derivation}
\end{equation}
By inserting \eqref{eq:h_estimated} and \eqref{eq:zqp} into \eqref{eq:mse_derivation}, 
and considering that $\mathbf{h}_{\text{R}}$ is uncorrelated with $\tilde{\mathbf{n}}_{{\text{R}}_p}$,
the error correlation matrix can be expressed as
\begin{align}
  E\left[    \mathbf{\epsilon}_{\text{R}}
\mathbf{\epsilon}_{\text{R}}^T  \right]
=  
    \mathbf{R_{h_{\text{R}}}} -\mathbf{R_{h_{\text{R}}}}\hat{\mathbf{\Phi}}_{\text{R}}^T  \mathbf{C}_{\mathbf{z}_{\mathcal{Q}_{p}}}^{-1}\hat{\mathbf{\Phi}}_{\text{R}} \mathbf{R_{h_{\text{R}}}} \text{.}
\end{align}
Note that the channel estimation correlation matrix can be taken into account for robust detection as presented in Section~\ref{sec:Robust_2}.  
The corresponding expression for the MSE is given by
\begin{equation}
\mathcal{M}_{\text{R, LRA-MMSE}} 
= \mathrm{tr}\left(\mathbf{R_{h_{\text{R}}}} -\mathbf{R_{h_{\text{R}}}}\hat{\mathbf{\Phi}}_{\text{R}}^T  \mathbf{C}_{\mathbf{z}_{\mathcal{Q}_{p}}}^{-1}\hat{\mathbf{\Phi}}_{\text{R}} \mathbf{R_{h_{\text{R}}}} \right) \text{.}
\label{eq:mse}
\end{equation}
The correlation matrix of the channel estimate is given by
\begin{align}
  E\left[    \hat{\mathbf{h}}_{\text{R, LRA-MMSE}}
\hat{\mathbf{h}}_{\text{R, LRA-MMSE}}^T  \right]
=  \mathbf{R_{h_{\text{R}}}}\hat{\mathbf{\Phi}}_{\text{R}}^T  \mathbf{C}_{\mathbf{z}_{\mathcal{Q}_{p}}}^{-1}\hat{\mathbf{\Phi}}_{\text{R}} \mathbf{R_{h_{\text{R}}}} \text{.}
\end{align}
By using the approximation in \eqref{eq:Czqp_approx} for $\mathbf{C}_{\mathbf{z}_{\mathcal{Q}_{p}}}^{-1}$ the correlation matrix can be expressed as
\begin{align}
  E\left[    \hat{\mathbf{h}}_{\text{R, LRA-MMSE}}
\hat{\mathbf{h}}_{\text{R, LRA-MMSE}}^T  \right]
= \frac{\pi }{  \pi-2 } \mathbf{R_{h_{\text{R}}}}\hat{\mathbf{\Phi}}_{\text{R}}^T  
\left( \frac{2}{\pi-2}
\mathbf{B}_{\text{eff}}
\mathbf{B}_{\text{eff}} ^T+  \mathbf{I}
\right)^{-1}
\hat{\mathbf{\Phi}}_{\text{R}} \mathbf{R_{h_{\text{R}}}} \text{.}
\end{align}
By considering $\mathbf{R_{h_{\text{R}}}}=\frac{1}{2}\mathbf{I}$
and $\tau=N_t$, \eqref{eq:bussgang_factor} can be applied and
the correlation matrix of the channel estimate reads as
\begin{align}
  E\left[    \hat{\mathbf{h}}_{\text{R, LRA-MMSE}}
\hat{\mathbf{h}}_{\text{R, LRA-MMSE}}^T  \right]
=  \frac{1}{4} 
\frac{2}{\pi}\frac{2}{N_t \sigma_x^2+\sigma_n^2}
\frac{\pi }{  \pi-2 } 
\tilde{\mathbf{\Phi}}_{\text{R}}^T
\mathbf{B}_{\text{eff}}^T
\left( \frac{2}{\pi-2}
\mathbf{B}_{\text{eff}}
\mathbf{B}_{\text{eff}} ^T+  \mathbf{I}
\right)^{-1} \mathbf{B}_{\text{eff}} \tilde{\mathbf{\Phi}}_{\text{R}}  \text{.}
\end{align}
Using the push-through identity one can write
\begin{align}
  E\left[    \hat{\mathbf{h}}_{\text{R, LRA-MMSE}}
\hat{\mathbf{h}}_{\text{R, LRA-MMSE}}^T  \right]
= 
\frac{1}{N_t \sigma_x^2+\sigma_n^2}
\frac{1}{  \pi-2 } 
\tilde{\mathbf{\Phi}}_{\text{R}}^T
\mathbf{B}_{\text{eff}}^T \mathbf{B}_{\text{eff}} 
\left( \frac{2}{\pi-2}
\mathbf{B}_{\text{eff}}^T  \mathbf{B}_{\text{eff}}+  \mathbf{I}
\right)^{-1} 
\tilde{\mathbf{\Phi}}_{\text{R}}  \text{.}
\end{align}
In the following it is considered that the diagonal elements in the matrix $\mathbf{B}_{\text{eff}}^T  \mathbf{B}_{\text{eff}}$ are much larger than the off diagonal elements. Considering
$\mathbf{B}_{\text{eff}}^T  \mathbf{B}_{\text{eff}} \approx \mathrm{diag} \left[ b_1, b_2 \ldots b_{\tau 2N_r}  \right]$ yields
\begin{align}
 & E\left[    \hat{\mathbf{h}}_{\text{R, LRA-MMSE}}
\hat{\mathbf{h}}_{\text{R, LRA-MMSE}}^T  \right]  
= \label{eq:diagonal_expression} \\
& \frac{1}{N_t \sigma_x^2+\sigma_n^2}
\frac{1}{2}
\tilde{\mathbf{\Phi}}_{\text{R}}^T
\left(
\mathbf{I}_{\tau 2N_r}-(\pi-2)
\mathrm{diag} \left[ \frac{1}{2 b_1 +\pi-2 }, \frac{1}{2 b_2 +\pi-2 } \ldots 
\frac{1}{2 b_{\tau 2N_r} +\pi-2 }
\right]
\right)
\tilde{\mathbf{\Phi}}_{\text{R}}  \text{.}
\notag
\end{align}
For a comparator network matrix with uniformly distributed non-zero entries, the probability of a non-zero at any position in matrix $B_{\text{eff}}^{\prime}$ is equal to $\frac{1}{N_r \tau}$. Accordingly it holds $E\left[  | B_{\text{eff},i,j}^{\prime}|^2    \right] = \frac{1}{N_r \tau} \frac{1}{2}  $, which implies
$ E\left[
\mathbf{B}_{\text{eff}}^T \mathbf{B}_{\text{eff}} 
\right] =  (1+\frac{\alpha}{2 N_r}) \mathbf{I}
 $ and $ E\left[b_i\right]=1+\frac{\alpha}{2 N_r}$.
 By replacing $b_i$ by its expected value,
  \eqref{eq:diagonal_expression} is approximated by
 \begin{align}
  E\left[    \hat{\mathbf{h}}_{\text{R, LRA-MMSE}}
\hat{\mathbf{h}}_{\text{R, LRA-MMSE}}^T  \right]  
= 
\frac{1}{N_t \sigma_x^2+\sigma_n^2}
\frac{1}{2}
\left(
1-\frac{\pi-2}{2(1+\frac{\alpha}{2 N_r}) +\pi-2 }
\right)
\tilde{\mathbf{\Phi}}_{\text{R}}^T
\tilde{\mathbf{\Phi}}_{\text{R}}  \text{.}
\end{align}
Finally, with $\tilde{\mathbf{\Phi}}_{\text{R}}^T
\tilde{\mathbf{\Phi}}_{\text{R}}=N_t \sigma_x^2 \mathbf{I} $, it holds
 \begin{align}
  E\left[    \hat{\mathbf{h}}_{\text{R, LRA-MMSE}}
\hat{\mathbf{h}}_{\text{R, LRA-MMSE}}^T  \right] 
= 
\frac{1}{2}
\frac{N_t \sigma_x^2}{N_t \sigma_x^2+\sigma_n^2}
\left(
1-\frac{\pi-2}{2(1+\frac{\alpha}{2 N_r}) +\pi-2 }
\right) \mathbf{I}_{2N_rN_t}  \text{.}
\label{eq:correlation_estimation}
\end{align}
In the following a short hand notation for the prefactor in \eqref{eq:correlation_estimation} will be used in terms of $\kappa$. With this, the sum MSE can be approximated as
$\mathcal{M}_{\text{R, LRA-MMSE}} =  N_r N_t (1 - 2 \kappa  )$.

\section{Linear Detection}
\label{sec:linear_detection}
In this section, we present linear detection techniques based on comparator networks. In particular, we describe the proposed LRA-LMMSE detector for perfect (or near perfect) CSI scenarios and the proposed robust LRA-LMMSE detector for scenarios with imperfect CSI. 
\textcolor{r3}{
To the best of the authors knowledge, the proposed study is the first to deal with a robust approach for 1-bit MIMO receivers.
}

\subsection{Proposed LRA-LMMSE Detector for Perfect CSI}
\label{sec:LRAMMSE_detector}
Based on the proposed system model in (\ref{new_system_model}), the corresponding linear receiver to estimate the transmitted symbols is derived as follows. The optimization problem that yields the linear optimal receiver reads as
\begin{equation}
\mathbf{G}_{\text{R, LRA-MMSE}} = \arg\min_{\mathbf{G}_{\text{R}}}\enspace E\left[\left\vert\left\vert \mathbf{x}_{\text{R}}-\mathbf{G}_{\text{R}}^T\mathbf{z}_{\mathcal{Q}}\right\vert\right\vert^2_2\right],
\label{LRA_MMSE_filter_Problem}
\end{equation}
where $\mathbf{G}_{\text{R}}\in\mathbb{R}^{(2N_r+\alpha)\times 2N_t}$. The solution for
\eqref{LRA_MMSE_filter_Problem} is given by
\begin{equation}
\mathbf{G}_{\text{R, LRA-MMSE}} = \mathbf{C}_{\mathbf{z}_{\mathcal{Q}}}^{-1}\mathbf{C}_{\mathbf{z}_{\mathcal{Q}}\mathbf{x}_{\text{R}}},
\label{equ_LRAMMSE}
\end{equation} 
where the involved covariance matrix $\mathbf{C}_{\mathbf{z}_{\mathcal{Q}}}$ is calculated as \cite{Jacovitti}
\begin{equation}
\mathbf{C}_{\mathbf{z}_\mathcal{Q}}=\frac{2}{\pi}\text{sin}^{-1}\left(\mathbf{K}_{\text{R}}\mathbf{C}_{\mathbf{z}_{\text{R}}}\mathbf{K}_{\text{R}}\right),
\end{equation}
and the cross-correlation matrix $\mathbf{C}_{\mathbf{z}_{\mathcal{Q}}\mathbf{x}_{\text{R}}}$ is based on the Bussgang theorem \cite{Bussgang}
\begin{equation}
\begin{split}
\mathbf{C}_{\mathbf{z}_\mathcal{Q}\mathbf{x}_{\text{R}}} =\sqrt{\frac{2}{\pi}} \frac{\sigma_x^2}{2} \mathbf{K}_{\text{R}}\mathbf{BH}_{\text{R}},
\label{eq:Czqxr}
\end{split}
\end{equation}
with $\mathbf{K}_{\text{R}}=\text{diag}(\mathbf{C}_{\mathbf{z}_{\text{R}}})^{-\frac{1}{2}}\in \mathbb{R}^{(2N_r+\alpha)\times (2N_r+\alpha)}$. The cross-correlation matrix between $\mathbf{z}_{\text{R}}$ and $\mathbf{x}_{\text{R}}$ is given by
\begin{equation}
\mathbf{C}_{\mathbf{z}_{\text{R}}\mathbf{x}_{\text{R}}} = \frac{1}{2} \sigma_x^2 \mathbf{B}\mathbf{H}_{\text{R}},
\label{eq:Czrxr}
\end{equation}
where due to the real-valued notation of the system the input correlation matrix is given by 
$\mathbf{C_{\mathbf{x}_{\text{R}}}} = E\left[\mathbf{x}_{\text{R}}\mathbf{x}_{\text{R}}^T\right] =\frac{1}{2} \sigma_x^2 \mathbf{I}_{2N_t}$.
Besides, the correlation matrix of $\mathbf{z}_{\text{R}}$ is given by  
\begin{equation}
\begin{split}
\mathbf{C}_{\mathbf{z}_{\text{R}}} = \frac{1}{2}\sigma_x^2  \mathbf{B}\mathbf{H}_{\text{R}}\mathbf{H}_{\text{R}}^T\mathbf{B}^T+\frac{1}{2} \sigma_n^2 \mathbf{B}\mathbf{B}^T \text{,}
\label{eq:Czr}
\end{split}
\end{equation}
where the noise correlation matrix is given by $\mathbf{C_{\mathbf{n}_{\text{R}}}} = E\left[\mathbf{n}_{{\text{R}}}\mathbf{n}_{{\text{R}}}^T\right] = \frac{1}{2} \sigma_n^2 \mathbf{I}_{2N_r}$, which is
due to the real-valued notation of the system.\\

\subsection{Robust LRA-LMMSE Detector for Imperfect CSI}
\label{sec:Robust_general}
In this subsection, a robust LRA-MMSE detector is proposed, which takes 
into account statistics of CSI imperfections.
In this context, it is considered that the channel can be described by two uncorrelated terms as  
$\mathbf{h}_{\text{R}}  =   \sqrt{\lambda} {\mathbf{h}}_{\text{R},1}    + \sqrt{1 - \lambda}  \mathbf{h}_{\text{R,2}}   $, where ${\mathbf{h}}_{\text{R},1}, \lambda $ 
and the correlation matrix  $ \mathbf{R_{h_{\text{R,2}}}}= E\left[\mathbf{h}_{\text{R,2}}\mathbf{h}_{\text{R,2}}^T\right]$
are considered to be known at the receiver.
The quality of the CSI is determined by parameter $0<\lambda<1$.

As an example, this model could represent a time-varying channel with wide-sense stationarity and known temporal autocorrelation function, similar to the assumptions used in \cite{Kim2019_2}. 
In this case, ${\mathbf{h}}_{\text{R},1}$ is an outdated channel vector and $\sqrt{\lambda}$ represents the value of the temporal channel autocorrelation function, where we assume the same properties for all users.

Based on this model, we redefine the received signal as
\begin{equation}
\begin{split}
\mathbf{y}_{\text{R}} &= \begin{bmatrix} 
\mathfrak{R}\{\tilde{\mathbf{X}}\} &-\mathfrak{I}\{\tilde{\mathbf{X}}\}\\
\mathfrak{I}\{\tilde{\mathbf{X}}\} & \mathfrak{R}\{\tilde{\mathbf{X}}\}
\end{bmatrix}\begin{bmatrix}
\mathfrak{R}\{\mathbf{h}\} \\
\mathfrak{I}\{\mathbf{h}\}
\end{bmatrix} +
\begin{bmatrix}
\mathfrak{R}\{\mathbf{n}_{d}\} \\
\mathfrak{I}\{\mathbf{n}_{d}\}
\end{bmatrix}\\
&=  \sqrt{\lambda} {\mathbf{H}}_{\text{R},1}\mathbf{x}_{\text{R}} + \sqrt{1 - \lambda}\tilde{\mathbf{X}}_{\text{R}}\mathbf{h}_{\text{R,2}} +\mathbf{n}_{{\text{R}}_d},
\end{split}
\end{equation}
where $\mathbf{y}_{\text{R}} \in \mathbb{R}^{2 N_r \times 1} $ and $\tilde{\mathbf{X}} = (\mathbf{x}^T\otimes\mathbf{I}_{N_r}) \in \mathbb{C}^{N_r \times N_t N_r}$. The subscript $d$ refers to data. Moreover, ${\mathbf{H}}_{\text{R},1} \in \mathbb{R}^{2 N_r \times 2 N_t}$ is the known part of the channel matrix, described by
\begin{equation}
\begin{split}
&{\mathbf{H}}_{\text{R},1} =  \begin{bmatrix} 
\mathfrak{R}\{{\mathbf{H}}_{1}\} &-\mathfrak{I}\{{\mathbf{H}}_{1}\}\\
\mathfrak{I}\{{\mathbf{H}}_{1}\} & \mathfrak{R}\{{\mathbf{H}}_{1}\}
\end{bmatrix}  \text{with} \enspace {\mathbf{H}}_{1} = \text{unvec}({\mathbf{h}}_{1}) \\
&\text{and} \enspace {\mathbf{h}}_{1} = \begin{bmatrix}
{\mathbf{h}}_{\text{R},1, \text{a}}  + j{\mathbf{h}}_{\text{R},1, \text{b}}
\end{bmatrix},
\label{eq:H_hat}
\end{split}
\end{equation}
where ${\mathbf{h}}_{\text{R},1, \text{a}}$ corresponds to the first half of the known part ${\mathbf{h}}_{\text{R},1}$ and ${\mathbf{h}}_{\text{R},1, \text{b}}$ to the second half. With this, the quantized signal of the comparator aided MIMO receiver can be expressed as
\begin{equation}
\begin{split}
\mathbf{z}_{\mathcal{Q}_r} &= \mathcal{Q} (\mathbf{z}_{{\text{R}}_r}) = 
\sqrt{\lambda}
\mathbf{A}_{{\text{R}}_r} \mathbf{B}{\mathbf{H}}_{\text{R},1} \mathbf{x}_{\text{R}} + 
\sqrt{1 - \lambda} 
\mathbf{A}_{{\text{R}}_r} \mathbf{B} \tilde{\mathbf{X}}_{\text{R}}\mathbf{h}_{\text{R,2}}  + \mathbf{A}_{{\text{R}}_r} \mathbf{B} \mathbf{n}_{{\text{R}}_d} + \mathbf{n}_{{\text{R}}_{q, d}},
\end{split}
\end{equation}
\textcolor{r3}{
of dimensions ${(2 N_r+\alpha)\times 1}$},
where $\mathbf{A}_{{\text{R}}_r}$ is the Bussgang-based linear operator for the transmit data, similarly as calculated in \eqref{eq:A_rp}. Then, the LRA-LMMSE filter is applied to $\mathbf{z}_{\mathcal{Q}_r}$, to obtain
\begin{equation}
\hat{\mathbf{x}}_{\text{R}} =\mathbf{G}_{\text{R}}^T \mathbf{z}_{\mathcal{Q}_r},
\end{equation}
where the matrix $\mathbf{G}_{\text{R}}\in \mathbb{R}^{(2N_r+\alpha)\times 2N_t}$ is chosen to minimize the MSE between the transmitted symbol $\mathbf{x}_{\text{R}}$
and the filter output, i.e.,
\begin{equation}
\begin{split}
\tilde{\mathbf{G}}_{\text{R, LRA-MMSE}} &= \arg\min_{\mathbf{G}_{\text{R}}}\enspace E\left[\left\vert\left\vert \mathbf{x}_{\text{R}}-\mathbf{G}_{\text{R}}^T\mathbf{z}_{\mathcal{Q}_r}\right\vert\right\vert^2_2\right]\\
&= \arg\min_{\mathbf{G}_{\text{R}}}\enspace -2\mathrm{tr} \left(\mathbf{G}_{\text{R}}^T \mathbf{C}_{\mathbf{z}_{\mathcal{Q}_r}\mathbf{x}_{\text{R}}} \right) + \mathrm{tr} \left(\mathbf{G}_{\text{R}}^T\mathbf{C}_{\mathbf{z}_{\mathcal{Q}_r}}\mathbf{G}_{\text{R}} \right),
\label{eq:G_LRAMMSE_2}
\end{split}
\end{equation}
where the latter describes an equivalent problem.
By differentiating \eqref{eq:G_LRAMMSE_2} with respect to $\mathbf{G}_{\text{R}}^T$ and equating it to zero, the solution of the LRA-LMMSE
receive filter is given by
\begin{equation}
\tilde{\mathbf{G}}_{\text{R, LRA-MMSE}} = \mathbf{C}_{\mathbf{z}_{\mathcal{Q}_r}}^{-1}\mathbf{C}_{\mathbf{z}_{\mathcal{Q}_r}\mathbf{x}_{\text{R}}}.
\label{eq:Robust_G}
\end{equation}
As in the previous cases, the covariance matrix, \textcolor{r3}{
of dimensions ${(2 N_r+\alpha)\times (2 N_r+\alpha)}$},  is calculated as \cite{Jacovitti}
\begin{equation}
\mathbf{C}_{\mathbf{z}_{\mathcal{Q}_r}} = \frac{2}{\pi}\text{sin}^{-1}\left(\mathbf{K}_{\text{R}_r}\mathbf{C}_{\mathbf{z}_{\text{R}_r}}\mathbf{K}_{\text{R}_r}\right),
\end{equation}
and the cross-correlation matrix, \textcolor{r3}{
of dimensions ${(2 N_r+\alpha)\times 2 N_t}$}, is based on the Bussgang theorem \cite{Bussgang}
\begin{equation}
\mathbf{C}_{\mathbf{z}_{\mathcal{Q}_r}\mathbf{x}_{\text{R}}} = \sqrt{\frac{2}{\pi}}\frac{\sigma_x^2}{2}\mathbf{K}_{\text{R}_r} \mathbf{B} {\mathbf{H}}_{\text{R},1},
\label{eq:Czqxr_2}
\end{equation}
with $\mathbf{K}_{\text{R}_r}=\text{diag}(\mathbf{C}_{\mathbf{z}_{\text{R}_r}})^{-\frac{1}{2}}$ and the cross-correlation matrix between $\mathbf{z}_{\text{R}_r}$ and $\mathbf{x}_{\text{R}}$,
of dimensions ${(2 N_r+\alpha)\times 2 N_t}$,
is given by 
$\mathbf{C}_{\mathbf{z}_{\text{R}_r}\mathbf{x}_{\text{R}}} = \frac{1}{2}    \sigma_x^2 \mathbf{B}{\mathbf{H}}_{\text{R},1},$
where due to the real-valued notation of the system the input correlation matrix $\mathbf{C_{\mathbf{x}_{\text{R}}}} = E\left[\mathbf{x}_{\text{R}}\mathbf{x}_{\text{R}}^T\right] =  \frac{1}{2} \sigma_x^2 \mathbf{I}_{2N_t}$ is considered.
Moreover, the correlation matrix of the unquantized output signal $\mathbf{z}_{\text{R}_r}$ is given by
\begin{equation}\label{eq:Czr_2}
\mathbf{C}_{\mathbf{z}_{\text{R}_r}} = \lambda \frac{1}{2}\sigma_x^2  \mathbf{B}{\mathbf{H}}_{\text{R},1}{\mathbf{H}}_{\text{R},1}^T\mathbf{B}^T + (1-\lambda) \mathbf{B}\mathbf{\Gamma}_{{\text{R}}}\mathbf{B}^T + \frac{1}{2}\sigma_n^2 \mathbf{B}\mathbf{B}^T.
\end{equation}
The matrix  $\mathbf{\Gamma}_{{\text{R}}}\in \mathbb{R}^{N_rN_t\times N_rN_t}$ in \eqref{eq:Czr_2} is calculated as follows:
\begin{equation}
\begin{split}
\mathbf{\Gamma}_{{\text{R}}} &= E\left[\tilde{\mathbf{X}}_{\text{R}}
\mathbf{R_{h_{\text{R},2}}}
\tilde{\mathbf{X}}_{\text{R}}^T\right] \text{.}
\label{eq:gamma_r}
\end{split}
\end{equation}
In the following it is considered that the transmit symbols are taken from an $M_x$-ary symbol alphabet and that the prior probabilities are known at the receiver. 
With this, the expected value in \eqref{eq:gamma_r} can be calculated as
\begin{equation}
\begin{split}
\mathbf{\Gamma}_{{\text{R}}} &= \sum_{i = 1}^{M_x^{N_t}} P\left( \tilde{\mathbf{X}}_{\text{R}, i} \right) \left(\tilde{\mathbf{X}}_{\text{R}, i}\mathbf{R_{h_{\text{R},2}}}\tilde{\mathbf{X}}_{\text{R}, i}^T \right) \text{,}
\end{split}
\end{equation}
where $   P\left( \tilde{\mathbf{X}}_{\text{R}, i} \right) =   \frac{1}{4^{N_t}}$ 
in case of uniformly distributed QPSK transmit symbols.

\subsection{Robust LRA-LMMSE Detector for Imperfect CSI Caused by Channel Estimation}
\label{sec:Robust_2}
The robust detector described in the previous section can be applied to the case of imperfect channel estimation caused by channel estimation.
In this case, we substitute the channel description from the previous section in terms of
$\sqrt{\lambda} {\mathbf{h}}_{\text{R},1} = \hat{\mathbf{h}}_{\text{R}} $
and
$\sqrt{1 - \lambda}  \mathbf{h}_{\text{R,2}} = \bm{\varepsilon}_{\text{R}}  $.
With this, the cross-correlation matrix, 
of dimensions ${(2 N_r+\alpha)\times 2 N_t}$,
is given by
\cite{Bussgang}
\begin{equation}
\mathbf{C}_{\mathbf{z}_{\mathcal{Q}_r}\mathbf{x}_{\text{R}}} = \sqrt{\frac{2}{\pi}}\frac{\sigma_x^2}{2}\mathbf{K}_{\text{R}_r} \mathbf{B} \hat{\mathbf{H}}_{\text{R}},
\label{eq:Czqxr_2}
\end{equation}
with $\mathbf{K}_{\text{R}_r}=\text{diag}(\mathbf{C}_{\mathbf{z}_{\text{R}_r}})^{-\frac{1}{2}}$.
The cross-correlation matrix between $\mathbf{z}_{\text{R}_r}$ and $\mathbf{x}_{\text{R}}$ reads as
$
\mathbf{C}_{\mathbf{z}_{\text{R}_r}\mathbf{x}_{\text{R}}} = \frac{1}{2}\sigma_x^2  \mathbf{B}\hat{\mathbf{H}}_{\text{R}}.    
$
Moreover, the correlation matrix of the unquantized output signal $\mathbf{z}_{\text{R}_r}$ is given by
\begin{equation}
\begin{split}
\mathbf{C}_{\mathbf{z}_{\text{R}_r}} =  \frac{1}{2}  \sigma_x^2\mathbf{B}\hat{\mathbf{H}}_{\text{R}}\hat{\mathbf{H}}_{\text{R}}^T\mathbf{B}^T + \mathbf{B}\mathbf{\Gamma}_{{\text{R}}}\mathbf{B}^T + \frac{1}{2} \sigma_n^2 \mathbf{B}\mathbf{B}^T.
\label{eq:Czr_channel_estimation}
\end{split}
\end{equation}
The matrix $\mathbf{\Gamma}_{{\text{R}}}$ in \eqref{eq:Czr_channel_estimation} is calculated as follows
\begin{equation}
\begin{split}
\mathbf{\Gamma}_{{\text{R}}} &= E\left[\tilde{\mathbf{X}}_{\text{R}}E\left[\bm{\varepsilon}_{\text{R}}\bm{\varepsilon}_{\text{R}}^T\right]\tilde{\mathbf{X}}_{\text{R}}^T\right]
= E\left[\tilde{\mathbf{X}}_{\text{R}}\mathbf{R_{h_{\text{R}}}}\tilde{\mathbf{X}}_{\text{R}}^T -\tilde{\mathbf{X}}_{\text{R}}\mathbf{R_{h_{\text{R}}}}\hat{\mathbf{\Phi}}_{\text{R}}^T  \mathbf{C}_{\mathbf{z}_{\mathcal{Q}_{p}}}^{-1}\hat{\mathbf{\Phi}}_{\text{R}} \mathbf{R_{h_{\text{R}}}} \tilde{\mathbf{X}}_{\text{R}}^T\right],
\label{eq:gamma_r_channel_estimation}
\end{split}
\end{equation}
where the matrix
$E\left[\bm{\varepsilon}_{\text{R}}\bm{\varepsilon}_{\text{R}}^T\right]$ 
depends on the considered channel estimation method. For the proposed channel estimation in the section \ref{sec:channel_estimation} the matrix is given by $E\left[\bm{\varepsilon}_{\text{R}}\bm{\varepsilon}_{\text{R}}^T\right]= \mathbf{R_{h_{\text{R}}}} -\mathbf{R_{h_{\text{R}}}}\hat{\mathbf{\Phi}}_{\text{R}}^T  \mathbf{C}_{\mathbf{z}_{\mathcal{Q}_{p}}}^{-1}\hat{\mathbf{\Phi}}_{\text{R}} \mathbf{R_{h_{\text{R}}}}$, which is related to previously calculated MSE expression \eqref{eq:mse}.

\section{Design of the Comparator Network}
\label{sec:design}

The matrix design of the comparator network in (\ref{B_prime}) is detailed in this section. Two types of networks are considered, namely, fully and partially connected networks.

\subsection{Fully Connected Network}
In this network, every two of the received signals are compared, which means that the number of comparators is $\alpha_f= \tbinom{2N_r}{2}=N_r(2N_r-1)$. 
For instance, if we consider a system with $N_r=2$ receive antennas, $\alpha_f=\tbinom{4}{2}=6$ comparators are needed in this network. Then, the corresponding matrix $\mathbf{B}'$ is described by
\begin{equation}
    \mathbf{B}'_{\alpha_f}= \frac{1}{\sqrt{2}}
    \begin{bsmallmatrix}
 1&-1&0&0\\ 1&0&-1&0\\ 1&0&0&-1\\ 0&1&-1&0\\ 0&1&0&-1\\ 0&0&1&-1\\
 \end{bsmallmatrix}   .
 \label{eq:fully_connected}
\end{equation}
The main drawback of the fully connected network is the massive use of the comparators, where the number of comparators $\alpha_f$ is approximately proportional to the square of the number of receive antennas $N_r$.  Therefore, in large-scale multiuser MIMO systems a much larger number of comparators will be required, which might be unrealistic.

\subsection{Partially Connected Network}

In order to increase the usage efficiency of the comparators, the partially connected network is proposed, where the number of utilized comparators $\alpha_p$ is only a small fraction of the number of comparators in the fully connected network. For the same scenario described in the previous subsection, one possibility for the corresponding matrix $\mathbf{B}'_{{\alpha}_p}$ can be described by
\begin{equation}
\mathbf{B}'_{{\alpha}_p}=
\frac{1}{\sqrt{2}}
\begin{bsmallmatrix}
1&-1&0&0\\ 1&0&-1&0\\ 1&0&0&-1\\ 0&1&-1&0\\
\end{bsmallmatrix}.
\label{eq:partially_connected}
\end{equation}
Different types of network design are considered, namely random, greedy search based and sequentially SINR search based. 
The random comparator network design randomly selects 
$\alpha_p$ out of $\alpha_f$ comparator input combinations.
The design of the partially connected comparator network in general is a combinatorial problem.
For a comparator network consisting of $\alpha_p$ comparators $\tbinom{N_r(2N_r-1)}{\alpha_p}$ different configurations exist.
Two sophisticated search based comparator network designs are developed in the following subsections.

\begin{algorithm}[H]
\footnotesize
		\caption{MMSE based Greedy Search}
		\label{alg:MSE}
		\begin{algorithmic}[1]
		
		\STATE Find the fully connected network $\mathbf{B}'$ in \eqref{eq:fully_connected} and get the number of rows, defined by $\alpha_f$ 
		\STATE Extract the first $\alpha_p$ rows of $\mathbf{B}'$, defined by $\mathbf{B}'_{\alpha_p}$
		\STATE Constitute $\mathbf{B}$ in \eqref{new_system_model} and calculate $\mathbf{G}_{{\text{R}}}$ in \eqref{equ_LRAMMSE}
		\STATE Compute the MSE with $E\left[\left\vert\left\vert \mathbf{x}_{\text{R}}-\mathbf{G}_{{\text{R}}}\mathbf{z}_{\mathcal{Q}}\right\vert\right\vert^2_2\right]$, defined by $l_\text{min}$
		\FOR{$i = 1 : \alpha_p$}
		    \STATE Take the $i$th row of $\mathbf{B}'_{\alpha_p}$ and freeze the other $\alpha_p-1$ rows
		    \FOR{$j = 1 : \alpha_f$}
		    \IF{the $j$th row of $\mathbf{B}'$ is already in $\mathbf{B'_{\alpha_p}}$}
		    \STATE $j=j+1$
		    \ELSE
		    \STATE Replace the $i$th row of $\mathbf{B}'_{\alpha_p}$ with the $j$th row of $\mathbf{B}'$ 
		    \STATE Constitute $\mathbf{B}$ and calculate $\mathbf{G}_{{\text{R}}}$
		    \STATE Compute the MSE value, defined by $l$
		    \IF{$l < l_\text{min}$}
		                \STATE $l_\text{min}=l$
		                \STATE Update $\mathbf{B}'_{\alpha_p}$
		    \ENDIF
		    \ENDIF
		    \ENDFOR
		\ENDFOR
        \end{algorithmic}
	\end{algorithm}

\subsection{Greedy Search Algorithm}
The proposed Greedy Search algorithm implies that in each iteration, the comparator input combination with the highest MSE reduction is selected.
In a nutshell, the proposed algorithm searches for each individual comparator sequentially over all possible combinations for the minimum MSE input combination. In this process, the comparators sequentially change their inputs according to the MSE criterion. The detailed process is summarized in Algorithm \ref{alg:MSE}.

\subsection{Sequential SINR based Search}
\label{sec:SINR_seq_search}
The SINR before the quantization operation for the output signal of comparator $l$ with respect to the $k$th real-valued channel input is described by
\begin{align}
\mathrm{SINR}_{l,k} = \frac{ \frac{1}{2} \sigma_x^2    \vert [\mathbf{B}' \mathbf{H}_{\text{R}}]_{l,k} \vert^2    }{   \frac{1}{2} \sigma_x^2 \sum_{j=1, j \ne k}^{ 2 N_t}      \vert [\mathbf{B}' \mathbf{H}_{\text{R}}]_{l,j} \vert^2       +       \sigma_n^2     }    \text{.}
\label{SINR_virtual}
\end{align}
Moreover, the MSE of the $k$th real valued signal stream after filtering with $\mathbf{G}_{\text{R}}$
reads as
\begin{align}
\mathrm{MSE}_k =  [ \mathbf{G}_{\text{R}}^T  \mathbf{C}_{\mathbf{z}_{\mathcal{Q}_r}}     \mathbf{G}_{\text{R}}  -2 \mathbf{G}_{\text{R}}^T \mathbf{C}_{\mathbf{z}_\mathcal{Q}\mathbf{x}_{\text{R}}}    +\mathbf{C_{\mathbf{x}_{\text{R}}}}  ]_{k,k} \text{.}
\label{MSE_stream}
\end{align}
The proposed sequential SINR search based comparator network with $\alpha_p$ comparators is sequentially constructed in $\alpha_p$ steps. Implicitly, the objective of the sequential construction is the minimization of the maximum MSE associated with the $2N_t$ signal streams.
In each of the $\alpha_p$ steps the algorithm computes the MSE values for each real-valued signal stream based on the current state of the comparator network in combination with an appropriate receive filter. Based on these MSE values, the weakest signal stream is identified where the corresponding index is denoted as $k_{max}$.
Subsequently,
the SINR values of all possible comparator output signals with respect to the signal stream $k_{\text{max}}$ are considered, namely $\mathrm{SINR}_{k_{\text{max}},l}$ for $l \in \mathcal L$, where $L$ denotes the set of comparators, that are not included in the sequentially constructed comparator network.
The index associated with the maximum SINR is denoted as
$l_{\text{max}} = \mathrm{arg max}_{l \in \mathcal L} \
\mathrm{SINR}_{k_{\text{max}},l}$.
  Then the comparator network is extended by the comparator combination with index $l_{\text{max}}$ which corresponds to the highest SINR value for the weakest stream $k_{\text{max}}$ in terms of MSE.
The detailed process for the construction of a network with $\alpha$ comparators  is summarized in Algorithm \ref{alg:SINR_Seq}.
	 \begin{algorithm}[H]
\footnotesize
		\caption{Sequential SINR based Search}
		\label{alg:SINR_Seq}
		\begin{algorithmic}[1]
		
		\STATE Define the matrix $\mathbf{B}'$ \eqref{eq:fully_connected}
		for the  fully connected network
		 and compute the number of rows $\alpha_f$ 
		 \STATE Define the set $\mathcal L = \{1, \ldots, \alpha_f  \}$
		\STATE Compute $\mathrm{SINR}_{l,k}$ for all $k$ and $l$ according to \eqref{SINR_virtual}.
		
		\STATE For initialization consider the system without the comparator network
		with $\mathbf{B}=\mathbf{I}_{2N_r}$
		
		\FOR{$i = 1 : \alpha_p$}
		    \STATE 
		    Compute $\mathbf{G}_{{\text{R}}}$   in \eqref{equ_LRAMMSE}
		    due to $\mathbf{B}$

		    \STATE Compute  $\mathrm{MSE}_k$ for the $2K$ signal streams with \eqref{MSE_stream} 
		    
		    \STATE Find the index of the weakest signal $k_{\text{max}} = \mathrm{arg max}_k \mathrm{MSE}_k  $
		    
		    \STATE Find the index of the unused comparator in $\mathbf{B}'$ with the maximum $\mathrm{SINR}$ w.r.t.\  $k_{\text{max}}$
		    \STATE
		    $l_{\text{max}} = \mathrm{arg max}_{l \in \mathcal L} \
\mathrm{SINR}_{k_{\text{max}},l}$
		    \STATE Extend the comparator network matrix with the 
		    $l_{\text{max}}$th row of $\mathbf{B}'$
		    
		    \STATE
		    $\mathbf{B}=\begin{bmatrix}
\mathbf{B}; \mathbf{B}'_{l_{\text{max}}}\end{bmatrix}$

		    \STATE Update the set of unused comparators $\mathcal L = \{1, \ldots,l_{\text{max}}-1,l_{\text{max}}+1   , \ldots,      \alpha_f  \}$
		\ENDFOR
        \end{algorithmic}
	\end{algorithm}

\section{Cost Analysis and Power Consumption}
\label{sec:complexity}
In this section, we carry out an analysis ot the cost and the power consumption of the proposed comparator network aided receivers.
\subsection{Cost Analysis}
The corresponding complexity orders are summarized in Table \ref{tab:complexity}, where $\mathcal{O}(\cdot)$ is the big O notation. Note that the complexity order can be calculated as functions of the numbers of receive and transmit antennas $N_r$ and $N_t$ and additional comparators $\alpha_f$ or $\alpha_p$.
\begin{table}[ht]
\centering
{\footnotesize{
  \caption{Computational Complexity}
  \label{tab:complexity}
  \begin{tabular}{ | c  | c  | c | }
  \hline
    Approach &  Network Design  & LRA-MMSE Detection  \\ \hline \hline
No Network  & -  &  $\mathcal{O}((2 N_r)^3+2N_t(2N_r)^2+4N_rN_t)$
    \\ \hline
    Full Connection  & - &  $\mathcal{O}((2N_r +\alpha_f)^3+2N_t(2N_r+\alpha_f)^2+2N_t(2N_r+\alpha_f))$  \\ \hline
 MMSE based Greedy Search    & $\mathcal{O}(\alpha_p  (\alpha_f-\alpha_p)(2 N_r +\alpha_p)^3     )$  & $\mathcal{O}((2N_r +\alpha_p)^3+2N_t(2N_r+\alpha_p)^2+2N_t(2N_r+\alpha_p))$ \\ \hline
 SINR Sequential Search    & $\mathcal{O}( \sum_{i=1}^{\alpha_p}    
 (2N_r +i-1)^3 )$
 & $\mathcal{O}((2N_r +\alpha_p)^3+2N_t(2N_r+\alpha_p)^2+2N_t(2N_r+\alpha_p))$ \\ \hline
 
    Random Selection    & -  & $\mathcal{O}((2N_r +\alpha_p)^3+2N_t(2N_r+\alpha_p)^2+2N_t(2N_r+\alpha_p))$ \\ \hline
  \end{tabular}
  }}
\end{table}
The required computational and hardware costs\footnote{The costs do not include the crossbar switch required to select which antennas will be used as inputs on each comparator.} in terms of additional comparators are approximately calculated in Table \ref{tab:complexity2}. The values presented for the computational cost (arithmetic operations) are calculated using Table \ref{tab:complexity} and considering the following parameters: $N_r = 10$, $N_t = 2$, $\alpha_f = 190$ and $\alpha_p = 20$.
\begin{table}[ht]
\centering
\caption{Computational and Hardware Costs}
  \label{tab:complexity2}
{\footnotesize{
\begin{tabular}{|c|c|c|}
\hline
        Approach  & Computational Cost  & Hardware Cost     \\ \hline \hline
No Network  & $\mathcal{O}(9,68 \times 10^3)$  & -      \\ \hline 
Full connection  & $\mathcal{O}(9,44  \times 10^6 )$  & 190   \\ \hline
MMSE based Greedy Search  & $\mathcal{O}(2,18 \times 10^8   )$  & 20 \\ \hline
 SINR Sequential Search    &
$\mathcal{O}(2,24 \times 10^5  )$  & 20 \\ \hline
Random Selection  & $\mathcal{O}(7,06 \times 10^4)$  & 20     \\ \hline
\end{tabular}
}}
\end{table}

\subsection{Power consumption}
\label{sec:power}
Compared to the conventional multi-bit systems, the advantage of 1-bit ADCs or comparators is that they do not require automatic gain control (AGC). Table \ref{Energy} shows the power consumption at the receiver as a function of the quantization bits, \textcolor{r4}{which is calculated similarly to \cite{Roth2017}, as
\begin{equation}
P_{\text{1-bit}} = P_{\text{LO}}+N_r(P_{\text{LNA}}+P_{\text{H}}+2P_{\text{M}})+2N_r( \text{FOM}\times 2f_{\text{Nyquist}})
\end{equation}
\begin{equation}
P_{\text{Traditional}} = P_{\text{LO}}+N_r(P_{\text{LNA}}+P_{\text{H}}+2P_{\text{M}})+2N_r(P_{\text{AGC}}  + \text{FOM}\times 2^{q}f_{\text{Nyquist}})
\end{equation}
\begin{equation}
	P_{\text{CN}} = P_{\text{LO}}+N_r(P_{\text{LNA}}+P_{\text{H}}+2P_{\text{M}}) + (2N_r+\alpha)(\text{FOM}\times 2f_{\text{Nyquist}}),
\end{equation}}
where $P_{\text{LO}}$, $P_{\text{LNA}}$, $P_{\text{H}}$, $P_{\text{M}}$ and $P_{\text{AGC}}$ denote the power consumption in the local oscillator (LO), low noise amplifier (LNA), $\frac{\pi}{2}$ hybrid and LO buffer, mixer and AGC, respectively. The parameter $q$ refers to the number of quantization bits and $\alpha$ is the number of comparators in the comparator network. The power consumption of different hardware components in the receiver for mmWave communications \cite{Roth2017} is given as $P_{\text{LO}}=22.5\text{ mW}$, $P_{\text{LNA}}=5.4\text{ mW}$, $P_{\text{H}}=3\text{ mW}$, $P_{\text{AGC}}=2\text{ mW}$, $P_{\text{M}}=0.3\text{ mW}$, $P_{\text{AGC}}=2\text{ mW}$, $f_{\text{Nyquist}}=2.5\text{ GHz}$ and $\text{FOM}=15\text{ fJ}$. From the results, it can be seen that the proposed network consumes much less power than common multi-bit systems. 
\begin{table}[!htbp]
	\centering
	\caption{Receiver power consumption among comparator network and multi-bit systems, where $N_r=16$ and $\alpha=2N_r$}
	\begin{tabular}{|c|c|c|c|c|c|c|c|c|c|c|c}
		\hline
		\multirow{2}{*}{}                                                & \multirow{2}{*}{1-bit} & \multirow{2}{*}{CN} &  \multicolumn{9}{c|}{Multi-bit systems}                                                      \\ \cline{4-12} 
		&             &        & 2-bit & 3-bit & 4-bit & 5-bit & 6-bit & 7-bit & 8-bit & 9-bit & \multicolumn{1}{c|}{10-bit} \\ \hline
		\begin{tabular}[c]{@{}c@{}}Power consumption\\ (mW)\end{tabular} & 168.9 & 171.3              & 235.3 & 240.1 & 249.7 & 268.9 & 307.3 & 384.1 & 537.7 & 844.9 & \multicolumn{1}{c|}{1459.3} \\ \hline
	\end{tabular}
	\label{Energy}
\end{table}

\section{Sum Rate}
\label{sec:achievable_rates}

In this subsection, we analyze the sum-rate of the multiuser MIMO system with the proposed comparator networks. In particular, we consider a data transmission model that relies on the channel estimates from the LRA-MMSE channel estimator together with its error statistics and linear filtering for the detection process.

\subsection{Data Transmission with Linear Receiver}
It is considered that in the data transmission stage the $N_t$ users simultaneously transmit their data symbols represented by the vector $\mathbf{x}_{\text{R}}$ to the BS, which is a stacked vector with real and imaginary parts, as introduced in \eqref{real_sys_compact}. In the present study, real and imaginary parts represent independent data symbols. After processing by the comparators and ADCs, the signal reads as
\begin{equation}
\begin{split}
\mathbf{z}_{\mathcal{Q}_d} &= \mathcal{Q} (\mathbf{z}_{{\text{R}}_d}) = 
\mathcal{Q}(\mathbf{B}\mathbf{y}_{{\text{R}}_d}) = \mathcal{Q} (\mathbf{B}\mathbf{H}_{\text{R}}\mathbf{x}_{\text{R}}+ \mathbf{B}\mathbf{n}_{{\text{R}}_d})\\
&= \mathbf{A}_{{\text{R}}_d} \mathbf{B}\mathbf{H}_{\text{R}}\mathbf{x}_{\text{R}} + \mathbf{A}_{{\text{R}}_d} \mathbf{B}\mathbf{n}_{{\text{R}}_d} + \mathbf{n}_{{\text{R}}_{q, d}},
\end{split}
\label{zqd}
\end{equation}
where the same definitions from Section \ref{sec:channel_estimation} apply.
Note that the subscript $d$ is associated with data transmission stage.
Then, the LRA-LMMSE channel estimate \eqref{eq:h_estimated} is used to compute a linear receiver which provides an estimate of the data symbols transmitted from the $N_t$ users. In this context, the quantized signal is separated into $2N_t$ streams by multiplying the signal with the receiver filter matrix.
An example for the receiver matrix is given by \eqref{equ_LRAMMSE}, which in this case is computed based on the estimated channel. Therefore, we obtain
\begin{equation}
\begin{split}
\hat{\mathbf{x}}_{\text{R}}&=\mathbf{G}_{\text{R}} \mathbf{z}_{\mathcal{Q}_d} =  \mathbf{G}_{\text{R}} (\mathbf{A}_{{\text{R}}_d} \mathbf{B}\mathbf{H}_{\text{R}}\mathbf{x}_{\text{R}} + \mathbf{A}_{{\text{R}}_d} \mathbf{B}\mathbf{n}_{{\text{R}}_d} + \mathbf{n}_{{\text{R}}_{q, d}})\\
&= \mathbf{G}_{\text{R}} \mathbf{A}_{{\text{R}}_d} \mathbf{B}(\hat{\mathbf{H}}_{\text{R}}\mathbf{x}_{\text{R}} + \mathbfcal{E}_{\text{R}} \mathbf{x}_{\text{R}}) + \mathbf{G}_{\text{R}} \mathbf{A}_{{\text{R}}_d} \mathbf{B}\mathbf{n}_{{\text{R}}_d} + \mathbf{G}_{\text{R}} \mathbf{n}_{{\text{R}}_{q, d}},
\end{split}
\end{equation}
where $\hat{\mathbf{H}}_{\text{R}}$ is the estimated channel matrix described by \eqref{eq:H_hat} and $\mathbfcal{E}_{\text{R}} = \mathbf{H}_{\text{R}} -\hat{\mathbf{H}}_{\text{R}}$ is the channel estimation error matrix. 
In the sum rate analysis it is considered that each user corresponds to two real-valued channels. Then, the $k$th element represents an estimate of the signal of the $k$th real-valued channel, similarly as in \cite{7931630}, with $k \in [1,2N_t]$, which reads as
\begin{equation}
\begin{split}
\hat{\mathbf{x}}_{\text{R}_{k}} = \quad &\underbrace{\mathbf{g}_{\text{R}_{k}}^T \mathbf{A}_{{\text{R}}_d} \mathbf{B} \hat{\mathbf{h}}_{\text{R}_{k}}\mathbf{x}_{\text{R}_{k}}}_{\text{desired signal}} + \underbrace{\mathbf{g}_{\text{R}_{k}}^T\sum_{i \neq k}^{K} \mathbf{A}_{{\text{R}}_d} \mathbf{B} \hat{\mathbf{h}}_{\text{R}_i}\mathbf{x}_{\text{R}_i}}_{\text{interference}} + \underbrace{\mathbf{g}_{\text{R}_{k}}^T\sum_{i=1}^{K} \mathbf{A}_{{\text{R}}_d} \mathbf{B} \bm{\varepsilon}_{\text{R}_i}\mathbf{x}_{\text{R}_i}}_{\text{channel estimation error}}\\
&+ \underbrace{\mathbf{g}_{\text{R}_{k}}^T\mathbf{A}_{{\text{R}}_d} \mathbf{B}\mathbf{n}_{{\text{R}}_d}}_{\text{AWGN noise}} + \underbrace{\mathbf{g}_{\text{R}_{k}}^T\mathbf{n}_{{\text{R}}_{q, d}}}_{\text{quant. noise}},
\end{split}
\label{eq:x_hat_k}
\end{equation}
where $\mathbf{g}_{\text{R}_{k}}^T$ is the ${k}$th row of $\mathbf{G}_{\text{R}}$ and $\hat{\mathbf{h}}_{\text{R}_{k}}$ is the ${k}$th column of $\hat{\mathbf{H}}_{\text{R}}$. Moreover, $\bm{\varepsilon}_{\text{R}_{i}}$ is the ${i}$th column of the matrix $\mathbfcal{E}_{\text{R}}$.

\subsection{Sum Rate Lower Bound}
Since the Gaussian noise case corresponds to the worst case scenario, we can find a lower bound for the achievable rate by interpreting the quantization noise as Gaussian \cite{Mezghani2012CapacityLB}. The equivalent noise covariance matrix reads as
\begin{equation}
\begin{split}
\mathbf{C}_{\mathbf{n}_{{\text{R}}_{q, d}}} = \mathbf{C}_{\mathbf{z}_{\mathcal{Q}_d}} - \mathbf{A}_{{\text{R}}_d}\mathbf{C}_{\mathbf{z}_{{\text{R}}_d}}\mathbf{A}_{{\text{R}}_d}^T,
\label{eq:Cnqd}
\end{split}
\end{equation}
where $\mathbf{C}_{\mathbf{z}_{\mathcal{Q}_d}} = E\left[ \mathbf{z}_{\mathcal{Q}_d} \mathbf{z}_{\mathcal{Q}_d}^T \right]$ is the auto-correlation matrix of the quantized data signal, like in \eqref{Czqp}, and
$\mathbf{C}_{\mathbf{z}_{{\text{R}}_d}} = E\left[\mathbf{z}_{{\text{R}}_d}\mathbf{z}_{{\text{R}}_d}^T \right] = \frac{1}{2} \sigma_x^2 \mathbf{BH}_{\text{R}}{\mathbf{H}_{\text{R}}^T}\mathbf{B}^T+\frac{1}{2} \sigma_n^2 \mathbf{BB}^T$ is the auto-correlation matrix of the received data signal, as calculated in \eqref{eq:Czr}. With this and by considering Gaussian signaling, the ergodic achievable rate per real-valued channel is lower bounded by 
\begin{equation}
\scalebox{0.99}{$
I_{\text{R}_{k}} = E\left[ \frac{1}{2} \log_2 \left ( 1 +  \frac{  \sigma_x^2 \left\vert \mathbf{d}_{\text{R}_{k}} \hat{\mathbf{h}}_{\text{R}_{k}}
 \right\vert^2}{ \sigma_x^2 \sum_{i \neq {k}}^{K} \left\vert  \mathbf{d}_{\text{R}_{k}} \hat{\mathbf{h}}_{\text{R}_i} \right\vert^2 +  \sigma_x^2 \sum_{i=1}^{K} \left\vert \mathbf{d}_{\text{R}_{k}}\bm{\varepsilon}_{\text{R}_i} \right\vert^2 + \sigma_n^2 \left\vert\left\vert\mathbf{d}_{\text{R}_{k}}\right\vert\right\vert^2_2 + 2\mathbf{g}_{\text{R}_{k}}^T \mathbf{C}_{\mathbf{n}_{{\text{R}}_{q, d}}} \mathbf{g}_{\text{R}_{k}}} \right) \right] $}  \ \text{[bpcu]}   , 
\label{eq:I_tilde_k}
\end{equation}
where $\mathbf{d}_{\text{R}_{k}} =\mathbf{g}_{\text{R}_{k}}^T\mathbf{A}_{{\text{R}}_d} \mathbf{B}$ and the expectation operator is taken with respect to channel realizations and channel estimates.
According to \cite{Mezghani2012CapacityLB}, this method provides an accurate lower bound especially for the low SNR regime. 
Finally, the sum rate is lower-bounded by $\sum_{k=1}^{K} I_{\text{R}_{k}}$, which can be computed via Monte Carlo Simulations.
A detailed description is shown in Appendix \ref{subsec:derivation_of_Irk}.

\section{Numerical Results}
\label{sec:numerical_results}

In this section, an uplink single-cell 1-bit MIMO system is considered. The average receive SNR per user per antenna is defined as $10\log\left(\frac{\sigma_x^2}{N_t N_r \sigma_n^2}  \mathrm{E}\{ \mathrm{trace}  \{ \mathbf{H} \mathbf{H}^H    \}  \}  \right)$
and the modulation scheme is QPSK. For the lower bound on the sum rate, Gaussian signaling is considered.
The considered channel matrix has in general the form
$\mathbf{H} = \mathbf{H}_{\text{w}} \mathrm{diag}[ \sqrt{\beta_1} \ldots \sqrt{\beta_{N_t}}]$, where $\beta_i$ represents large scale coefficients and $[\mathbf{H}_{\text{w}}]_{n,m}$ are i.i.d. complex Gaussian random variables.

\subsection{Perfect CSI}
\label{subsec:perfect_CSI}

\subsubsection{\emph{Proof of Concept}: More Antennas versus the Comparator Network}
\begin{figure}[t]
\centering
\usetikzlibrary{positioning,calc}

\definecolor{mycolor1}{rgb}{0.00000,1.00000,1.00000}%
\definecolor{mycolor2}{rgb}{1.00000,0.00000,1.00000}%

\definecolor{mustard}{rgb}{0.92941,0.69020,0.12941}%

\definecolor{newpurple}{rgb}{0.5, 0 ,1}%

\definecolor{darkblue}{rgb}{0, 0.4470, 0.7410}

\pgfplotsset{every axis label/.append style={font=\footnotesize},
every tick label/.append style={font=\footnotesize},
every plot/.append style={ultra thick} 
}

\begin{tikzpicture}[font=\footnotesize] 

\begin{axis}[%
name=ber2,
ymode=log,
width  = 0.8\columnwidth,
height = 0.5\columnwidth,
scale only axis,
xmin  = -10,
xmax  = 30,
xlabel= {SNR [dB]},
xmajorgrids,
ymin=0.0001,
ymax=1,
ylabel={BER},
ymajorgrids,
legend entries={{No comparator network},
                {No comparator network plus antenna, $N_r + 1$},
				{Random selected comparator network},	
				{MMSE search based comparator network},
                {SINR search based comparator network},	},
legend style={fill=white, fill opacity=0.6, draw opacity=1,
text opacity =1,at={(0.99,0.99)}, anchor=north east,draw=black,fill=white,legend cell align=left,font=\footnotesize}
]

\addlegendimage{smooth,color=gray,solid, thick, mark=triangle,
y filter/.code={\pgfmathparse{\pgfmathresult-0}\pgfmathresult}}
\addlegendimage{smooth,color=darkblue,solid, thick, mark=none,
y filter/.code={\pgfmathparse{\pgfmathresult-0}\pgfmathresult}}
\addlegendimage{smooth,color=mustard,solid, thick, mark=o,
y filter/.code={\pgfmathparse{\pgfmathresult-0}\pgfmathresult}}
\addlegendimage{smooth,color=newpurple,solid, thick, mark=diamond,
y filter/.code={\pgfmathparse{\pgfmathresult-0}\pgfmathresult}}
\addlegendimage{smooth,color=green,solid, thick, mark=square,
y filter/.code={\pgfmathparse{\pgfmathresult-0}\pgfmathresult}}

\addplot+[smooth,color=gray,thick,solid, every mark/.append style={solid} ,mark=triangle,
y filter/.code={\pgfmathparse{\pgfmathresult-0}\pgfmathresult}]
  table[row sep=crcr]{%
  -10.0000    0.2561\\
   -5.0000    0.1521\\
         0    0.0753\\
    5.0000    0.0389\\
   10.0000    0.0258\\
   15.0000    0.0216\\
   20.0000    0.0203\\
   25.0000    0.0200\\
   30.0000    0.0199\\
   };

\addplot+[smooth,color=darkblue,thick,solid, thick, every mark/.append style={solid} ,mark=none,
y filter/.code={\pgfmathparse{\pgfmathresult-0}\pgfmathresult}]
  table[row sep=crcr]{%
  -10.0000    0.2432\\
   -5.0000    0.1371\\
         0    0.0620\\
    5.0000    0.0290\\
   10.0000    0.0179\\
   15.0000    0.0145\\
   20.0000    0.0134\\
   25.0000    0.0131\\
   30.0000    0.0131\\
};

\addplot+[smooth,color=mustard,thick,solid, thick, every mark/.append style={solid} ,mark=o,
y filter/.code={\pgfmathparse{\pgfmathresult-0}\pgfmathresult}]
  table[row sep=crcr]{%
  -10.0000    0.2537\\
   -5.0000    0.1488\\
         0    0.0713\\
    5.0000    0.0351\\
   10.0000    0.0222\\
   15.0000    0.0185\\
   20.0000    0.0174\\
   25.0000    0.0170\\
   30.0000    0.0169\\
};

\addplot+[smooth,color=newpurple,thick,solid, thick, every mark/.append style={solid} ,mark=diamond,
y filter/.code={\pgfmathparse{\pgfmathresult-0}\pgfmathresult}]
  table[row sep=crcr]{%
  -10.0000    0.2512\\
   -5.0000    0.1451\\
         0    0.0656\\
    5.0000    0.0285\\
   10.0000    0.0156\\
   15.0000    0.0119\\
   20.0000    0.0107\\
   25.0000    0.0104\\
   30.0000    0.0103\\
};

\addplot+[smooth,color=green,thick,solid, thick, every mark/.append style={solid} ,mark=square,
y filter/.code={\pgfmathparse{\pgfmathresult-0}\pgfmathresult}]
  table[row sep=crcr]{%
 -10.0000    0.2522\\
   -5.0000    0.1462\\
         0    0.0673\\
    5.0000    0.0292\\
   10.0000    0.0161\\
   15.0000    0.0121\\
   20.0000    0.0109\\
   25.0000    0.0106\\
   30.0000    0.0104\\
   };


\addplot+[smooth,color=gray,thick,solid, every mark/.append style={solid} ,mark=triangle,
y filter/.code={\pgfmathparse{\pgfmathresult-0}\pgfmathresult}]
  table[row sep=crcr]{%
  -10.0000    0.1760\\
   -5.0000    0.0693\\
         0    0.0173\\
    5.0000    0.0043\\
   10.0000    0.0017\\
   15.0000    0.0011\\
   20.0000    0.000955\\
   25.0000    0.000931\\
   30.0000    0.00093733\\
};

\addplot+[smooth,color=darkblue,thick,solid, thick, every mark/.append style={solid} ,mark=none,
y filter/.code={\pgfmathparse{\pgfmathresult-0}\pgfmathresult}]
  table[row sep=crcr]{%
 -10.0000    0.1687\\
   -5.0000    0.0632\\
         0    0.0145\\
    5.0000    0.0033\\
   10.0000    0.0012\\
   15.0000    0.000779\\
   20.0000    0.0006645\\
   25.0000    0.0006510\\
   30.0000    0.00062833\\
   };

\addplot+[smooth,color=mustard,thick,solid, thick, every mark/.append style={solid} ,mark=o,
y filter/.code={\pgfmathparse{\pgfmathresult-0}\pgfmathresult}]
  table[row sep=crcr]{%
  -10.0000    0.1745\\
   -5.0000    0.0679\\
         0    0.0164\\
    5.0000    0.0039\\
   10.0000    0.0015\\
   15.0000    0.000922\\
   20.0000    0.0007705\\
   25.0000    0.00074583\\
   30.0000    0.00072733\\
};

\addplot+[smooth,color=newpurple,thick,solid, thick, every mark/.append style={solid} ,mark=diamond,
y filter/.code={\pgfmathparse{\pgfmathresult-0}\pgfmathresult}]
  table[row sep=crcr]{%
 -10.0000    0.1726\\
   -5.0000    0.0654\\
         0    0.0144\\
    5.0000    0.0028\\
   10.0000    0.0008\\
   15.0000    0.000458\\
   20.0000    0.00035933\\
   25.0000    0.00031117\\
   30.0000    0.00031367\\
   };

\addplot+[smooth,color=green,thick,solid, thick, every mark/.append style={solid} ,mark=square,
y filter/.code={\pgfmathparse{\pgfmathresult-0}\pgfmathresult}]
  table[row sep=crcr]{%
 -10.0000    0.1731\\
   -5.0000    0.0660\\
         0    0.0146\\
    5.0000    0.0027\\
   10.0000    0.0007\\
   15.0000    0.0003545\\
   20.0000    0.0002586\\
   25.0000    0.00024083\\
   30.0000    0.0002445\\
   };


\draw (7.5,0.025) ellipse (0.2cm and 0.5cm);
\draw[dspconn]    (6.8,0.025) -- (-1,0.007) ;
\draw (-6.4,0.005) node [anchor=north west][inner sep=0.75pt]  [font=\footnotesize]  {$N_{r} = 8$};

\draw (7.5,0.0018) ellipse (0.2cm and 0.5cm);
\draw[dspconn]    (6.9,0.0015) -- (5,0.0009) ;
\draw (-1.5,0.0007) node [anchor=north west][inner sep=0.75pt]  [font=\footnotesize]  {$N_{r} = 16$};

\end{axis}
\end{tikzpicture}
\caption{Comparison between extra physical channels and comparator network with $\alpha_p=2$.
}
\label{fig:1MORE_ANTENNA}    
\end{figure}
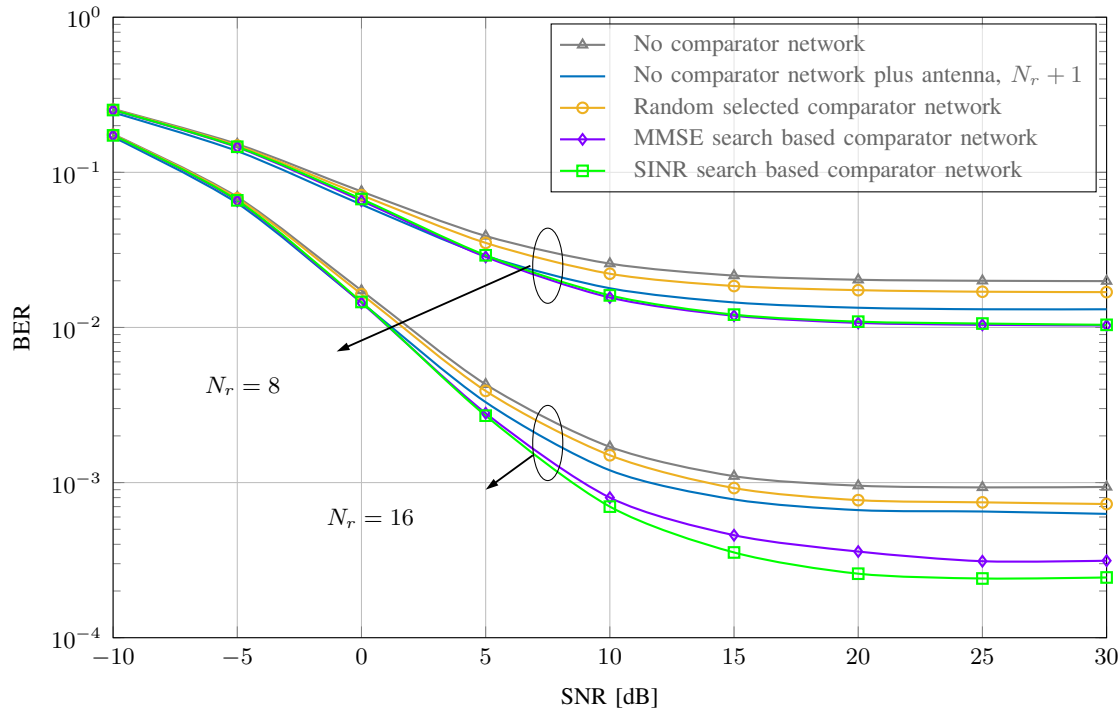

$\\$
In this subsection, systems with \textcolor{r4}{$N_t=3$} and different numbers of receive antennas $N_r$ are considered.
In this experiment Rayleigh fading is considered which implies $\beta_i=1$.
While performing the signal detection, the LRA-LMMSE detector from Subsection \ref{sec:LRAMMSE_detector} is applied in the system.
The BER performance is simulated based on 4000 random channel realizations and 1000 noise realizations. In order to verify the advantage of the additional comparator network, the performance of systems with partially connected networks are compared to systems with additional receive antennas. The configuration schemes with $N_r$ and $N_r + 1$ receive antennas are considered, which means the addition of a single antenna element. Since each extra antenna would correspond to two comparators, to make a fair comparison, the total number of extra comparators utilized by the partially connected networks is always $\alpha_p = 2$ in the presented simulations. 
As we can see from Fig. \ref{fig:1MORE_ANTENNA}, the random selected comparator network only has a benefit in comparison to the system without comparator network in a scenario with the same number of antennas. However, the optimized comparator networks, MMSE based Greedy Search and SINR search based networks, outperform the larger arrays in BER in the high SNR regime. 
Note that an increasing number of receive antennas improves the conditions for comparator network optimization in terms of possible input combinations, which can yield a performance gain in BER.
Finally, this experiment illustrates for different array sizes $N_r$, that 
extra virtual channels can be more beneficial than
an extra physical channel.

\vspace{1em}
\subsubsection{Proposed LRA-LMMSE Detector}
\begin{figure}[t]
\centering
%
%
%
\usetikzlibrary{positioning,calc}
\usetikzlibrary{spy}

\definecolor{mycolor1}{rgb}{0.00000,1.00000,1.00000}%
\definecolor{mycolor2}{rgb}{1.00000,0.00000,1.00000}%

\definecolor{mustard}{rgb}{0.92941,0.69020,0.12941}%

\definecolor{newpurple}{rgb}{0.5, 0 ,1}%

\definecolor{darkblue}{rgb}{0, 0.4470, 0.7410}

\definecolor{mycolor1}{rgb}{0.00000,1.00000,1.00000}%
\definecolor{mycolor2}{rgb}{1.00000,0.00000,1.00000}%

\pgfplotsset{every axis label/.append style={font=\footnotesize},
every tick label/.append style={font=\footnotesize}
}
\hspace{-2em}
\begin{tikzpicture}[spy using outlines={rectangle, magnification=8,connect spies}]

\begin{axis}[%
name=A,
ymode=log,
width  = 0.8\columnwidth,
height = 0.5\columnwidth,
scale only axis,
xmin  = -10,
xmax  = 30,
xlabel= {SNR [dB]},
xmajorgrids,
ymin=0.00001,
ymax=0.5,
ylabel={BER},
ymajorgrids,
legend style={legend columns=2,fill=white, fill opacity=0.6, draw opacity=1,
text opacity =1,
at={(0.99,0.99)},anchor=north east,draw=black,fill=white,legend cell align=left,font=\tiny}
]

\addplot+[smooth,color=gray,thick,thick, every mark/.append style={solid} ,mark=triangle,
y filter/.code={\pgfmathparse{\pgfmathresult-0}\pgfmathresult}]
  table[row sep=crcr]{%
  -10.0000    0.1838\\
   -5.0000    0.0819\\
         0    0.0282\\
    5.0000    0.0118\\
   10.0000    0.0070\\
   15.0000    0.0059\\
   20.0000    0.0053\\
   25.0000    0.0051\\
   30.0000    0.0051\\
};
\addlegendentry{No comp.\ network}

\addplot+[smooth,color=mustard,thick, thick, every mark/.append style={solid} ,mark=o,
y filter/.code={\pgfmathparse{\pgfmathresult-0}\pgfmathresult}]
  table[row sep=crcr]{%
  -10.0000    0.1653\\
   -5.0000    0.0625\\
         0    0.0143\\
    5.0000    0.0032\\
   10.0000    0.0014\\
   15.0000    0.0007\\
   20.0000    0.0006\\
   25.0000    0.0006\\
   30.0000    0.0006\\
   };
\addlegendentry{Random comp.\ network}

\addplot+[smooth,color=green,thick, thick, every mark/.append style={solid} ,mark=square,
y filter/.code={\pgfmathparse{\pgfmathresult-0}\pgfmathresult}]
  table[row sep=crcr]{%
  -10.0000    0.1647\\
   -5.0000    0.0613\\
         0    0.0123\\
    5.0000    0.0017\\
   10.0000    0.0003\\
   15.0000    0.0001\\
   20.0000    0.0001\\
   25.0000    0.0001\\
   30.0000    0.0001\\
   };
\addlegendentry{Seq.\ SINR search based comp.\ network}

\addplot+[smooth,color=newpurple,thick, thick, every mark/.append style={solid} ,mark=diamond,
y filter/.code={\pgfmathparse{\pgfmathresult-0}\pgfmathresult}]
  table[row sep=crcr]{%
  -10.0000    0.1579\\
   -5.0000    0.0534\\
         0    0.0082\\
    5.0000    0.0007\\
   10.0000    0.0000722\\
   15.0000    0.0000273\\
   20.0000    0.0000195\\
   25.0000    0.0000195\\
   30.0000    0.0000195\\
   };
\addlegendentry{MMSE search based comp.\ network}

\addplot+[smooth,color=darkblue,thick,thick, every mark/.append style={solid} ,mark=pentagon,
y filter/.code={\pgfmathparse{\pgfmathresult-0}\pgfmathresult}]
  table[row sep=crcr]{%
-10.0000    0.1310\\
   -5.0000    0.0311\\
         0    0.0016\\
    5.0000    0.0000117\\
};
\addlegendentry{Fully connected comp.\ network, $\alpha_f=496$}

\addplot+[smooth,color=black, solid, every mark/.append style={solid} ,mark=star,
y filter/.code={\pgfmathparse{\pgfmathresult-0}\pgfmathresult}]
  table[row sep=crcr]{%
  -10.0000    0.1521\\
   -5.0000    0.0425\\
         0    0.0037\\
    5.0000    0.000123\\
   10.0000    0.00000586\\
};
\addlegendentry{2-bit ADCs \cite{Bayes_joint_det_est}}

\addplot+[smooth,color=black,solid, every mark/.append style={solid} ,mark=asterisk,
y filter/.code={\pgfmathparse{\pgfmathresult-0}\pgfmathresult}]
  table[row sep=crcr]{%
  -10.0000    0.1389\\
   -5.0000    0.0316\\
         0    0.0013\\
    5.0000    0.00000195\\
};
\addlegendentry{3-bit ADCs \cite{Bayes_joint_det_est}}

\addplot+[smooth,color=black,solid, every mark/.append style={solid} ,mark=10-pointed star,
y filter/.code={\pgfmathparse{\pgfmathresult-0}\pgfmathresult}]
  table[row sep=crcr]{%
  -10.0000    0.1216\\
   -5.0000    0.0241\\
         0    0.0006\\
    5.0000    0.00000195\\
};
\addlegendentry{LMMSE, unquantized \cite{M_Joham_ZF}}

\end{axis}

\end{tikzpicture}%
\caption{BER performance with ($\alpha_p=2N_r$) and without comparator network with $N_t=4$, $N_r=16$.}
\label{fig:BER3x16}    
\end{figure}
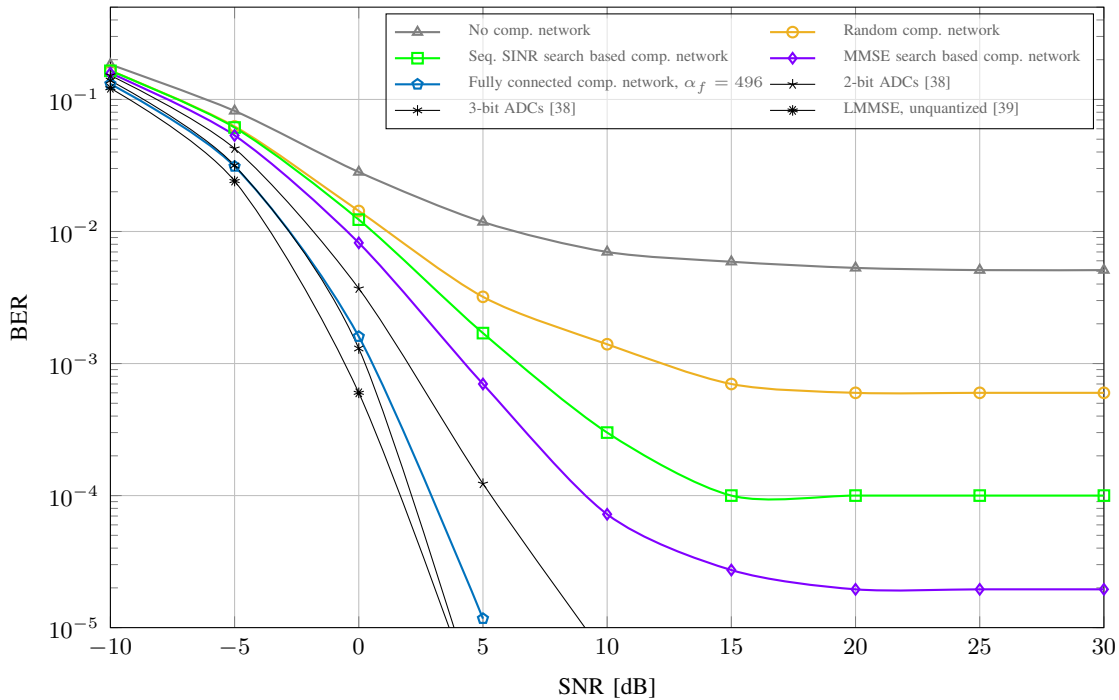

$\\$
In this subsection, numerical results on signal detection are shown.
The BER performance for the Rayleigh fading channel of fully and partially connected networks with dimensions $N_r=16$ and $N_t=4$ are shown in Fig.~\ref{fig:BER3x16}, where partially connected refers to comparator networks with $\alpha_p=2N_r$ comparators. The case of fully connected networks refers to comparator networks with, in this case, $\alpha_f=\tbinom{2N_r}{2}$ comparators.
 As additional reference methods 2-bit and 3-bit detection methods \cite{Bayes_joint_det_est} are considered which rely on message passing.
 For illustration of the high resolution case we consider linear MMSE detection \cite{M_Joham_ZF}.\footnote{For the 2-bit ADCs we consider thresholds $(0,\pm 1)$ and for the 3-bit ADCs we consider thresholds $(0,\pm 0.5, \pm 1, \pm 1.75 )$ in conjunction with channel normalization $\mathrm{tr}(\mathbf{H} \mathbf{H}^H )=N_r N_t$ and $\sigma_x^2=1$.}
Simulation results show that the system with fully connected network achieves the best BER performance with the cost of a large number of comparators and high computational complexity. In the partially connected networks, the MMSE based Greedy Search outperforms the random selection approach especially at high SNR, where the error floor is eliminated. A surprising observation is that the Greedy Search approach has almost the same BER performance as the fully connected method but with much less comparators. This shows great advantages of the greedy search based partially connected network. However, also the approach with the comparator network using random selected inputs is beneficial in terms of BER. While making comparison with the approach without additional comparator network, it can be seen that by adding extra $32$ comparators the performance gain is significant and the error floor goes down largely.

However, it should be mentioned that although the greedy search approach yields comparable good BER performance with less required comparators, its computational complexity is the highest among all the approaches. A reasonable performance-complexity trade-off can be achieved by considering the proposed Sequential SINR based Search.

\textcolor{r4}{
Fig.~\ref{fig:BER_Large_Scale_MIMO} shows results with larger dimensions where only approaches with moderate computational complexity are considered. Simulation results show that the error floor can be significantly reduced by taking into account a comparator network for detection.}

\begin{figure}[t]
\centering
%
%
%
\usetikzlibrary{positioning,calc}
\usetikzlibrary{spy}

\definecolor{mycolor1}{rgb}{0.00000,1.00000,1.00000}%
\definecolor{mycolor2}{rgb}{1.00000,0.00000,1.00000}%

\definecolor{mustard}{rgb}{0.92941,0.69020,0.12941}%

\definecolor{newpurple}{rgb}{0.5, 0 ,1}%

\definecolor{darkblue}{rgb}{0, 0.4470, 0.7410}

\definecolor{mycolor1}{rgb}{0.00000,1.00000,1.00000}%
\definecolor{mycolor2}{rgb}{1.00000,0.00000,1.00000}%

\pgfplotsset{every axis label/.append style={font=\footnotesize},
every tick label/.append style={font=\footnotesize}
}
\hspace{-2em}
\begin{tikzpicture}[spy using outlines={rectangle, magnification=8,connect spies}]

\begin{axis}[%
name=A,
ymode=log,
width  = 0.4\columnwidth,
height = 0.4\columnwidth,
scale only axis,
xmin  = -10,
xmax  = 30,
xlabel= {SNR [dB]},
xmajorgrids,
ymin=0.0001,
ymax=0.15,
ylabel={BER},
ymajorgrids,
legend entries={ 
},
legend columns=1,
legend style={fill=white, fill opacity=0.6, draw opacity=1,
text opacity =1,
at={(0.99,0.99)},anchor=north east,draw=black,fill=white,legend cell align=left,font=\tiny}
]


\definecolor{dark_red}{rgb}{0.5,0,0}
\definecolor{dark_blue}{rgb}{0,0,0.5}
\definecolor{dark_green}{rgb}{0, 0.33, 0.13}
\definecolor{purple}{rgb}{0.4940 0.1840 0.5560}

\addplot+[smooth,color=gray,thick,thick, every mark/.append style={solid} ,mark=triangle,
y filter/.code={\pgfmathparse{\pgfmathresult-0}\pgfmathresult}]
  table[row sep=crcr]{%
  -10.0000    0.1204\\
   -5.0000    0.0462\\
         0    0.0182\\
    5.0000    0.0103\\
   10.0000    0.0081\\
   15.0000    0.0075\\
   20.0000    0.0073\\
   25.0000    0.0072\\
   30.0000    0.0072\\
};
\addlegendentry{No comp.\ network}

\addplot+[smooth,color=mustard,thick, thick, every mark/.append style={solid} ,mark=o,
y filter/.code={\pgfmathparse{\pgfmathresult-0}\pgfmathresult}]
  table[row sep=crcr]{%
 -10.0000    0.1002\\
   -5.0000    0.0285\\
         0    0.0066\\
    5.0000    0.0023\\
   10.0000    0.0014\\
   15.0000    0.0012\\
   20.0000    0.0011\\
   25.0000    0.0011\\
   30.0000    0.0011\\
};
\addlegendentry{Random comp.\ network}

\addplot+[smooth,color=green,thick, thick, every mark/.append style={solid} ,mark=square,
y filter/.code={\pgfmathparse{\pgfmathresult-0}\pgfmathresult}]
  table[row sep=crcr]{%
  -10.0000    0.1022\\
   -5.0000    0.0292\\
         0    0.0060\\
    5.0000    0.0016\\
   10.0000    0.0008\\
   15.0000    0.0006\\
   20.0000    0.0005\\
   25.0000    0.0005\\
   30.0000    0.0005\\
};
\addlegendentry{Seq.\ SINR search based comp.\ network}

\addplot+[smooth,color=darkblue,thick, thick, every mark/.append style={solid} ,mark=pentagon,
y filter/.code={\pgfmathparse{\pgfmathresult-0}\pgfmathresult}]
  table[row sep=crcr]{%
   -10.0000    0.0606\\
   -5.0000    0.0052\\
         0    0.00003244\\
    5.0000    0.0000000305\\
         };
\addlegendentry{Fully connected comp.\ network, $\alpha_f=2016$}

\addplot+[smooth,color=black,solid, every mark/.append style={solid} ,mark=10-pointed star,
y filter/.code={\pgfmathparse{\pgfmathresult-0}\pgfmathresult}]
  table[row sep=crcr]{%
  -10.0000    0.0534\\
   -5.0000    0.0031\\
         0    0.00000357\\
};
\addlegendentry{LMMSE, unquantized \cite{M_Joham_ZF}}

\end{axis}

\begin{axis}[%
name=B,
at={($(A.east)+(50,0em)$)},
                anchor= west,
ymode=log,
width  = 0.4\columnwidth,
height = 0.4\columnwidth,
scale only axis,
xmin  = -15,
xmax  = 20,
xlabel= {SNR [dB]},
xmajorgrids,
ymin=0.0001,
ymax=0.5,
ylabel={BER},
ymajorgrids,
legend style={fill=white, fill opacity=0.6, draw opacity=1,
text opacity =1,
at={(0.99,0.99)},anchor=north east,draw=black,fill=white,legend cell align=left,font=\tiny}
]

\addplot+[smooth,color=gray,thick,thick, every mark/.append style={solid} ,mark=triangle,
y filter/.code={\pgfmathparse{\pgfmathresult-0}\pgfmathresult}]
  table[row sep=crcr]{%
  -15.0000    0.1611\\
  -10.0000    0.0671\\
   -5.0000    0.0236\\
         0    0.0110\\
    5.0000    0.0077\\
   10.0000    0.0067\\
   15.0000    0.0064\\
   20.0000    0.0063\\
};
\addlegendentry{No comp.\ network}

\addplot+[smooth,color=mustard,thick, thick, every mark/.append style={solid} ,mark=o,
y filter/.code={\pgfmathparse{\pgfmathresult-0}\pgfmathresult}]
  table[row sep=crcr]{%
  -15.0000    0.1422\\
  -10.0000    0.0480\\
   -5.0000    0.0105\\
         0    0.0028\\
    5.0000    0.0014\\
   10.0000    0.0010\\
   15.0000    0.0009\\
   20.0000    0.0009\\
   };
\addlegendentry{Random comp.\ network}

\addplot+[smooth,color=green,thick, thick, every mark/.append style={solid} ,mark=square,
y filter/.code={\pgfmathparse{\pgfmathresult-0}\pgfmathresult}]
  table[row sep=crcr]{%
  -15.0000    0.1457\\
  -10.0000    0.0510\\
   -5.0000    0.0114\\
         0    0.0027\\
    5.0000    0.0011\\
   10.0000    0.0008\\
   15.0000    0.0007\\
   20.0000    0.0007\\
   };
\addlegendentry{Seq.\ SINR search based comp.\ network}

\addplot+[smooth,color=black,solid, every mark/.append style={solid} ,mark=10-pointed star,
y filter/.code={\pgfmathparse{\pgfmathresult-0}\pgfmathresult}]
  table[row sep=crcr]{%
  -15.0000    0.0946\\
  -10.0000    0.0121\\
   -5.0000    0.0001\\
   };
\addlegendentry{LMMSE, unquantized \cite{M_Joham_ZF}}

\end{axis}

\end{tikzpicture}%
\caption{BER comparison with large scale MIMO system ($\alpha_p=2N_r$) , $N_r=32$, $N_t=8$ (left); $N_r=64$, $N_t=15$ (right).
}
\label{fig:BER_Large_Scale_MIMO}    
\end{figure}
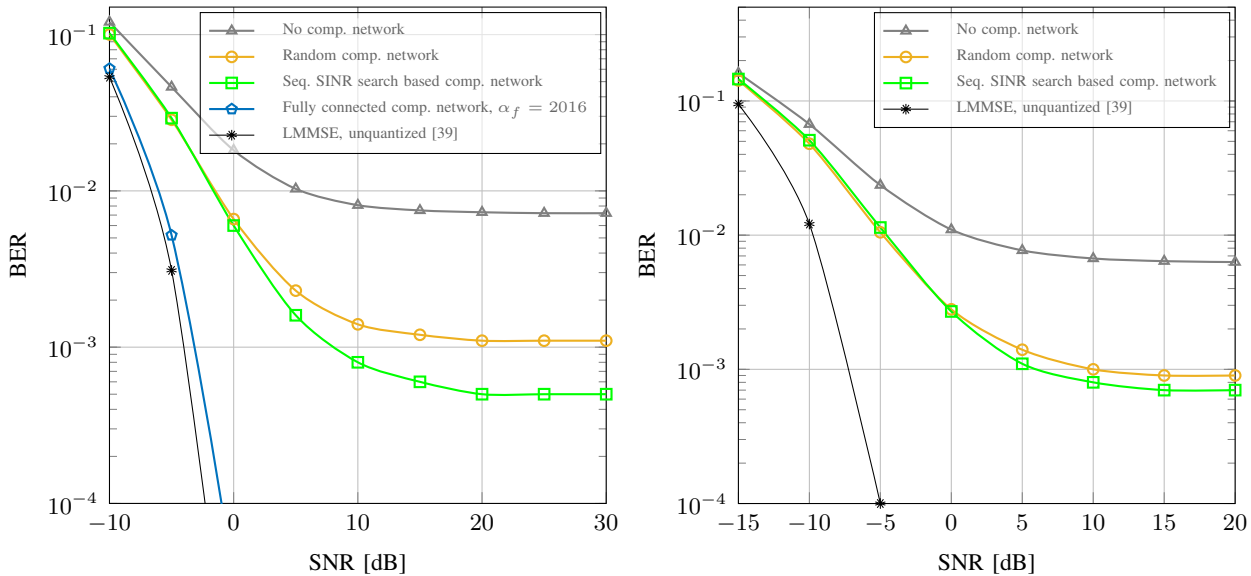
\begin{figure}[t]
\centering
%
%
%
\usetikzlibrary{positioning,calc}
\usetikzlibrary{spy}

\definecolor{mycolor1}{rgb}{0.00000,1.00000,1.00000}%
\definecolor{mycolor2}{rgb}{1.00000,0.00000,1.00000}%

\definecolor{mustard}{rgb}{0.92941,0.69020,0.12941}%

\definecolor{newpurple}{rgb}{0.5, 0 ,1}%

\definecolor{darkblue}{rgb}{0, 0.4470, 0.7410}

\definecolor{mycolor1}{rgb}{0.00000,1.00000,1.00000}%
\definecolor{mycolor2}{rgb}{1.00000,0.00000,1.00000}%

\pgfplotsset{every axis label/.append style={font=\footnotesize},
every tick label/.append style={font=\footnotesize}
}
\hspace{-2em}
\begin{tikzpicture}[spy using outlines={rectangle, magnification=8,connect spies}]

\begin{axis}[%
ymode=log,
width  = 0.8\columnwidth,
height = 0.5\columnwidth,
scale only axis,
xmin  = -10,
xmax  = 30,
xlabel= {SNR [dB]},
xmajorgrids,
ymin=0.0001,
ymax=0.5,
ylabel={BER},
ymajorgrids,
legend style={fill=white, fill opacity=0.6, draw opacity=1,
text opacity =1,
at={(0.01,0.01)},anchor=south west,draw=black,fill=white,legend cell align=left,font=\footnotesize}
]

\addplot+[smooth,color=gray,thick,thick, every mark/.append style={solid} ,mark=triangle,
y filter/.code={\pgfmathparse{\pgfmathresult-0}\pgfmathresult}]
  table[row sep=crcr]{%
  -10.0000    0.2335\\
   -5.0000    0.1529\\
         0    0.1015\\
    5.0000    0.0766\\
   10.0000    0.0637\\
   15.0000    0.0590\\
   20.0000    0.0571\\
   25.0000    0.0561\\
   30.0000    0.0560\\
};
\addlegendentry{No comp.\ network}

\addplot+[smooth,color=mustard,thick, thick, every mark/.append style={solid} ,mark=o,
y filter/.code={\pgfmathparse{\pgfmathresult-0}\pgfmathresult}]
  table[row sep=crcr]{%
   -10.0000    0.2187\\
   -5.0000    0.1353\\
         0    0.0812\\
    5.0000    0.0533\\
   10.0000    0.0394\\
   15.0000    0.0333\\
   20.0000    0.0311\\
   25.0000    0.0297\\
   30.0000    0.0295\\
   };
\addlegendentry{Random comp.\ network}

\addplot+[smooth,color=green,thick, thick, every mark/.append style={solid} ,mark=square,
y filter/.code={\pgfmathparse{\pgfmathresult-0}\pgfmathresult}]
  table[row sep=crcr]{%
  -10.0000    0.2180\\
   -5.0000    0.1323\\
         0    0.0724\\
    5.0000    0.0389\\
   10.0000    0.0218\\
   15.0000    0.0142\\
   20.0000    0.0109\\
   25.0000    0.0094\\
   30.0000    0.0090\\
   };
\addlegendentry{Seq.\ SINR search based comp.\ network}

\addplot+[smooth,color=newpurple,thick, thick, every mark/.append style={solid} ,mark=diamond,
y filter/.code={\pgfmathparse{\pgfmathresult-0}\pgfmathresult}]
  table[row sep=crcr]{%
  -10.0000    0.2122\\
   -5.0000    0.1225\\
         0    0.0627\\
    5.0000    0.0318\\
   10.0000    0.0177\\
   15.0000    0.0112\\
   20.0000    0.0085\\
   25.0000    0.0072\\
   30.0000    0.0069\\
   };
\addlegendentry{MMSE search based comp.\ network}

\addplot+[smooth,color=darkblue,thick,thick, every mark/.append style={solid} ,mark=pentagon,
y filter/.code={\pgfmathparse{\pgfmathresult-0}\pgfmathresult}]
  table[row sep=crcr]{%
  -10.0000    0.1922\\
   -5.0000    0.1037\\
         0    0.0485\\
    5.0000    0.0225\\
   10.0000    0.0105\\
   15.0000    0.0055\\
   20.0000    0.0036\\
   25.0000    0.0028\\
   30.0000    0.0025\\
};
\addlegendentry{Fully connected comp.\ network, $\alpha_f=496$}

\addplot+[smooth,color=black, solid, every mark/.append style={solid} ,mark=star,
y filter/.code={\pgfmathparse{\pgfmathresult-0}\pgfmathresult}]
  table[row sep=crcr]{%
  -10.0000    0.2092\\
   -5.0000    0.1172\\
         0    0.0592\\
    5.0000    0.0316\\
   10.0000    0.0195\\
   15.0000    0.0146\\
   20.0000    0.0126\\
   25.0000    0.0113\\
   30.0000    0.0110\\
};
\addlegendentry{2-bit ADCs \cite{Bayes_joint_det_est}}

\addplot+[smooth,color=black,solid, every mark/.append style={solid} ,mark=asterisk,
y filter/.code={\pgfmathparse{\pgfmathresult-0}\pgfmathresult}]
  table[row sep=crcr]{%
  -10.0000    0.1983\\
   -5.0000    0.1048\\
         0    0.0477\\
    5.0000    0.0209\\
   10.0000    0.0100\\
   15.0000    0.0057\\
   20.0000    0.0042\\
   25.0000    0.0035\\
   30.0000    0.0031\\
};
\addlegendentry{3-bit ADCs \cite{Bayes_joint_det_est}}

\addplot+[smooth,color=black,solid, every mark/.append style={solid} ,mark=10-pointed star,
y filter/.code={\pgfmathparse{\pgfmathresult-0}\pgfmathresult}]
  table[row sep=crcr]{%
  -10.0000    0.1842\\
   -5.0000    0.0953\\
         0    0.0405\\
    5.0000    0.0160\\
   10.0000    0.0056\\
   15.0000    0.0018\\
   20.0000    0.0005\\
   25.0000    0.0000977\\
   30         0.00000977\\
   };
\addlegendentry{LMMSE, unquantized \cite{M_Joham_ZF}}

\end{axis}
\end{tikzpicture}
\caption{BER comparison with large scale fading  ($\alpha_p=2N_r$), $N_r=16$, $N_t=4$.
}
\label{fig:Large_scale_fading}    
\end{figure}
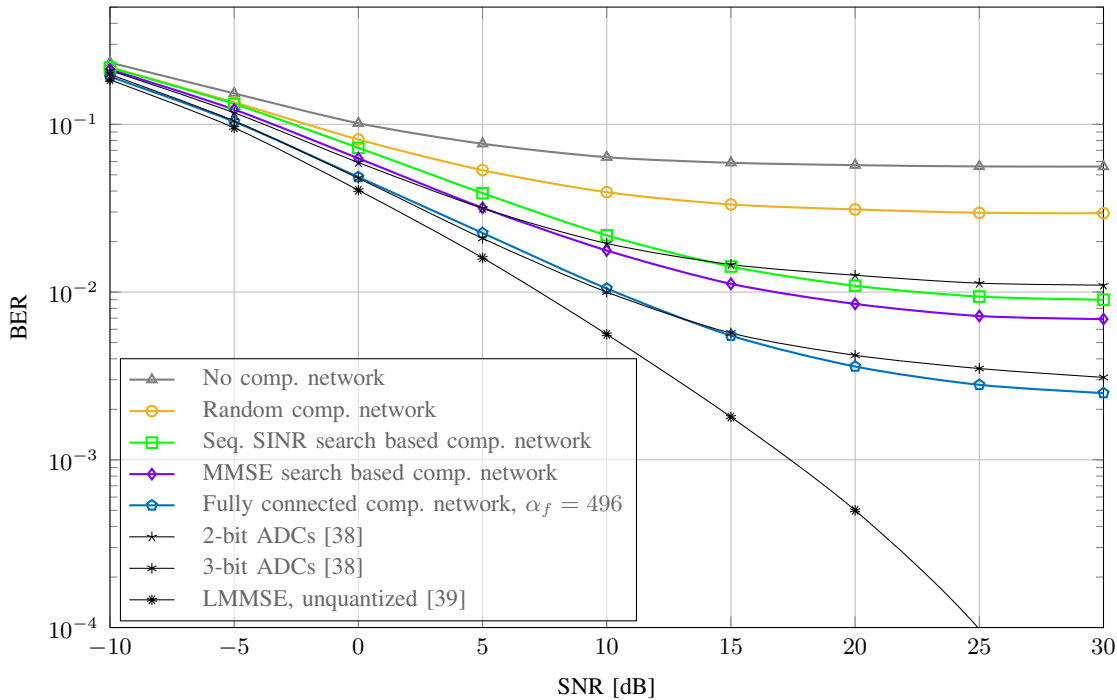
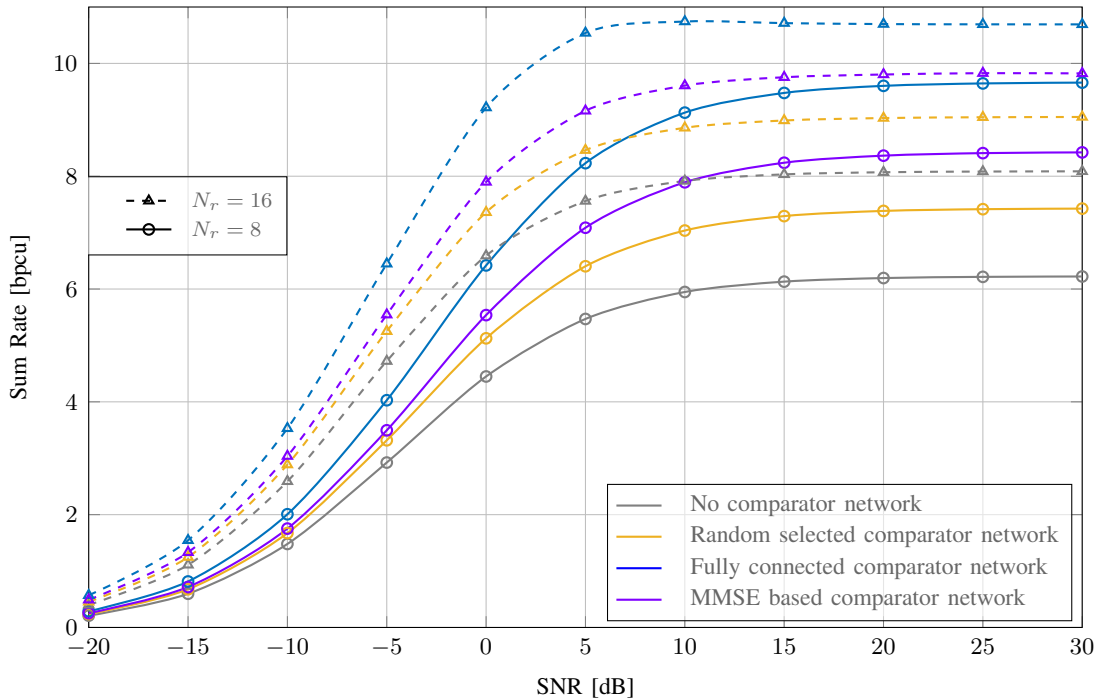
\begin{figure}[t]
\centering
%
%
%
\usetikzlibrary{positioning,calc}

\definecolor{mycolor1}{rgb}{0.00000,1.00000,1.00000}%
\definecolor{mycolor2}{rgb}{1.00000,0.00000,1.00000}%

\definecolor{mustard}{rgb}{0.92941,0.69020,0.12941}%

\definecolor{newpurple}{rgb}{0.5, 0 ,1}%

\definecolor{darkblue}{rgb}{0, 0.4470, 0.7410}

\pgfplotsset{every axis label/.append style={font=\footnotesize},
every tick label/.append style={font=\footnotesize},
every plot/.append style={ultra thick} 
}

\begin{tikzpicture}[font=\footnotesize] 

\begin{axis}[%
name=mse,
width  = 0.8\columnwidth,
height = 0.5\columnwidth,
scale only axis,
xmin  = -20,
xmax  = 30,
xlabel= {SNR [dB]},
xmajorgrids,
ymin=0,
ymax=11,
ylabel={Sum Rate [bpcu]},
ymajorgrids,
legend entries={No comparator network,
                Random selected comparator network,
				Fully connected comparator network,	
				MMSE based comparator network,															},
legend style={fill=white, fill opacity=0.6, draw opacity=1,
text opacity =1,at={(0.99,0.01)}, anchor= south east,draw=black,fill=white,legend cell align=left,font=\footnotesize}
]

\addlegendimage{smooth,color=gray,solid, thick, mark=none,
y filter/.code={\pgfmathparse{\pgfmathresult-0}\pgfmathresult}}
\addlegendimage{smooth,color=mustard,solid, thick, mark=none,
y filter/.code={\pgfmathparse{\pgfmathresult-0}\pgfmathresult}}
\addlegendimage{smooth,color=blue,solid, thick, mark=none,
y filter/.code={\pgfmathparse{\pgfmathresult-0}\pgfmathresult}}
\addlegendimage{smooth,color=newpurple,solid, thick, mark=none,
y filter/.code={\pgfmathparse{\pgfmathresult-0}\pgfmathresult}}

\addplot+[smooth,color=gray,thick,solid, every mark/.append style={solid} ,mark=o,
y filter/.code={\pgfmathparse{\pgfmathresult-0}\pgfmathresult}]
  table[row sep=crcr]{%
  -20.0000    0.2090\\
  -15.0000    0.5974\\
  -10.0000    1.4805\\
   -5.0000    2.9237\\
         0    4.4497\\
    5.0000    5.4692\\
   10.0000    5.9474\\
   15.0000    6.1311\\
   20.0000    6.1947\\
   25.0000    6.2156\\
   30.0000    6.2223\\
};

\addplot+[smooth,color=mustard,thick,solid, thick, every mark/.append style={solid} ,mark=o,
y filter/.code={\pgfmathparse{\pgfmathresult-0}\pgfmathresult}]
  table[row sep=crcr]{%
  -20.0000    0.2377\\
  -15.0000    0.6770\\
  -10.0000    1.6728\\
   -5.0000    3.3183\\
         0    5.1260\\
    5.0000    6.4034\\
   10.0000    7.0371\\
   15.0000    7.2930\\
   20.0000    7.3849\\
   25.0000    7.4157\\
   30.0000    7.4256\\
   };

\addplot+[smooth,color=darkblue,thick,solid, thick, every mark/.append style={solid} ,mark=o,
y filter/.code={\pgfmathparse{\pgfmathresult-0}\pgfmathresult}]
  table[row sep=crcr]{%
  -20.0000    0.2877\\
  -15.0000    0.8153\\
  -10.0000    2.0072\\
   -5.0000    4.0286\\
         0    6.4185\\
    5.0000    8.2314\\
   10.0000    9.1274\\
   15.0000    9.4764\\
   20.0000    9.6021\\
   25.0000    9.6452\\
   30.0000    9.6594\\
   };

\addplot+[smooth,color=newpurple,thick,solid, thick, every mark/.append style={solid} ,mark=o,
y filter/.code={\pgfmathparse{\pgfmathresult-0}\pgfmathresult}]
  table[row sep=crcr]{%
  -20.0000    0.2541\\
  -15.0000    0.7177\\
  -10.0000    1.7529\\
   -5.0000    3.4954\\
         0    5.5377\\
    5.0000    7.0865\\
   10.0000    7.8920\\
   15.0000    8.2390\\
   20.0000    8.3647\\
   25.0000    8.4096\\
   30.0000    8.4223\\
};


\addplot+[smooth,color=gray,thick,dashed, every mark/.append style={solid} ,mark=triangle,
y filter/.code={\pgfmathparse{\pgfmathresult-0}\pgfmathresult}]
  table[row sep=crcr]{%
  -20.0000    0.4016\\
  -15.0000    1.1102\\
  -10.0000    2.5906\\
   -5.0000    4.7215\\
         0    6.5909\\
    5.0000    7.5608\\
   10.0000    7.9160\\
   15.0000    8.0328\\
   20.0000    8.0706\\
   25.0000    8.0828\\
   30.0000    8.0867\\
};

\addplot+[smooth,color=mustard,thick,dashed, thick, every mark/.append style={solid} ,mark=triangle,
y filter/.code={\pgfmathparse{\pgfmathresult-0}\pgfmathresult}]
  table[row sep=crcr]{%
  -20.0000    0.4544\\
  -15.0000    1.2473\\
  -10.0000    2.8872\\
   -5.0000    5.2512\\
         0    7.3598\\
    5.0000    8.4616\\
   10.0000    8.8580\\
   15.0000    8.9880\\
   20.0000    9.0313\\
   25.0000    9.0455\\
   30.0000    9.0502\\
};

\addplot+[smooth,color=darkblue,thick,dashed, thick, every mark/.append style={solid} ,mark=triangle,
y filter/.code={\pgfmathparse{\pgfmathresult-0}\pgfmathresult}]
  table[row sep=crcr]{%
  -20.0000    0.5704\\
  -15.0000    1.5450\\
  -10.0000    3.5297\\
   -5.0000    6.4476\\
         0    9.2169\\
    5.0000   10.5397\\
   10.0000   10.7423\\
   15.0000   10.7140\\
   20.0000   10.6972\\
   25.0000   10.6929\\
   30.0000   10.6920\\
};

\addplot+[smooth,color=newpurple,thick,dashed, thick, every mark/.append style={solid} ,mark=triangle,
y filter/.code={\pgfmathparse{\pgfmathresult-0}\pgfmathresult}]
  table[row sep=crcr]{%
  -20.0000    0.4912\\
  -15.0000    1.3322\\
  -10.0000    3.0379\\
   -5.0000    5.5446\\
         0    7.8982\\
    5.0000    9.1592\\
   10.0000    9.6068\\
   15.0000    9.7547\\
   20.0000    9.8032\\
   25.0000    9.8270\\
   30.0000    9.8226\\
};


\addplot[smooth,color=black, thick, every mark/.append style={solid} ,mark=o,
y filter/.code={\pgfmathparse{\pgfmathresult-0}\pgfmathresult}]
  table[row sep=crcr]{%
	1 -2\\
};\label{P72}

\addplot[smooth,color=black,thick,dashed, every mark/.append style={solid} ,mark=triangle, 
y filter/.code={\pgfmathparse{\pgfmathresult-0}\pgfmathresult}]
  table[row sep=crcr]{%
	0 -2\\
};\label{P73}

\node [draw,fill=white, fill opacity=0.6,draw opacity=1,
text opacity =1,at ={(-10,8)}, anchor= north east,draw=black,fill=white,font=\scriptsize]  {
\setlength{\tabcolsep}{0.5mm}
\renewcommand{\arraystretch}{.8}
\begin{tabular}{l}
\ref{P73}{\hspace{0.15cm} $N_r=16$}\\
\ref{P72}{\hspace{0.15cm} $N_r=8$}\\
\end{tabular}
};

\end{axis}
\end{tikzpicture}
\caption{Sum Rate comparisons with and without comparator network, under perfect CSI.} 
\label{fig:AR_2x4_Perfect}    
\end{figure}

In addition, a channel with large scale fading is considered in Fig.~\ref{fig:Large_scale_fading}.
In this context, the log-distance path loss model \cite[p.~104]{Rappaport_1996} is considered which implies that $\beta_i$ is proportional to $  \left( \frac{d_0}{d_i} \right)^{n_{\text{PL}}}$, where $d_0$ is a reference distance and $d_i$ is the user's distance to the BS and $n_{\text{PL}}$ is the path loss exponent chosen equal to 3. The users are uniformly distributed on a disc with 500m radius with the BS in the center.
It can be observed that receivers with optimized comparator network show a relatively good BER performance in the presence of large scale fading.
\vspace{1em}
\subsubsection{Sum Rate Analysis}
$\\$
In this subsection, \textcolor{r4}{$N_t=3$ users and different numbers of base station antennas are considered.} The sum rates are obtained by averaging over $2000$ different Rayleigh fading channels and $2000$ noise realizations per channel. The comparison between the lower bound of the ergodic sum rate with and without the comparator network under perfect CSI is presented in Fig. \ref{fig:AR_2x4_Perfect}, where the system that utilizes the additional comparator networks shows a significant benefit. In this experiment, the randomly connected network has $\alpha_p=2N_r$ comparators, while the fully connected have $\alpha_f=\tbinom{2N_r}{2}$ comparators. We can see that the system with fully connected network achieves the best sum rate performance, followed by the system with random selected inputs. \textcolor{r4}{Receivers with a comparator network of size $\alpha_p=2N_r$ are comparable with receivers with twice the number of antennas.
For random comparator networks, the performance of the 
receivers with twice the number of antennas serves as an upper bound, as verified in Fig.~\ref{fig:AR_2x4_Perfect}. 
}

\subsection{Imperfect CSI}
\label{subsec:imperfect_CSI}

In this subsection, the pilot sequences are column-wise orthogonal with length $ \tau= N_t$, i.e., $ \mathbf{\Phi}^T\mathbf{\Phi} =\tau \sigma_x^2 \mathbf{I}_{N_t}$. Two types of network design are considered: random selected and fully connected networks. The former randomly selects $\alpha_p$ out of $\alpha_f$ ($\alpha_p<<\alpha_f$) comparators. The fully connected networks refers to comparator networks where all possible combinations are considered, meaning $\alpha_f=\tbinom{2N_r}{2}$ comparators. In all experiments Rayleigh fading channels are considered. 
\begin{figure}[ht]
\centering
%
%
%
\usetikzlibrary{positioning,calc}

\definecolor{mycolor1}{rgb}{0.00000,1.00000,1.00000}%
\definecolor{mycolor2}{rgb}{1.00000,0.00000,1.00000}%

\definecolor{mustard}{rgb}{0.92941,0.69020,0.12941}%

\definecolor{darkblue}{rgb}{0, 0.4470, 0.7410}

\pgfplotsset{every axis label/.append style={font=\footnotesize},
every tick label/.append style={font=\footnotesize},
every plot/.append style={ultra thick} 
}

\begin{tikzpicture}[font=\footnotesize] 

\begin{axis}[%
name=mse,
width  = 0.8\columnwidth,
height = 0.5\columnwidth,
scale only axis,
xmin  = -10,
xmax  = 30,
xlabel= {SNR [dB]},
xmajorgrids,
ymin=0,
ymax=55,
ylabel={MSE},
ymajorgrids,
legend entries={No comparator network,
                Random selected comparator network,
									2-bit ADCs \cite{Bayes_joint_det_est},
									3-bit ADCs \cite{Bayes_joint_det_est},
								Fully connected comparator network,	
								LMMSE unquantized \cite{Biguesh_2006}},
legend style={fill=white, fill opacity=0.6, draw opacity=1,
text opacity =1,at={(0.99,0.99)}, anchor=north east,draw=black,fill=white,legend cell align=left,font=\footnotesize}
]

\addlegendimage{smooth,color=gray,solid, thick, mark=none,
y filter/.code={\pgfmathparse{\pgfmathresult-0}\pgfmathresult}}
\addlegendimage{smooth,color=mustard,solid, thick, mark=none,
y filter/.code={\pgfmathparse{\pgfmathresult-0}\pgfmathresult}}
\addlegendimage{smooth,color=magenta,solid, thick, mark=none,
y filter/.code={\pgfmathparse{\pgfmathresult-0}\pgfmathresult}}
\addlegendimage{smooth,color=green,solid, thick, mark=none,
y filter/.code={\pgfmathparse{\pgfmathresult-0}\pgfmathresult}}
\addlegendimage{smooth,color=darkblue,solid, thick, mark=none,
y filter/.code={\pgfmathparse{\pgfmathresult-0}\pgfmathresult}}
\addlegendimage{smooth,color=black,solid, thick, mark=none,
y filter/.code={\pgfmathparse{\pgfmathresult-0}\pgfmathresult}}

\addplot+[smooth,color=gray,thick, every mark/.append style={solid} ,mark=none,
y filter/.code={\pgfmathparse{\pgfmathresult-0}\pgfmathresult}]
  table[row sep=crcr]{%
  -10.0000   52.3590\\
   -5.0000   41.2454\\
         0   31.4051\\
    5.0000   26.2414\\
   10.0000   24.2501\\
   15.0000   23.5759\\
   20.0000   23.3579\\
   25.0000   23.2885\\
   30.0000   23.2665\\
};

\addplot+[only marks,color=gray, thick, every mark/.append style={solid} ,mark=o,
y filter/.code={\pgfmathparse{\pgfmathresult-0}\pgfmathresult}]
  table[row sep=crcr]{%
  -10.0000   52.4181\\
   -5.0000   41.3427\\
         0   31.5184\\
    5.0000   26.3690\\
   10.0000   24.3793\\
   15.0000   23.7035\\
   20.0000   23.4880\\
   25.0000   23.4164\\
   30.0000   23.3941\\
   };

\addplot+[smooth,color=mustard,thick, thick, every mark/.append style={solid} ,mark=none,
y filter/.code={\pgfmathparse{\pgfmathresult-0}\pgfmathresult}]
  table[row sep=crcr]{%
-10.0000   50.7058\\
   -5.0000   38.0140\\
         0   26.7763\\
    5.0000   20.8793\\
   10.0000   18.6052\\
   15.0000   17.8353\\
   20.0000   17.5864\\
   25.0000   17.5071\\
   30.0000   17.4820\\
};

\addplot+[only marks,color=mustard,thick, every mark/.append style={solid} ,mark=o,
y filter/.code={\pgfmathparse{\pgfmathresult-0}\pgfmathresult}]
  table[row sep=crcr]{%
 -10.0000   50.8003\\
   -5.0000   38.1597\\
         0   27.0195\\
    5.0000   21.1365\\
   10.0000   18.8553\\
   15.0000   18.0892\\
   20.0000   17.8489\\
   25.0000   17.7921\\
   30.0000   17.7556\\
};

\addplot+[smooth,color=mustard,thick, thick, dashed, every mark/.append style={solid} ,mark=none,
y filter/.code={\pgfmathparse{\pgfmathresult-0}\pgfmathresult}]
  table[row sep=crcr]{%
  -10.0000   49.0366\\
   -5.0000   34.7513\\
         0   22.1025\\
    5.0000   15.4652\\
   10.0000   12.9055\\
   15.0000   12.0390\\
   20.0000   11.7588\\
   25.0000   11.6695\\
   30.0000   11.6413\\
};

\addplot+[smooth,color=darkblue, solid,thick, thick, every mark/.append style={solid} ,mark=none,
y filter/.code={\pgfmathparse{\pgfmathresult-0}\pgfmathresult}]
  table[row sep=crcr]{%
  -10.0000   47.1993\\
   -5.0000   31.1599\\
         0   16.9580\\
    5.0000    9.5057\\
   10.0000    6.6317\\
   15.0000    5.6588\\
   20.0000    5.3442\\
   25.0000    5.2440\\
   30.0000    5.2122\\
};

\addplot+[only marks,color=darkblue,thick, every mark/.append style={solid} ,mark=o,
y filter/.code={\pgfmathparse{\pgfmathresult-0}\pgfmathresult}]
  table[row sep=crcr]{%
  -10.0000   47.2317\\
   -5.0000   31.1593\\
         0   17.0276\\
    5.0000    9.5123\\
   10.0000    6.6306\\
   15.0000    5.6699\\
   20.0000    5.3561\\
   25.0000    5.2619\\
   30.0000    5.2214\\
};

\addplot+[only marks,color=magenta,thick, every mark/.append style={solid} ,mark=o,
y filter/.code={\pgfmathparse{\pgfmathresult-0}\pgfmathresult}]
  table[row sep=crcr]{%
  -10.0000   49.3438\\
   -5.0000   34.0580\\
         0   20.3853\\
    5.0000   13.3718\\
   10.0000   11.0504\\
   15.0000   10.3474\\
   20.0000   10.1022\\
   25.0000    9.9817\\
   30.0000    9.9794\\
};

\addplot+[only marks,color=green,thick, every mark/.append style={solid} ,mark=o,
y filter/.code={\pgfmathparse{\pgfmathresult-0}\pgfmathresult}]
  table[row sep=crcr]{%
  -10.0000   47.8483\\
   -5.0000   30.9537\\
         0   16.0971\\
    5.0000    9.1852\\
   10.0000    7.2086\\
   15.0000    6.5925\\
   20.0000    6.4344\\
   25.0000    6.3611\\
   30.0000    6.3959\\
};

\addplot+[smooth,color=black,solid, thick, thick, every mark/.append style={solid} ,mark=none,
y filter/.code={\pgfmathparse{\pgfmathresult-0}\pgfmathresult}]
  table[row sep=crcr]{%
  -10.0000   45.7143\\
   -5.0000   28.2572\\
         0   12.8000\\
    5.0000    4.6890\\
   10.0000    1.5610\\
   15.0000    0.5020\\
   20.0000    0.1596\\
   25.0000    0.0506\\
   30.0000    0.0160\\
};

\addplot+[only marks,color=black,thick, every mark/.append style={solid} ,mark=o,
y filter/.code={\pgfmathparse{\pgfmathresult-0}\pgfmathresult}]
  table[row sep=crcr]{%
  -10.0000   45.8498\\
   -5.0000   28.2182\\
         0   12.8292\\
    5.0000    4.6646\\
   10.0000    1.5500\\
   15.0000    0.5011\\
   20.0000    0.1598\\
   25.0000    0.0505\\
   30.0000    0.0161\\
};

\addplot[smooth,color=black, thick, every mark/.append style={solid} ,mark=none,
y filter/.code={\pgfmathparse{\pgfmathresult-0}\pgfmathresult}]
  table[row sep=crcr]{%
	1 -10\\
};\label{P31}

\addplot[only marks,color=black,thick, every mark/.append style={solid} ,mark=o, 
y filter/.code={\pgfmathparse{\pgfmathresult-0}\pgfmathresult}]
  table[row sep=crcr]{%
	0 -10\\
};\label{P32}

\addplot[smooth, dashed, color=black,thick, every mark/.append style={solid} ,mark=none, 
y filter/.code={\pgfmathparse{\pgfmathresult-0}\pgfmathresult}]
  table[row sep=crcr]{%
	0 -10\\
};\label{P99}

\node [draw,fill=white, fill opacity=0.6,draw opacity=1,
text opacity =1,at ={(29.5,25)}, anchor= south east,draw=black,fill=white,font=\footnotesize]  {
\setlength{\tabcolsep}{0.5mm}
\renewcommand{\arraystretch}{.8}
\begin{tabular}{l}
\ref{P31}{\hspace{0.15cm} Analytical Result \eqref{eq:mse} }\\
\hspace{0.15cm} \ref{P32}{\hspace{0.4cm} Numerical Result}\\
 \ref{P99}{\hspace{0.15cm} Approximation \eqref{eq:correlation_estimation}}
  \\
\end{tabular}
};

\end{axis}
\end{tikzpicture}
\caption{MSE comparisons of LRA-LMMSE channel estimators  with and without comparator network in $4\times16$ MIMO systems.} 
\label{fig:MSE_3x16}    
\end{figure}
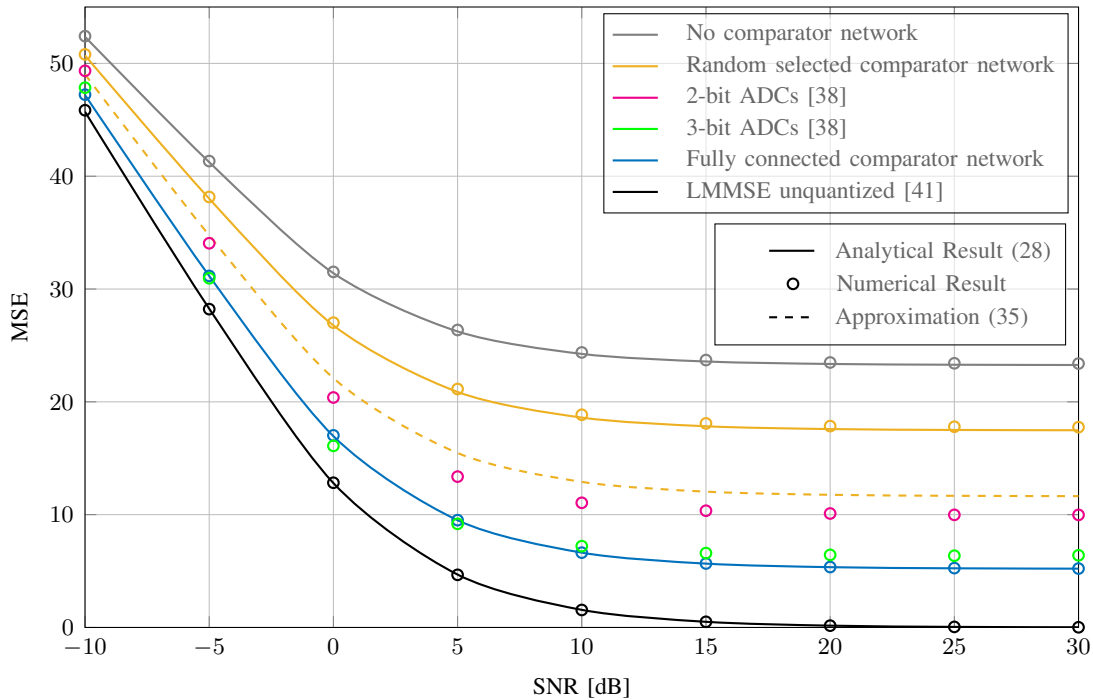
\subsubsection{Channel Estimation}
$\\$
In this subsection, an uplink single-cell 1-bit MIMO system with comparator network, $N_r=16$ is considered. The presented performance plots are obtained by taking the average over $4000$ different Rayleigh fading channel realizations and $100$ noise realizations per channel. Moreover, the lines labeled as ``Analytical Result'' are obtained with \eqref{eq:mse} while the marks labeled as ``Numerical Result'' are obtained with the MSE of the simulated channel estimator in (\ref{eq:h_estimated}). In the experiment, the partially connected networks have $\alpha_p=2 N_r=32$ comparators, while the fully connected have $\alpha_f=\tbinom{2N_r}{2}=496$ comparators.
As a performance reference a channel estimator with 2-bit quantization and 3-bit quantization is considered \cite{Bayes_joint_det_est}, where the thresholds are those used before in the detection part.
For illustration of the high-resolution case we consider linear MMSE channel estimation \cite{Biguesh_2006}.

The MSE comparison between the LRA-LMMSE channel estimators with fully and partially connected comparator networks are shown in Fig. \ref{fig:MSE_3x16}. The numerical and analytical results are aligned, which confirms the accuracy of the proposed model. As expected, the  system with the fully connected method  achieves  the  best MSE performance. However, it can be seen that the approach with the comparator network using random selected inputs is also beneficial in terms of MSE in comparison to the case without a comparator network. 
\begin{figure}[ht]
\centering
\input{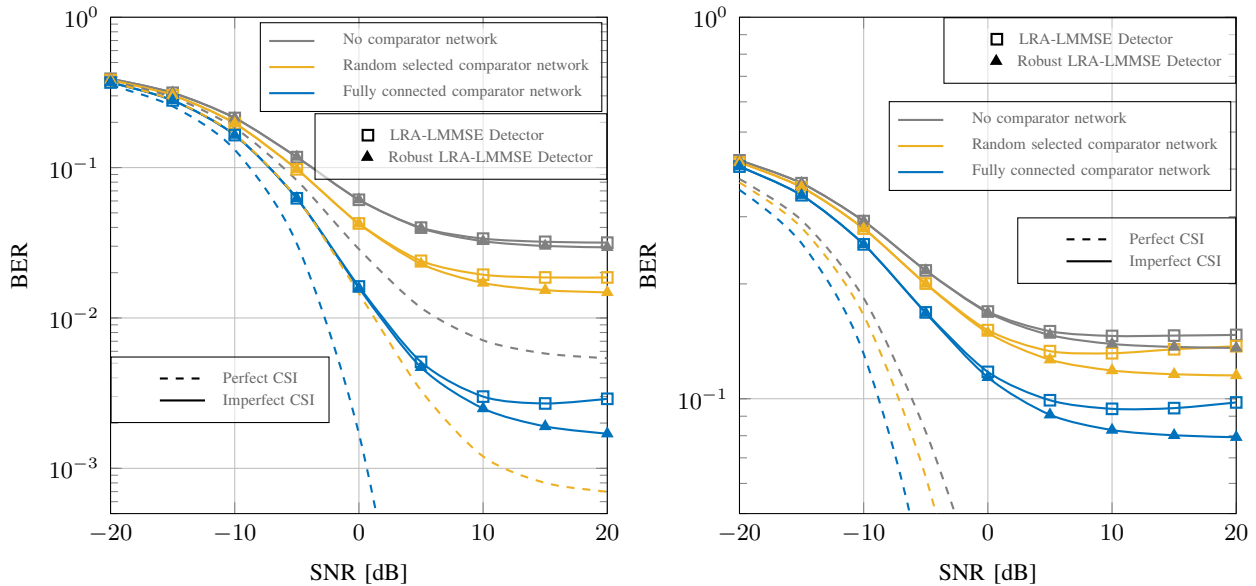}
\caption{Robust detectors with and without comparator network in $4\times16$ MIMO systems, $\lambda=0.8$ (left) and $\lambda=0.4$ (right) } 
\label{fig:BER_4x16_ROBUST_outdated_CSI}    
\end{figure}
\vspace{1em}
\subsubsection{Robust LRA-LMMSE Detector}
$\\$
In this subsection, it is considered a system with $N_r=16$. The BER performance plots are obtained by taking the average over $4000$ different channels and $100$ noise realizations per channel.
In these experiments, the partially connected networks have $\alpha_p=2N_r=32$ comparators, while the fully connected have $\alpha_f=\tbinom{2N_r}{2}=496$ comparators.



\textcolor{r3}{
In this experiment the receiver has imperfect CSI due to the model described in Sec.~\ref{sec:Robust_general}.
While the LRA-LMMSE detector relies only on the known part of the channel ${\mathbf{h}}_{\text{R},1}$, the robust LRA-LMMSE detector also takes into account the corresponding CSI mismatch statistics $ \mathbf{R_{h_{\text{R,2}}}}= E\left[\mathbf{h}_{\text{R,2}}\mathbf{h}_{\text{R,2}}^T\right]$.
For the experiment it is considered that 
$ \mathbf{R_{h_{\text{R,2}}}}=\mathbf{R_{h_{\text{R}}}}$.
The BER for the proposed robust LRA-LMMSE detector is simulated for a case with marginal CSI mismatch with $\lambda=0.8$, shown on the left hand side of Fig.~\ref{fig:BER_4x16_ROBUST_outdated_CSI} and for a case with significant CSI mismatch with $\lambda=0.4$, shown on the right hand side of Fig.~\ref{fig:BER_4x16_ROBUST_outdated_CSI}.
 Simulation results show that the robust LRA-LMMSE detector yields a significant performance advantage in comparison to the non-robust detector.
Moreover, we have noticed that for other CSI mismatch scenarios such as those described in Sec.~\ref{sec:Robust_2} the proposed robust detection also yields an improvement in BER. 
}
\vspace{1em}
\subsubsection{Sum Rate Analysis}
$\\$
In this subsection, it is considered a system with \textcolor{r4}{$N_t=3$} and different values for $N_r$. The sum rate results are obtained by taking the average over $2000$ different channels and $2000$ noise realizations per channel.
\begin{figure}[ht]
\centering
%
%
%
\usetikzlibrary{positioning,calc}

\definecolor{mycolor1}{rgb}{0.00000,1.00000,1.00000}%
\definecolor{mycolor2}{rgb}{1.00000,0.00000,1.00000}%

\definecolor{mustard}{rgb}{0.92941,0.69020,0.12941}%

\definecolor{newpurple}{rgb}{0.5, 0 ,1}%

\definecolor{darkblue}{rgb}{0, 0.4470, 0.7410}

\pgfplotsset{every axis label/.append style={font=\footnotesize},
every tick label/.append style={font=\footnotesize},
every plot/.append style={ultra thick} 
}

\begin{tikzpicture}[font=\footnotesize] 

\begin{axis}[%
name=mse,
width  = 0.8\columnwidth,
height = 0.5\columnwidth,
scale only axis,
xmin  = -20,
xmax  = 30,
xlabel= {SNR [dB]},
xmajorgrids,
ymin=0,
ymax=11,
ylabel={Sum Rate [bpcu]},
ymajorgrids,
legend entries={No comparator network,
                Random selected comparator network,
				Fully connected comparator network,	
				},
legend style={fill=white, fill opacity=0.6, draw opacity=1,
text opacity =1,at={(0.99,0.01)}, anchor= south east,draw=black,fill=white,legend cell align=left,font=\footnotesize}
]

\addlegendimage{smooth,color=gray,solid, thick, mark=none,
y filter/.code={\pgfmathparse{\pgfmathresult-0}\pgfmathresult}}
\addlegendimage{smooth,color=mustard,solid, thick, mark=none,
y filter/.code={\pgfmathparse{\pgfmathresult-0}\pgfmathresult}}
\addlegendimage{smooth,color=blue,solid, thick, mark=none,
y filter/.code={\pgfmathparse{\pgfmathresult-0}\pgfmathresult}}
\addlegendimage{smooth,color=newpurple,solid, thick, mark=none,
y filter/.code={\pgfmathparse{\pgfmathresult-0}\pgfmathresult}}

\addplot+[smooth,color=gray,solid,thick, every mark/.append style={solid} ,mark=o,
y filter/.code={\pgfmathparse{\pgfmathresult-0}\pgfmathresult}]
  table[row sep=crcr]{%
  -20.0000    0.0040\\
  -15.0000    0.0351\\
  -10.0000    0.2478\\
   -5.0000    1.0713\\
         0    2.4468\\
    5.0000    3.5004\\
   10.0000    3.9673\\
   15.0000    4.1127\\
   20.0000    4.1565\\
   25.0000    4.1552\\
   30.0000    4.1497\\
   };

\addplot+[smooth,color=mustard,solid, thick, every mark/.append style={solid} ,mark=o,
y filter/.code={\pgfmathparse{\pgfmathresult-0}\pgfmathresult}]
  table[row sep=crcr]{%
  -20.0000    0.0052\\
  -15.0000    0.0456\\
  -10.0000    0.3192\\
   -5.0000    1.3479\\
         0    3.0119\\
    5.0000    4.2880\\
   10.0000    4.8488\\
   15.0000    5.0352\\
   20.0000    5.0792\\
   25.0000    5.0845\\
   30.0000    5.0854\\
   };

\addplot+[smooth,color=darkblue,solid, thick, every mark/.append style={solid} ,mark=o,
y filter/.code={\pgfmathparse{\pgfmathresult-0}\pgfmathresult}]
  table[row sep=crcr]{%
  -20.0000    0.0076\\
  -15.0000    0.0673\\
  -10.0000    0.4651\\
   -5.0000    1.8992\\
         0    4.2040\\
    5.0000    6.1752\\
   10.0000    7.2419\\
   15.0000    7.6690\\
   20.0000    7.8042\\
   25.0000    7.8466\\
   30.0000    7.8589\\
};



\addplot+[smooth,color=gray,dashed,thick, every mark/.append style={solid} ,mark=triangle,
y filter/.code={\pgfmathparse{\pgfmathresult-0}\pgfmathresult}]
  table[row sep=crcr]{%
  -20.0000    0.0079\\
  -15.0000    0.0701\\
  -10.0000    0.4835\\
   -5.0000    1.9482\\
         0    4.0602\\
    5.0000    5.3857\\
   10.0000    5.8787\\
   15.0000    5.9728\\
   20.0000    5.9891\\
   25.0000    5.9860\\
   30.0000    5.9830\\
   };

\addplot+[smooth,color=mustard,dashed, thick, every mark/.append style={solid} ,mark=triangle,
y filter/.code={\pgfmathparse{\pgfmathresult-0}\pgfmathresult}]
  table[row sep=crcr]{%
  -20.0000    0.0103\\
  -15.0000    0.0906\\
  -10.0000    0.6175\\
   -5.0000    2.3970\\
         0    4.8189\\
    5.0000    6.3169\\
   10.0000    6.8281\\
   15.0000    6.9416\\
   20.0000    6.9617\\
   25.0000    6.9501\\
   30.0000    6.9487\\
   };

\addplot+[smooth,color=darkblue,dashed, thick, every mark/.append style={solid} ,mark=triangle,
y filter/.code={\pgfmathparse{\pgfmathresult-0}\pgfmathresult}]
  table[row sep=crcr]{%
  -20.0000    0.0165\\
  -15.0000    0.1447\\
  -10.0000    0.9567\\
   -5.0000    3.4765\\
         0    6.7537\\
    5.0000    8.8792\\
   10.0000    9.6481\\
   15.0000    9.8957\\
   20.0000    9.9600\\
   25.0000    9.9720\\
   30.0000    9.9725\\
};



\addplot[smooth,color=black, thick, every mark/.append style={solid} ,mark=o,
y filter/.code={\pgfmathparse{\pgfmathresult-0}\pgfmathresult}]
  table[row sep=crcr]{%
	-1 -2\\
};\label{P33}

\addplot[smooth,color=black,thick,dashed, every mark/.append style={solid} ,mark=triangle, 
y filter/.code={\pgfmathparse{\pgfmathresult-0}\pgfmathresult}]
  table[row sep=crcr]{%
	-1 -2\\
};\label{P34}

\node [draw,fill=white, fill opacity=0.6,draw opacity=1,
text opacity =1,at ={(-20,4.75)}, anchor= north west,draw=black,fill=white,font=\scriptsize]  {
\setlength{\tabcolsep}{0.5mm}
\renewcommand{\arraystretch}{.8}
\begin{tabular}{l}
\ref{P33}{\hspace{0.15cm} $N_r=16$}\\
\ref{P34}{\hspace{0.15cm} $N_r=8$}\\
\end{tabular}
};

\end{axis}
\end{tikzpicture}
\caption{Sum Rate comparisons of MIMO systems with and without comparator network, under imperfect CSI.} 
\label{fig:AR_2x4_Imperfect}    
\end{figure}
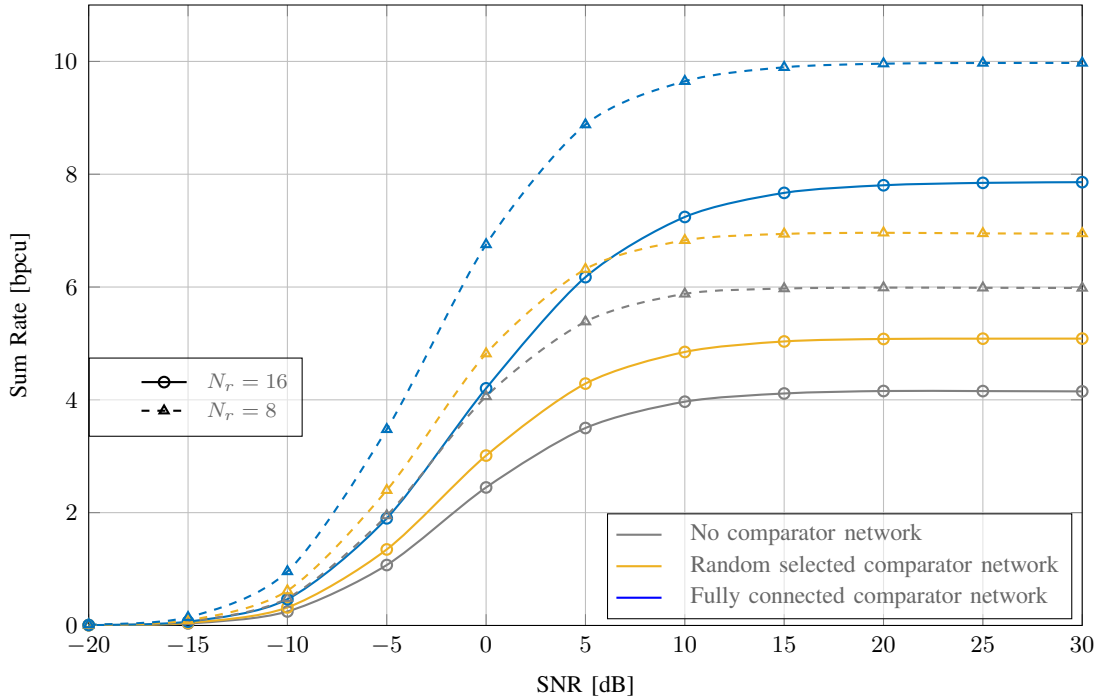
The sum rate versus SNR for the systems under imperfect CSI is shown in Fig. \ref{fig:AR_2x4_Imperfect}, which indicates a significant benefit for the system that utilizes the additional comparator networks. We can see that the system with fully connected network achieves the best sum rate performance, followed by the system with random selected inputs.
Note that the increased sum rate is not only due to the comparator network aided receive processing but also due to a more accurate channel estimation.
\section{Conclusions}
\label{sec:conclusion}

This study proposes a novel MIMO receiver architecture with the use of 1-bit quantization ADCs. Different from conventional systems, the proposed MIMO receiver includes a comparator network with binary outputs which can compare signals from different antennas. The resulting extension can be interpreted as a number of additional virtual channels. This additional virtual channels contain additional information which aid the detection and channel estimation processes with only a slight increase in hardware cost and computational complexity. In this context, channel estimation, signal detection and sum rate schemes are developed for the proposed system. 

Two types of comparator networks are proposed, fully and partially connected networks. Simulation results show that the proposed partially connected networks, especially the MMSE based greedy search approach
and Sequential SINR based Search, require less comparators while introducing only small performance degradation compared with the fully connected networks.

Simulation results show that
the proposed multiuser MIMO receiver architectures
in conjunction with the proposed channel estimation and detection algorithms
are superior to the conventional 1-bit quantization receiver methods in terms of BER and MSE. 
Moreover, numerical results show that adding virtual channels by using comparator networks can be better than adding extra physical channels which corresponds to additional receive antennas in terms of BER. 
\textcolor{r3}{
In this context, the proposed comparator networks are attractive for receivers with limited numbers of antennas due to physical space constraints.
Furthermore, a robust LRA-LMMSE detector has been developed, which takes into account the mismatch statistics of imperfect CSI at the receiver.}
Simulation results indicate the advantage in terms of BER performance and MSE in comparison to the non-robust detector. Moreover, by relying on linear receive processing and considering the proposed channel estimate and the corresponding estimation error, it has been derived an expression for lower bounding the ergodic sum rate. Numerical results show that the corresponding sum rate increases significantly when adding a comparator network to the system.

\appendix

\subsection{Derivation of the LRA-LMMSE Channel Estimator}
\label{subsec:derivation_of_W_LMMSE}
Recalling the optimization problem from \eqref{eq:WLMMSE} and
taking the partial derivative with respect to $\mathbf{W}^T$, we obtain
\begin{equation}
\frac{\partial{E\left[\left\vert\left\vert \mathbf{h}_{\text{R}}-\mathbf{W}\mathbf{z}_{\mathcal{Q}_p} \right\vert\right\vert^2_2\right]}}{\partial{\mathbf{W}^T}} = - E\left[ \mathbf{h}_{\text{R}}\mathbf{z}_{\mathcal{Q}_p}^T\right] + \mathbf{W}E\left[\mathbf{z}_{\mathcal{Q}_p}\mathbf{z}_{\mathcal{Q}_p}^T \right].
\label{eq:derived_WLMMSE}
\end{equation} 

Equating \eqref{eq:derived_WLMMSE} to zero and inserting \eqref{eq:zqp}, the LRA-LMMSE filter is
\begin{equation}
\begin{split}
\mathbf{W}_{\text{R, LRA-LMMSE}} &=
E\left[ \mathbf{h}_{\text{R}}\mathbf{z}_{\mathcal{Q}_p}^T\right]
E\left[\mathbf{z}_{\mathcal{Q}_p}\mathbf{z}_{\mathcal{Q}_p}^T \right]^{-1} = \mathbf{R_{h_{\text{R}}}}\hat{\mathbf{\Phi}}_{\text{R}}^T\mathbf{C}_{\mathbf{z}_{\mathcal{Q}_p}}^{-1},
\end{split}
\end{equation}
where it is considered that $\mathbf{h}_{\text{R}}$ is uncorrelated with $\tilde{\mathbf{n}}_{{\text{R}}_p}$.

\subsection{Derivation of the Ergodic Achievable Rate per Sub-Channel}
\label{subsec:derivation_of_Irk}
First, let us recall the received signal of the index $k$, with $k \in [1,2N_t]$, given by \eqref{eq:x_hat_k}
\begin{equation}
\begin{split}
\hat{\mathbf{x}}_{\text{R}_{k}} = \quad &\underbrace{\mathbf{g}_{\text{R}_{k}}^T \mathbf{A}_{{\text{R}}_d} \mathbf{B} \hat{\mathbf{h}}_{\text{R}_{k}}\mathbf{x}_{\text{R}_{k}}}_{\text{desired signal}} + \underbrace{\mathbf{g}_{\text{R}_{k}}^T\sum_{i \neq k}^{K} \mathbf{A}_{{\text{R}}_d} \mathbf{B} \hat{\mathbf{h}}_{\text{R}_i}\mathbf{x}_{\text{R}_i}}_{\text{interference}} + \underbrace{\mathbf{g}_{\text{R}_{k}}^T\sum_{i=1}^{K} \mathbf{A}_{{\text{R}}_d} \mathbf{B} \bm{\varepsilon}_{\text{R}_i}\mathbf{x}_{\text{R}_i}}_{\text{channel estimation error}} \\
&+ \underbrace{\mathbf{g}_{\text{R}_{k}}^T\mathbf{A}_{{\text{R}}_d} \mathbf{B}\mathbf{n}_{{\text{R}}_d}}_{\text{AWGN noise}} + \underbrace{\mathbf{g}_{\text{R}_{k}}^T\mathbf{n}_{{\text{R}}_{q, d}}}_{\text{quant. noise}},
\end{split}
\end{equation}
where $\mathbf{g}_{\text{R}_{k}}^T$ is the ${k}$th row of $\mathbf{G}_{\text{R}}$ and $\hat{\mathbf{h}}_{\text{R}_{k}}$ is the ${k}$th column of $\hat{\mathbf{H}}_{\text{R}}$. Moreover, $\bm{\varepsilon}_{\text{R}_{i}}$ is
the ${i}$th column of the matrix $\mathbfcal{E}_{\text{R}}$.
Assuming Gaussian signalling, the ergodic achievable rate per channel $k$, for the real-valued notation, is lower bounded by
\begin{equation}
I_{\text{R}_{k}} = E\left[ \frac{1}{2} \log_2 \left ( 1 + \text{SINR}_{k} \right)\right] \ \text{[bpcu]},
\label{eq:derivation_Irk}
\end{equation}
where
\begin{equation}
\text{SINR}_{k} = \frac{E\left[\vert T_{1} \vert^2\right]}{\sum_{i\neq k}^{K}E\left[\vert T_{2,i} \vert^2 \right] + \sum_{i=1}^{K}E\left[\vert T_{3,i} \vert^2 \right] + E\left[\vert T_{4} \vert^2 \right] + E\left[\vert T_{5} \vert^2 \right]}.
\label{eq:SINR_k}
\end{equation}
Note that this SINR definition is associated with the receiver output which should not be confused with the SINR expression in \eqref{SINR_virtual} that refers to the individual virtual subchannels of the comparator networks.
Then, the sum-rate is reads as
$
I_{\text{R}} = \sum_{k=1}^{K} I_{\text{R}_{k}} $.
In \eqref{eq:SINR_k}, the term
\begin{equation}
T_{1} = \mathbf{g}_{\text{R}_{k}}^T \mathbf{A}_{{\text{R}}_d} \mathbf{B} \hat{\mathbf{h}}_{\text{R}_{k}}\mathbf{x}_{\text{R}_{k}}
\end{equation}
represents the desired signal, while the parameter
\begin{equation}
T_{2,i} = \mathbf{g}_{\text{R}_{k}}^T\mathbf{A}_{{\text{R}}_d} \mathbf{B} \hat{\mathbf{h}}_{\text{R}_i}\mathbf{x}_{\text{R}_i} \text{, for} \enspace i \neq k \text{, } i=1\text{,}\hdots\text{,} K,
\end{equation}
is the interference caused by user $i$ to user $k$. The quantity
\begin{equation}
T_{3,i} = \mathbf{g}_{\text{R}_{k}}^T\mathbf{A}_{{\text{R}}_d} \mathbf{B} \bm{\varepsilon}_{\text{R}_i}\mathbf{x}_{\text{R}_i} \text{, for} \enspace i = k \text{, } i=1\text{,}\hdots\text{,} K,
\end{equation}
refers to the channel estimation error, whilst
\begin{equation}
T_{4} = \mathbf{g}_{\text{R}_{k}}^T\mathbf{A}_{{\text{R}}_d} \mathbf{B}\mathbf{n}_{{\text{R}}_d}
\end{equation}
corresponds to the AWGN noise and 
\begin{equation}
T_{5} = \mathbf{g}_{\text{R}_{k}}^T\mathbf{n}_{{\text{R}}_{q, d}}
\end{equation}
relates to the quantizer noise.
By considering
$E\left[\mathbf{x}_{\text{R}}\mathbf{x}_{\text{R}}^T\right] =\frac{\sigma_x^2}{2}\mathbf{I}_{2N_t}$
and
$\mathbf{C_{\mathbf{n}_{{\text{R}}_{d}}}} = E\left[\mathbf{n}_{{\text{R}}_{d}}\mathbf{n}_{{\text{R}}_{d}}^T\right] = \frac{\sigma_n^2}{2}\mathbf{I}_{2N_r}$ and $\mathbf{d}_{\text{R}_{k}} =\mathbf{g}_{\text{R}_{k}}^T\mathbf{A}_{{\text{R}}_d} \mathbf{B}$
 and
$E\left[\vert T_{5} \vert^2\right] = \mathbf{g}_{\text{R}_{k}}^T\mathbf{C}_{\mathbf{n}_{{\text{R}}_{q, d}}}\mathbf{g}_{\text{R}_{k}}$
 we get the following expression
\begin{equation}
\text{SINR}_{k} = \frac{  \sigma_x^2 \left\vert\mathbf{d}_{\text{R}_{k}} \hat{\mathbf{h}}_{\text{R}_{k}}\right\vert^2}{  \sigma_x^2 \sum_{i\neq k}^{K}\left\vert\mathbf{d}_{\text{R}_{k}}\hat{\mathbf{h}}_{\text{R}_i} \right\vert^2 + \sigma_x^2 \sum_{i=1}^{K}\left\vert\mathbf{d}_{\text{R}_{k}}\bm{\varepsilon}_{\text{R}_i}\right\vert^2 + \sigma_n^2\left\vert\left\vert\mathbf{d}_{\text{R}_{k}}\right\vert\right\vert^2_2 + 2\mathbf{g}_{\text{R}_{k}}^T\mathbf{C}_{\mathbf{n}_{{\text{R}}_{q, d}}}\mathbf{g}_{\text{R}_{k}}}.
\label{eq:SINRk}
\end{equation}
Finally, inserting \eqref{eq:SINRk} into \eqref{eq:derivation_Irk} yields the ergodic achievable rate per real-valued channel \eqref{eq:I_tilde_k}.

\textcolor{blue}{
\subsection{Sum Rate for Matched Filter Receiver}
The corresponding matched filter for the $k$th substream for the system with comparator network can be expressed as  
$\mathbf{g}_{\text{R}_{k}}^T=\hat{\mathbf{h}}_{\text{R}_{k}}^T \mathbf{B}^T$. For description of the equivalent linear model the matrix
$\mathbf{A}_{{\text{R}}_d}=\sqrt{\frac{2}{\pi}} \mathrm{diag}(\mathbf{C}_{\mathbf{z}_{{\text{R}}_d}})^{-\frac{1}{2}}$ is given by
\begin{align}
\mathrm{E}[\mathbf{A}_{{\text{R}}_d}]=
\sqrt{\frac{2}{\pi}} \mathrm{diag}  \left(   \mathrm{E}\left[
\frac{\sigma_x^2}{2}\mathbf{BH}_{\text{R}}{\mathbf{H}_{\text{R}}^T}\mathbf{B}^T+\frac{\sigma_n^2}{2}\mathbf{BB}^T \right]
\right)  ^{-\frac{1}{2}}
\end{align}
Considering a generic channel with $\mathrm{E}[\mathbf{H}_{\text{R}}{\mathbf{H}_{\text{R}}^T}]= N_t \mathbf{I} $
\begin{align}
\mathrm{E}[\mathbf{A}_{{\text{R}}_d}]=
\sqrt{\frac{2}{\pi} \frac{2}{N_t \sigma_x^2+\sigma_n^2}}
\mathbf{I}_{2 N_r +\alpha} \text{.}
\label{eq:Bussgang_Operator_data}
\end{align}
In the following, the prefactor in \eqref{eq:Bussgang_Operator_data} is termed $\zeta$.
The received signal after filtering can be expressed alternatively as
\begin{align}
\hat{\mathbf{x}}_{\text{R}_{k}} = \quad & \mathrm{E} [  \mathbf{g}_{\text{R}_{k}}^T \mathbf{A}_{{\text{R}}_d} \mathbf{B} {\mathbf{h}}_{\text{R}_{k}}]\mathbf{x}_{\text{R}_{k}} + n_{\text{eff},k} \text{,}
\end{align}
where $n_{\text{eff},k}$ represents some effective noise term.
For the matched filtering approach under the mentioned assumptions it holds
\begin{align}
\mathrm{E} [  \mathbf{g}_{\text{R}_{k}}^T \mathbf{A}_{{\text{R}}_d} \mathbf{B} {\mathbf{h}}_{\text{R}_{k}}] = 
\zeta
\mathrm{E} [  \hat{\mathbf{h}}_{\text{R}_{k}}^T \mathbf{B}^T
\mathbf{B}  \hat{\mathbf{h}}_{\text{R}_{k}}
]
\text{.}
\end{align}
In the next step it is considered that $\mathbf{B}^T
\mathbf{B}$ can be approximated by $(1+\frac{\alpha}{2 N_r}) \mathbf{I}$
\begin{align}
\mathrm{E} [  \mathbf{g}_{\text{R}_{k}}^T \mathbf{A}_{{\text{R}}_d} \mathbf{B} {\mathbf{h}}_{\text{R}_{k}}] = 
\zeta
\left(1+\frac{\alpha}{2 N_r}\right)
 \mathrm{trace} [  \mathrm{E} [   \hat{\mathbf{h}}_{\text{R}_{k}}
\hat{\mathbf{h}}_{\text{R}_{k}}^T  ]  ]
\text{.}
\end{align}
The correlation matrix of the channel estimate is given by 
\begin{align}
  E\left[    \hat{\mathbf{h}}_{\text{R, LRA-MMSE}}
\hat{\mathbf{h}}_{\text{R, LRA-MMSE}}^T  \right] 
= 
\frac{1}{2}
\frac{N_t \sigma_x^2}{N_t \sigma_x^2+\sigma_n^2}
\left(
1-\frac{\pi-2}{2(1+\frac{\alpha}{2 N_r}) +\pi-2 }
\right) \mathbf{I}_{2N_rN_t}  \text{,}
\end{align}
where we term the prefactor $\kappa$.
With this, it follows
\begin{align}
\mathrm{E} [  \mathbf{g}_{\text{R}_{k}}^T \mathbf{A}_{{\text{R}}_d} \mathbf{B} {\mathbf{h}}_{\text{R}_{k}}] = 
\zeta
\left(1+\frac{\alpha}{2 N_r}\right)
2N_r \kappa
\label{eq:MF_exp_value}
\text{.}
\end{align}
The corresponding variance is given by
\begin{align}
\mathrm{Var} [  \mathbf{g}_{\text{R}_{k}}^T \mathbf{A}_{{\text{R}}_d} \mathbf{B} {\mathbf{h}}_{\text{R}_{k}}]=
\mathrm{E} [  \mathbf{g}_{\text{R}_{k}}^T \mathbf{A}_{{\text{R}}_d} \mathbf{B} {\mathbf{h}}_{\text{R}_{k}}
{\mathbf{h}}_{\text{R}_{k}}^T
\mathbf{B}^T
\mathbf{A}_{{\text{R}}_d}^T
\mathbf{g}_{\text{R}_{k}}
]-\mathrm{E} [  \mathbf{g}_{\text{R}_{k}}^T \mathbf{A}_{{\text{R}}_d} \mathbf{B} {\mathbf{h}}_{\text{R}_{k}}]^2 \text{.}
\end{align}
By considering $\mathbf{B}^T
\mathbf{B}=(1+\frac{\alpha}{2N_r}) \mathbf{I} $ the variance can be rewritten as
\begin{align}
\mathrm{Var} [  \mathbf{g}_{\text{R}_{k}}^T \mathbf{A}_{{\text{R}}_d} \mathbf{B} {\mathbf{h}}_{\text{R}_{k}}]=
\zeta^2
\left(   1+\frac{\alpha}{2 N_r}  \right)^2
\mathrm{E} [    | \hat{\mathbf{h}}_{\text{R}_{k}}^T  {\mathbf{h}}_{\text{R}_{k}}   |^2
]-\mathrm{E} [  \mathbf{g}_{\text{R}_{k}}^T \mathbf{A}_{{\text{R}}_d} \mathbf{B} {\mathbf{h}}_{\text{R}_{k}}]^2  \text{,}
\end{align}
which can be expressed as
\begin{align}
& \mathrm{Var} [  \mathbf{g}_{\text{R}_{k}}^T \mathbf{A}_{{\text{R}}_d} \mathbf{B} {\mathbf{h}}_{\text{R}_{k}}]=  \\
& \zeta^2
\left(   1+\frac{\alpha}{2 N_r}  \right)^2
\left(
\mathrm{E} [     \hat{\mathbf{h}}_{\text{R}_{k}}^T  {\mathbf{h}}_{\text{R}_{k}}   
]^2
+
\mathrm{Var} [     \hat{\mathbf{h}}_{\text{R}_{k}}^T  {\mathbf{h}}_{\text{R}_{k}}  
]
\right)
-\mathrm{E} [  \mathbf{g}_{\text{R}_{k}}^T \mathbf{A}_{{\text{R}}_d} \mathbf{B} {\mathbf{h}}_{\text{R}_{k}}]^2  \text{.} \notag
\end{align}
By considering $\mathrm{E} [     \hat{\mathbf{h}}_{\text{R}_{k}}^T  {\mathbf{h}}_{\text{R}_{k}}   
] =  \mathrm{E} [     \hat{\mathbf{h}}_{\text{R}_{k}}^T   \hat{{\mathbf{h}}}_{\text{R}_{k}}   
]   $ and \eqref{eq:MF_exp_value} the variance reduces to
\begin{align}
\mathrm{Var} [  \mathbf{g}_{\text{R}_{k}}^T \mathbf{A}_{{\text{R}}_d} \mathbf{B} {\mathbf{h}}_{\text{R}_{k}}]=  
 \zeta^2
\left(   1+\frac{\alpha}{2 N_r}  \right)^2
\left(
\mathrm{Var} [     \hat{\mathbf{h}}_{\text{R}_{k}}^T  \hat{\mathbf{h}}_{\text{R}_{k}}  
]
+
\mathrm{Var} [     \hat{\mathbf{h}}_{\text{R}_{k}}^T  {\mathbf{\epsilon}}_{\text{R}_{k}}   
]
\right)
\end{align}
Considering that the entries in $\hat{\mathbf{h}}_{\text{R}_{k}} $ are Gaussian distributed, it follows that $\|  \hat{\mathbf{h}}_{\text{R}_{k}} \|^2$ is chi squared distributed with $2N_r$ degrees of freedom and $ \mathrm{E}[ \|  \hat{\mathbf{h}}_{\text{R}_{k}} \|^2]= 2 N_r \kappa$ and $ \mathrm{Var}[ \|  \hat{\mathbf{h}}_{\text{R}_{k}} \|^2]= 4 N_r \kappa^2$. 
Considering that the entries in
${\mathbf{\epsilon}}_{\text{R}_{k}}$ are uncorrelated with the channel estimate and Gaussian distributed, 
the variance is given by
\begin{align}
\mathrm{Var} [  \mathbf{g}_{\text{R}_{k}}^T \mathbf{A}_{{\text{R}}_d} \mathbf{B} {\mathbf{h}}_{\text{R}_{k}}]=  
 \zeta^2
\left(   1+\frac{\alpha}{2 N_r}  \right)^2
\left(
4 N_r \kappa^2
+
N_r \kappa(1-2\kappa)
\right)
\end{align}
This finally yields
\begin{align}
\mathrm{Var}[  \mathbf{g}_{\text{R}_{k}}^T \mathbf{A}_{{\text{R}}_d} \mathbf{B} {\mathbf{h}}_{\text{R}_{k}}]=
\zeta^2
\left(   1+\frac{\alpha}{2 N_r}  \right)^2
N_r \kappa    (2 \kappa  +1     )   \text{.}
\end{align}
The interference variance from the $i$th signal stream can be expressed as
\begin{align}
\mathrm{E}[ |
\mathbf{g}_{\text{R}_{k}}^T\mathbf{A}_{{\text{R}}_d} \mathbf{B} {\mathbf{h}}_{\text{R}_i}\mathbf{x}_{\text{R}_i}   |^2 ] =
\frac{\sigma_x^2}{2}
\zeta^2
\mathrm{E}[ |
\hat{\mathbf{h}}_{\text{R}_{k}}^T      \mathbf{B}^T\mathbf{B} {\mathbf{h}}_{\text{R}_i}   |^2 ] 
 \text{.}
\end{align}
Considering that $\hat{\mathbf{h}}_{\text{R}_{k}}$ and ${\mathbf{h}}_{\text{R}_i}$ are independent yields
\begin{align}
\mathrm{E}[ |
\mathbf{g}_{\text{R}_{k}}^T\mathbf{A}_{{\text{R}}_d} \mathbf{B} {\mathbf{h}}_{\text{R}_i}\mathbf{x}_{\text{R}_i}   |^2 ] =
\frac{\sigma_x^2}{4}
\zeta^2
\mathrm{E}[ |
\hat{\mathbf{h}}_{\text{R}_{k}}^T      \mathbf{B}^T\mathbf{B}   |^2 ] 
 \text{.}
\end{align}
By assuming that the entries in $\mathbf{B}'$ are binary distributed independent variables, the following approximation holds $E[ \mathbf{B}^T  \mathbf{B}
\mathbf{B}^T \mathbf{B}
   ]= E[\mathbf{B}^T
\mathbf{B}]^2+  \mathrm{Cov}[\mathbf{B}^T \mathbf{B}]  \approx  (1+\frac{\alpha}{2N_r})^2  \mathrm{I}  +\frac{\alpha}{4 N_r^2}(N_r-1) \mathrm{I}=\delta_B \mathrm{I}$.
With this, the interference variance reads as
\begin{align}
\mathrm{E}[ |
\mathbf{g}_{\text{R}_{k}}^T\mathbf{A}_{{\text{R}}_d} \mathbf{B} {\mathbf{h}}_{\text{R}_i}\mathbf{x}_{\text{R}_i}   |^2 ] =
\frac{\sigma_x^2}{2}
\zeta^2 \delta_B
N_r \kappa  
 \text{.}
\end{align}
The correlation matrix of the quantization noise is described by
\begin{align}
\mathbf{C}_{\mathbf{n}_{{\text{R}}_{q, d}}} = 
\frac{  \pi-2 }{\pi }
\left( \frac{2}{\pi-2}
\mathbf{B}
\mathbf{B}^T+  \mathbf{I}
\right)
- \zeta^2
\frac{ N_t \sigma_x^2+\sigma_n^2}{2}
\mathbf{B}
\mathbf{B} ^T  \text{.}
\end{align}
This can be expressed as
\begin{align}
\mathbf{C}_{\mathbf{n}_{{\text{R}}_{q, d}}} = 
\frac{  \pi-2 }{\pi }
\mathbf{I}
+
\left(
\frac{2}{\pi}
- \zeta^2
\frac{ N_t \sigma_x^2+\sigma_n^2}{2}   \right)
\mathbf{B}
\mathbf{B} ^T  \text{.}
\end{align}
The quantization noise variance is determined by
\begin{align}
\mathrm{E} [\mathbf{g}_{\text{R}_{k}}^T
\mathbf{C}_{\mathbf{n}_{{\text{R}}_{q, d}}} \mathbf{g}_{\text{R}_{k}}
]
= 
\frac{  \pi-2 }{\pi }
\mathrm{E} [\mathbf{g}_{\text{R}_{k}}^T
 \mathbf{g}_{\text{R}_{k}}
]
+
\left(
\frac{2}{\pi}
- \zeta^2
\frac{ N_t \sigma_x^2+\sigma_n^2}{2}   \right)
\mathrm{E} [\mathbf{g}_{\text{R}_{k}}^T \mathbf{B}
\mathbf{B}^T
 \mathbf{g}_{\text{R}_{k}}
]
  \text{.}
\end{align}
This can be expressed as
\begin{align}
\mathrm{E} [\mathbf{g}_{\text{R}_{k}}^T
\mathbf{C}_{\mathbf{n}_{{\text{R}}_{q, d}}} \mathbf{g}_{\text{R}_{k}}
]
= 
\frac{  \pi-2 }{\pi }  \left(1+\frac{\alpha}{2N_r}  \right)2N_r \kappa   
+ \delta_B
\left(
\frac{2}{\pi}
- \zeta^2
\frac{ N_t \sigma_x^2+\sigma_n^2}{2}   \right)
2N_r \kappa
  \text{.}
\end{align}
The variance of the thermal noise is given by
\begin{align}
    \mathrm{E} [  \mathbf{g}_{\text{R}_{k}}^T   \mathbf{A}_{{\text{R}}_d} \mathbf{B}
\mathbf{C}_{\mathbf{n}_{{\text{R}}_{d}}}   \mathbf{B}^T \mathbf{A}_{{\text{R}}_d}^T \mathbf{g}_{\text{R}_{k}}   ]  =  \frac{\sigma_n^2}{2} \zeta^2  \delta_B 2N_r \kappa
   \text{.}
\end{align}
The resulting rate per real valued substream is given by
\begin{equation}
\scalebox{0.99}{$
I_{\text{R}_{k}} =  
\frac{1}{2} \log_2 \left ( 1 +  \frac{   \frac{ \sigma_x^2}{2} \zeta^2 \left(1+\frac{\alpha}{2 N_r}\right)^2 4N_r^2 \kappa^2}
{ \frac{\sigma_x^2}{2} \zeta^2
\delta_B
 N_r  \kappa (K-1)
+  \frac{ \sigma_x^2}{2} 
\zeta^2
\left(   1+\frac{\alpha}{2 N_r}  \right)^2
N_r \kappa    (2 \kappa  +1     )
+
\frac{\sigma_n^2}{2} \zeta^2  \delta_B 2N_r \kappa
+
\frac{  \pi-2 }{\pi }  (1+\frac{\alpha}{2N_r})2N_r \kappa 
+  \delta_B
\left(
\frac{2}{\pi}
- \zeta^2
\frac{ N_t \sigma_x^2+\sigma_n^2}{2}   \right)
2N_r \kappa}
\right)   
$}    \text{.}   \notag 
\label{eq:I_analytical}
\end{equation}
This can be written more compact as
\begin{equation}
\scalebox{0.99}{$
I_{\text{R}_{k}} = 
\frac{1}{2} \log_2 \left ( 1 +  \frac{  \sigma_x^2 
\left(1+\frac{\alpha}{2 N_r}\right)^2
2N_r \kappa
}{ \frac{\sigma_x^2}{2}
\delta_B
    (K-1)
+  \frac{ \sigma_x^2}{2} 
\left(   1+\frac{\alpha}{2 N_r}  \right)^2
     (2 \kappa  +1     )
+
\sigma_n^2 \delta_B 
+
2\frac{  \pi-2 }{\pi } \zeta^{-2} (1+\frac{\alpha}{2N_r})    
+ 2 \zeta^{-2} \delta_B
\left(
\frac{2}{\pi}
- \zeta^2
\frac{ N_t \sigma_x^2+\sigma_n^2}{2}   \right)
 }
\right)
$}  \ \text{[bpcu]}   \text{.}
\label{eq:I_analytical_short}
\end{equation}
For the case of perfect CSI it holds $\kappa=\frac{1}{2}$ and the variance term is equal to zero, which yields
\begin{equation}
\scalebox{0.99}{$
I_{\text{R}_{k}} = 
\frac{1}{2} \log_2 \left ( 1 +  \frac{  \sigma_x^2 
\left(1+\frac{\alpha}{2 N_r}\right)^2
N_r 
}{ \frac{\sigma_x^2}{2}
\delta_B
    (K-1)
+
\sigma_n^2 \delta_B 
+
2\frac{  \pi-2 }{\pi } \zeta^{-2} (1+\frac{\alpha}{2N_r})    
+2 \zeta^{-2} \delta_B
\left(
\frac{2}{\pi}
-  \zeta^2
\frac{ N_t \sigma_x^2+\sigma_n^2}{2}   \right)
 }
\right)
$}  \ \text{[bpcu]}   \text{.} \label{eq:I_analytical_short_perfect CSI}
\end{equation}
}

\ifCLASSOPTIONcaptionsoff
  \newpage
\fi

\bibliographystyle{IEEEtran}
\bibliography{bib-refs}
\end{document}